\documentclass[a4paper,fleqn,usenatbib]{mnras}
\usepackage{mathptmx}
\usepackage[T1]{fontenc}
\usepackage{ae,aecompl}
\usepackage{times}
\usepackage{amssymb}
\usepackage{amsmath}
\usepackage{upgreek}
\usepackage{color}
\usepackage{graphicx}
\usepackage{pdflscape}

\newcommand{\ppmap}{{\sc ppmap}}

\title[Characteristic scale of star formation I.]{Characteristic scale of star formation. I. Clump formation efficiency on local scales}
\author[D.J. Eden et al.]{D.J. Eden,$^{1}$\thanks{E-mail: D.J. Eden@ljmu.ac.uk} T.J.T. Moore,$^{1}$ R. Plume,$^{2}$ A.J. Rigby,$^{3}$ J.S. Urquhart,$^{4}$  K.A. Marsh,$^{3,5}$ \newauthor C.H. Pe\~{n}aloza,$^{3,6}$ P.C. Clark,$^{3}$ M.W.L. Smith,$^{3}$ K. Tahani,$^{7}$ S.E. Ragan,$^{3}$ \newauthor M.A. Thompson,$^{8}$ D. Johnstone,$^{9,10}$ H. Parsons,$^{9}$ R. Rani,$^{1}$ \\
$^{1}$Astrophysics Research Institute, Liverpool John Moores University, IC2, Liverpool Science Park, 146 Brownlow Hill, Liverpool, L3 5RF, UK\\
$^{2}$Department of Physics and Astronomy, University of Calgary, 2500 University Drive NW, Calgary, Alberta T2N 1N4, Canada\\
$^{3}$School of Physics and Astronomy, Cardiff University, Cardiff CF24 3AA, UK\\
$^{4}$School of Physical Sciences, Ingram Building, University of Kent, Canterbury, Kent CT2 7NH, UK\\
$^{5}$Infrared Processing and Analysis Center, California Institute of Technology, Pasadena, California 91125, USA\\
$^{6}$SUPA, School of Physics and Astronomy, University of St. Andrews, North Haugh, St. Andrews KY16 9SS, UK\\
$^{7}$Department of Physics \& Astronomy, Kwantlen Polytechnic University, 12666 72nd Avenue, Surrey, BC, V3W 2M8, Canada\\
$^{8}$Centre for Astrophysics Research, School of Physics Astronomy \& Mathematics, University of Hertfordshire, College Lane, Hatfield, Herts AL10 9AB, UK\\
$^{9}$NRC Herzberg Astronomy and Astrophysics Centre, 5071 West Saanich Road, Victoria, BC V9E 2E7, Canada\\
$^{10}$Department of Physics and Astronomy, University of Victoria, Victoria, BC V8P 5C2, Canada\\
$^{11}$East Asian Observatory, 660 North A'ohoku Place, Hilo, Hawaii 96720, USA}

\date{Accepted XXX. Received YYY; in original form ZZZ}

\pubyear{2020}

\begin{document}
\label{firstpage}
\pagerange{\pageref{firstpage}--\pageref{lastpage}}
\maketitle

\begin{abstract}

We have used the ratio of column densities (CDR) derived independently from the 850-$\upmu$m continuum JCMT Plane Survey (JPS) and the  $^{13}$CO/C$^{18}$O $(J=3\rightarrow2)$ Heterodyne Inner Milky Way Plane Survey (CHIMPS) to produce maps of the dense-gas mass fraction (DGMF) in two slices of the Galactic Plane centred at $\ell$\,=\,30$\degr$ and $\ell$\,=\,40$\degr$. The observed DGMF is a metric for the instantaneous clump-formation efficiency (CFE) in the molecular gas. We split the two fields into velocity components corresponding to the spiral arms that cross them, and a two-dimensional power-spectrum analysis of the spiral arm DGMF maps reveals a break in slope at the approximate size scale of molecular clouds.  We interpret this as the characteristic scale of the amplitude of variations in the CFE and a constraint on the dominant mechanism regulating the CFE and, hence, the star-formation efficiency in CO-traced clouds.

\end{abstract}

\begin{keywords}

Stars: formation -- ISM: individual objects: W43 -- ISM: kinematics and dynamics -- submillimetre: ISM

\end{keywords}

\section{Introduction}

Star formation occurs in the densest regions of molecular clouds. These dense regions, especially at the distances associated with the Galactic Plane, are known as clumps, with typical radii and masses of 1.25\,pc and 1500\,M$_{\sun}$, respectively \citep[e.g.,][]{Urquhart15,Urquhart18}. With the star formation occurring within these structures, it is crucial to understand how they form in and from the more diffuse molecular gas or, at least, the efficiency of the process, and how that efficiency varies with large-scale and local environment.  This efficiency contributes to the global star-formation efficiency, the conversion of gas into stars, along with that of the formation of molecular clouds from neutral gas in the interstellar medium and stars from clumps.  The Schmidt-Kennicutt relation finds a linear relationship between the star-formation rate and the gas surface density \citep{Kennicutt98}; however, the dense gas is a crucial component of the star-formation process. The dense-gas abundance correlates with the star-formation rate \citep[e.g.][]{Gao04,Lada12} and linear correlations between massive-star tracers and molecular-gas tracers (e.g. $L_{FIR}-L_{CO}$) imply that dense-gas mass fractions (DGMF) are constant on average across all extragalactic systems \citep{Greve14}.

The dense clumps form due to supersonic turbulence within molecular clouds. This turbulence fragments the clouds into clumps \citep{Padoan02}. The distribution of clump masses is determined by the velocity power spectrum, with different forms of collapse or turbulent support giving different clump mass functions \citep{Klessen00,Klessen00a}. However, the turbulence probability distribution is intermittent, therefore the efficiency of clump formation is naturally limited \citep{Padoan02}.

The DGMF is computed by measuring the amount of the dense gas with respect to that of the more diffuse molecular gas. Different methods include measuring dense-gas molecular tracers such as HCN \citep{Wu05} or the sub-millimetre dust continuum compared to the molecular component, such as the $J=1\rightarrow0$ transition of CO \citep{Eden12,Eden13,Battisti14,Csengeri16}.

The clump formation efficiency (CFE) is considered to be analogous to the DGMF and is inferred from the following equation:

\begin{equation}
{\rm CFE} = \frac{1}{M_{\rm cloud}} \int^t_0 \frac{dM_{\rm dense}}{dt}\,dt
\label{equation}
\end{equation}

\noindent where $dM_{\rm dense}/dt$ is the instantaneous dense gas/clump formation rate. Therefore an elevated CFE either indicates a long time scale for clump formation (and that the cloud lasts longer than the clumps within it and continues to form clumps) or a high formation rate. However, the observed timescale for clump formation is found to be very short, a few $\times 10^5$\,yr \citep{Mottram11,Ginsburg12}, and the lack of starless clumps in the Galaxy rules out a long timescale \citep{Ginsburg12}, with only 12 per cent of ATLASGAL clumps found to be quiescent, a ratio that decreases with clump mass \citep{Urquhart18}.

In the Milky Way, where individual clumps can be studied, recent progress has found that, on kiloparsec scales, there is very little variation in the CFE or DGMF. On these scales, the mean value of CFE/DGMF is found to be $\sim$ 8 per cent \citep{NguyenLuong11,Eden12,Eden13,Battisti14}, consistent with the low efficiency found in simulations \citep{Padoan02}. However on much smaller scales there are CFE variations of more than two orders of magnitude with a lognormal distribution. The smaller scales correspond to the size and separation of molecular clouds \citep{Eden12}, or of molecular clumps in the case of star-formation efficiency traced by infrared emission \citep{Urquhart14,Eden15,Elia17}. These results imply that it is the conditions within individual molecular clouds and clumps that are most important in regulating the star-formation efficiency. The internal physics within molecular clouds may determine the form of the mass function of the dense, star-forming clumps within clouds \citep[e.g][]{Klessen07,Urban10}. It would follow, therefore, that internal cloud conditions would also determine the amount of gas converted to dense clumps.

The aims of this project are to map an analogue of the line-of-sight DGMF or CFE across a significant portion of the inner plane of the Milky Way, using column-density maps from the JCMT Plane Survey (JPS: \citealt{Moore15}; \citealt{Eden17}) and the $^{13}$CO/C$^{18}$O $(J=3\rightarrow2)$ Heterodyne Inner Milky Way Plane Survey (CHIMPS; \citealt{Rigby16}). These data will be used to determine the dominant or characteristic spatial scale of CFE variations, thereby constraining the primary regulating mechanism. In order to do this, we first need to establish our method of estimating changes in DGMF/CFE from the column-density ratio (CDR), using the sub-mm continuum as a tracer for dense clumps and CO $J=3\rightarrow2$ more strongly tracing the ambient molecular gas. Having thus spatially sampled the molecular gas, we then use a 2D power-spectrum analysis to identify the characteristic scale on which the ratio of these two quantities varies. Previous studies have applied this method to investigate the dynamics of the interstellar medium. These studies use different tracers over different size scales, from Galactic \ion{H}{i} \citep[e.g.][]{Crovisier83,Green93}, Galactic molecular gas \citep[e.g.][]{Pingel18,Feddersen19}, and Galactic dust \citep[e.g.][]{Schlegel98} to \ion{H}{i}, dust and star-formation maps in extragalactic systems \citep[e.g.][]{Goldman00,Stanimirovic00,Elmegreen03,Combes12}.

The paper is organised as follows: the data used is presented in Section 2, with the observed column density calculations in Section 3, and the methods used discussed in Section 4. The simulated CDR maps, and the discussion of them, are presented in Section 5 with the observed DGMF results in Section 6. A power-spectrum analysis of the observed maps is produced in Section 7, with a discussion of the results in Section 8 and Summary and Conclusions presented in Section 9.

\section{Data}
\label{data}

The two areas studied are the $\ell$\,=\,30$\degr$ and $\ell$\,=\,40$\degr$ fields of the James Clerk Maxwell Telescope (JCMT) Plane Survey (JPS; \citealt{Moore15,Eden17}). These two fields form one third of the JPS, which mapped 850-$\upmu$m continuum emission at 14.5-arcsec resolution, with a pixel-to-pixel rms noise of 29.89 and 27.89\,mJy\,beam$^{-1}$ for the $\ell$\,=\,30$\degr$ and 40$\degr$ fields, respectively. This corresponds to a mass sensitivity of $\sim$100\,M$_{\sun}$ at a distance of 20\,kpc. The JPS data do not trace structure on large scales, due to the observing method, and are effectively subject to a high-pass spatial filter.

The $\ell$\,=\,30$^{\circ}$ and $\ell$\,=\,40$^{\circ}$ fields are the only sections of the JPS that lie within the longitude limits of CHIMPS (\citealt{Rigby16}) and so are the only fields used in this study. CHIMPS covers 18 square degrees in the longitude range $\ell = 28\degr - 46\degr$. The latitude coverage is $|\,\emph{b}\,|$\,$\leq$\,0$\fdg$5. The CHIMPS data have 15-arcsec angular resolution, matching that of the JPS, a spectral resolution of 0.5\,km\,s$^{-1}$, and a median rms of 0.6\,K at these resolutions.

The $\ell$\,=\,30$\degr$ region contains a significant star-forming region in W43, which is at a key location in the Galaxy at the near end of the Long Bar \citep{NguyenLuong11}. The $\ell$\,=\,40$\degr$ field  has multiple spiral arms running across it but is away from the confusion of the end of the bar.

\section{Column Density Determinations}

\subsection{JPS}

The JPS column-density maps were produced using temperatures derived from an adapted version of the \ppmap\ method \citep{Marsh15}, a \emph{point process} that was applied to the entire \emph{Herschel} Galactic Plane Survey (Hi-GAL) data set in the 70--500\,$\upmu$m wavelengths \citep{Molinari10,Molinari16,Marsh17}. A full description of the \ppmap\ method can be found in \citet{Marsh15} and the adaptation to include the JPS 850-$\upmu$m data will be described in a later paper (Eden et al., in preparation).

\ppmap\ is a Bayesian procedure designed for estimating  column densities of diffuse dusty structures in multi-wavelength continuum data, a key feature being that it predicts line-of-sight variations in dust temperature, $T$, and opacity index, $\beta$. It does this by regarding $T$ and $\beta$ as extra dimensions of the mapping problem in addition to the usual 2D angular coordinates (Galactic longitude and latitude, $\ell,b$, for example). The original version of the algorithm \citep{Marsh15} yielded 3D image cubes of differential column density as a function of $\ell$, $b$, and $T$, but we now include $\beta$ as an additional variable \citep{Marsh18}, enabling the generation of 4D hypercubes ($\ell, b, T, \beta$). This is in contrast to conventional techniques which typically generate 2D maps of column density and temperature, assuming that $T$ and $\beta$ are constant everywhere along the line of sight \citep[see, for example,][]{Konyves10,Peretto10,Bernard10}. 

The key inputs to \ppmap\ are the observed images at a set of different wavelengths, the corresponding point-spread functions (PSFs; from \citealt{Poglitsch10,Holland13,Griffin13}), the measurement noise values, and a ``dilution" parameter, $\eta$, whose purpose is essentially to produce the simplest image that fits all of the data. \ppmap\ uses an iterative technique, based on the Point Process formalism \citep{Marsh15}, to generate a density function representing the expectation value of differential column density, starting from an initially smooth distribution. The outputs include a 4D hypercube of differential column density, a corresponding hypercube of uncertainty values, a 2D map of integrated line-of-sight column density, and 2D maps of density-weighted mean line-of-sight temperature and opacity index. In addition to being able to dispense with the ``constant line-of-sight $T$ and $\beta$'' assumptions, \ppmap\ has the advantage that it is not necessary to smooth all of the input images to the same spatial resolution. All observed images are used at their native resolution based on knowledge of the PSFs, thus providing higher spatial resolution than is possible with conventional techniques.

For this work, we make use of the column-density-weighted, line-of-sight mean temperature maps combined with JPS 850-$\umu$m intensities. We do not use the {\sc PPMAP} column densities as they contain the extended emission from $\emph{Planck}$ \citep{Planck16} and Hi-GAL, and we are aiming to trace the emission from the densest structures detected in JPS by making use of the spatial filtering inherent to SCUBA-2 data \citep{Holland13}.  This results in suppression of extended emission and therefore a bias towards steep-gradient and so spatially compact, high-column-density and, hence, high-volume-density sources. We use the following formula:

\begin{equation}
N_{\rmn{H_{2}}} = \frac{S_{\nu, \rmn{peak}}}{B_{\nu}(T_{d})\kappa_{\nu}m_{\rmn{H}}\upmu},
\end{equation}

\noindent where $S_{\nu, \rmn{peak}}$ is the JPS pixel intensity, $\kappa_{\nu}$ is the mass absorption coefficient taken to be 0.01\, cm$^{2}$\,g$^{-1}$ \citep{Mitchell01} accounting for a gas-to-dust ratio of 100, $B_{\nu}(T_{d})$ is the Planck function evaluated at temperature $T_{d}$, where $T_{d}$ is the density-weighted mean dust temperature as derived by \ppmap\ at that pixel, $m_{\rmn{H}}$ is the mass of a hydrogen atom, and $\upmu$ is the mean mass per hydrogen molecule, taken to be 2.8 \citep{Kauffmann08}. We filtered the JPS maps to include only pixels with intensity greater than three times the pixel-to-pixel rms noise, an approximate column density threshold of 6\,$\times$\,10$^{20}$\,cm$^{-2}$.

\subsection{CHIMPS}

\begin{figure*}
\begin{tabular}{ll}
\includegraphics[width=0.49\linewidth]{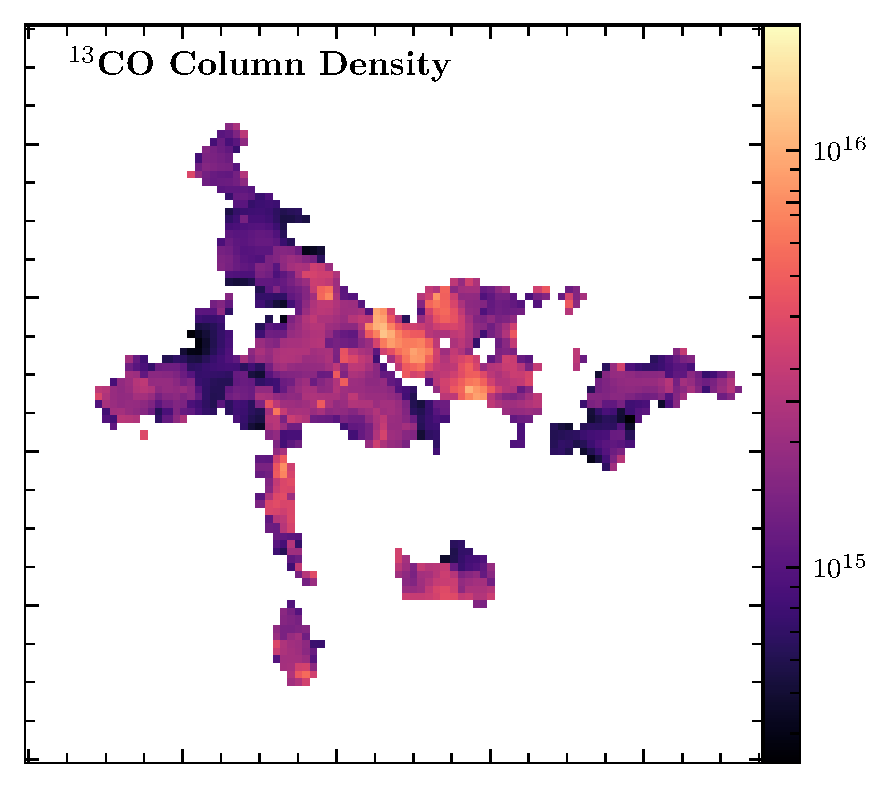} & \includegraphics[width=0.49\linewidth]{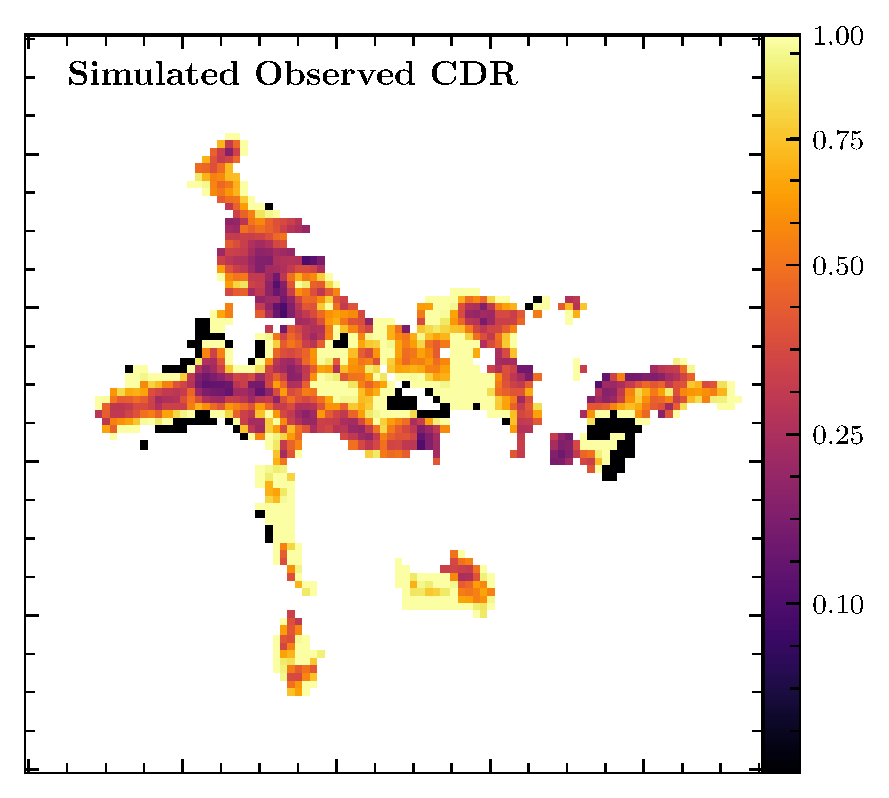} \\
\includegraphics[width=0.49\linewidth]{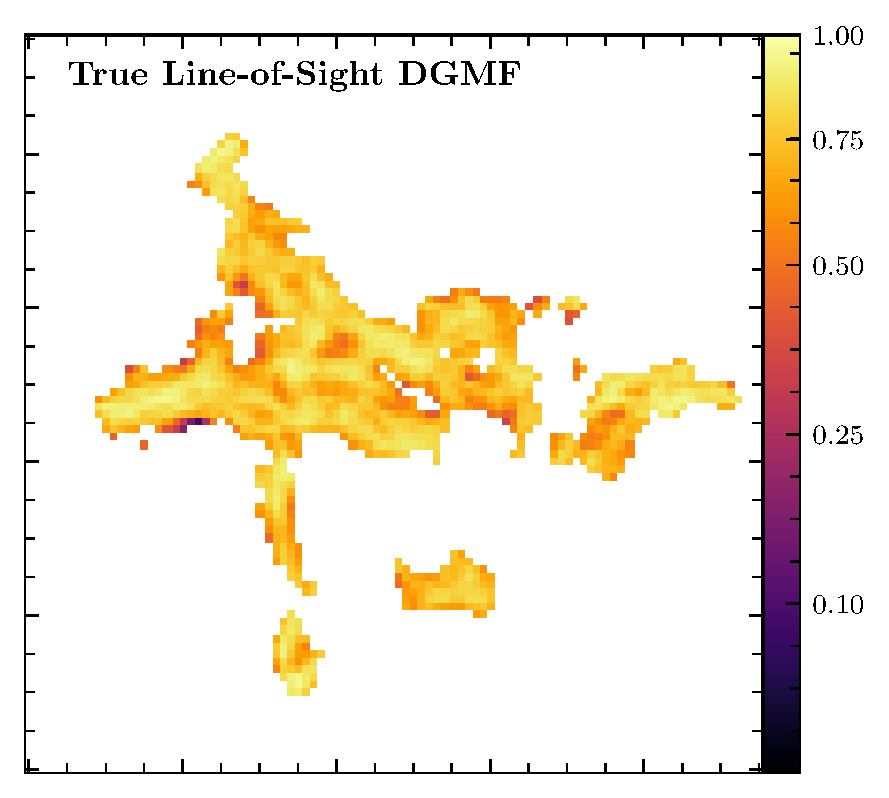} & \includegraphics[width=0.49\linewidth]{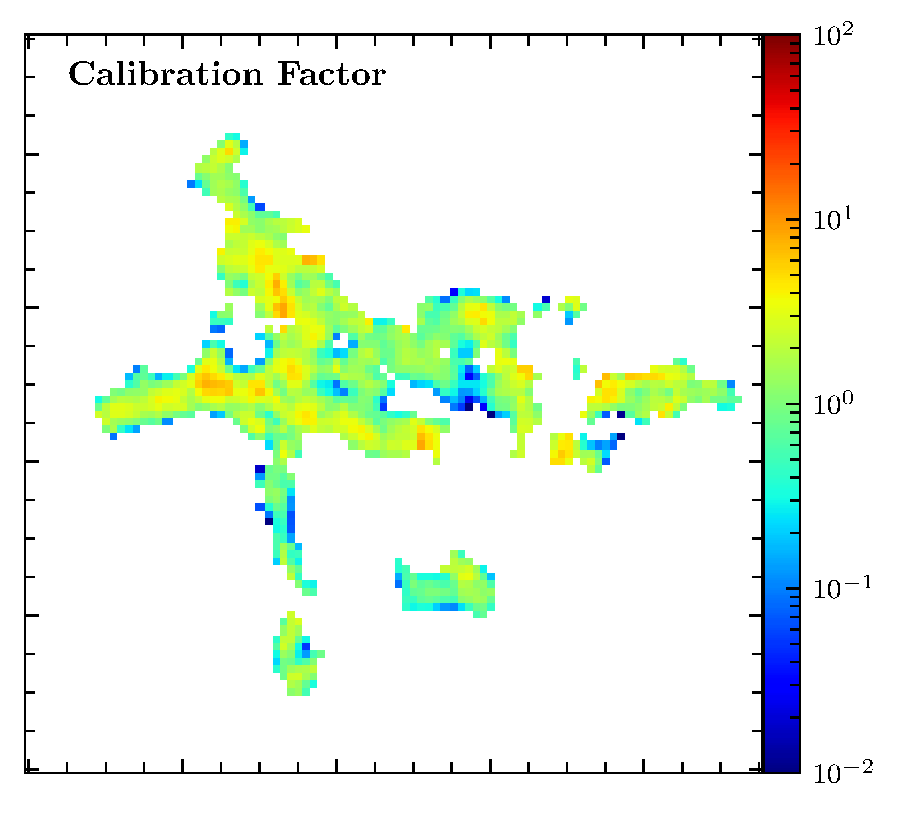} \\
\end{tabular}
\caption{A series of maps derived from the simulated cloud of \citet{Penaloza17} and \citet{Penaloza18}. Top left: $^{13}$CO column densities calculated as in the method of \citet{Rigby19} in units of cm$^{-2}$. Top right: CDR map derived by simulating the JPS and CHIMPS observations. Bottom left: ``true'' line-of-sight DGMF, calculated as the fraction of dense gas along each line of sight. Bottom right: ratio map of the ``true'' line-of-sight DGMF with the CDR derived from the simulated observations, the calibration factor.}
\label{fakemap}
\end{figure*}

The CHIMPS column-density maps are produced using an approximate local thermodynamic equilibrium (LTE) analysis \citep[e.g.][]{Wilson13}, following the procedure outlined in \citet{Rigby19}. In brief, after resampling the $^{12}$CO $J=3\rightarrow2$ CO High-Resolution Survey (COHRS; \citealt{Dempsey13}) observations to match the CHIMPS pixel grid, excitation temperatures were calculated directly from the COHRS $^{12}$CO $J=3\rightarrow2$ emission, under the assumption that the emission is optically thick. The optical depth of $^{13}$CO $J=3\rightarrow2$ emission was then calculated from the brightness temperature of $^{13}$CO $J=3\rightarrow2$ emission measured from CHIMPS data, by assuming the same excitation temperature, and thereby the column density of $^{13}$CO was calculated. 
This was converted to column densities of molecular hydrogen by adopting an abundance ratio of $^{12}$CO to $^{13}$CO of 50, assuming a source Galactocentric distance of 5.5\,kpc \citep{Milam05}, and an abundance ratio of $^{12}$CO to H$_2$ of $8.5 \times 10^{-5}$ \citep{Frerking82}.

The CHIMPS excitation temperatures and, hence, column densities are calculated at a resolution of 27.4\,arcsec. The CHIMPS maps are also filtered to include only pixels that have $N({\rm 13CO})$ above 3\,$\times$\,10$^{15}$\,cm$^{-2}$, which corresponds to $N({\rm H}_{2})$ above 3\,$\times$\,10$^{21}$\,cm$^{-2}$ \citep{Frerking82,Wilson94}.

\section{Methods}

As a proxy for the CFE and DGMF (see above), we will use the ratio of the column densities determined separately from JPS and CHIMPS data, a quantity we name the column-density ratio (CDR). Taking the ratio of two column-density distributions (i.e., maps of $N({\rm H_{2}})$), if produced from observations with different selection effects, will yield the distribution of the ratio of the two mass components traced in each case, in the form of a running average over the beam area or, in this case, the 27-arcsec smoothing area required to calculate the CHIMPS column densities. Since the beam sizes of JPS and CHIMPS are identical, the smoothing applied to both surveys is also identical, and thus the CDR is directly equivalent to a DGMF value measured along each line of sight.
 
The JPS continuum data, like all SCUBA-2 results, are spatially filtered with a cutoff at 8\,arcmin and so preferentially detect compact sources with high column density.  They are therefore more sensitive to dense, star-forming sources and filaments and we use them to trace the dense-clump component of the gas.  CHIMPS CO data are not spatially filtered and therefore trace the ambient large-scale molecular gas component. This approach may seem counter-intuitive, since the optically thin sub-mm continuum should trace total column density, while CO $J=3\rightarrow2$ emission has a volume-density threshold related to the critical density ($\sim 10^4$\,cm$^{-3}$).  Hence the method, and particularly the choice of tracers, needs to be tested with the aid of some reference data with a known spatial density distribution.  For this, we use a simulated molecular cloud, which will be discussed in detail in the following section. This process will also provide a calibration for CDR values derived from the observational data, something that would be required whatever tracers of total and dense gas were used, if estimates of the true CDR/DGMF were sought.

Note, however, that our primary aim in this Section is not to measure the absolute or true CFE or DGMF from the CDR but to determine how well the chosen tracers of 850-$\upmu$m continuum and $^{13}$CO $J=3\rightarrow2$ are estimating the relative values, and so detecting region-to-region variations. For this, all we need is a well-defined relationship between the CDR values obtained from the observational method and the actual values in the reference data.
We will be spatially sampling the CDR in the molecular gas and so will make no attempt at this stage to make discrete measurements of `cloud-integrated’ DGMF. In principle, we should be able to detect correlated variations in CDR on angular scales ranging from the resolution of the data to the scale of the survey area and infer variations in the line-of-sight DGMF and CFE.  Thus we aim to identify the dominant scale associated with any hypothetical mechanism that may be regulating the amount of dense gas present and hence constrain the mechanism itself.

\section{Simulated Column-Density Ratio Maps}

The simulated molecular cloud used is that described in \citet{Penaloza17}. This cloud model uses an adapted version of the {\sc gadget-2} code \citep{Springel05}, modified to include the chemistry of the formation and destruction of molecular species \citep{Glover12}. The simulations produce synthetic CO observations at multiple rotational transitions and, to synthesise these observations, the publicly available {\sc radmc-3d} radiative transfer code \citep{Dullemond12} is used. A full description of the simulations, initial conditions, and chemical evolution can be found in \citet{Penaloza17}, with the cosmic-ray ionisation rates and initial column densities and masses in \citet{Penaloza18}. Cloud GC16-Z1-G10 is adopted as the test data for this study. The parameter values of this cloud can be found in \cite{Glover16}.

\subsection{Simulating the observations}

The simulated molecular cloud is capable of having all quantities of interest calculated for each voxel. To simulate the column-density maps observed by the JCMT in the JPS and CHIMPS data, the instrumental conditions of SCUBA-2 and HARP needed to be simulated.

To replicate JPS SCUBA-2 continuum data, the column-density information in the model cloud was collapsed along one dimension and resampled to account for the JPS beam of 14.4\,arcsec \citep{Eden17} and the modal distance of 5.5\,kpc within the $\ell$ = 30$\degr$ field \citep{Russeil11,Rigby19}. To simulate the filtering of large-scale structure that occurs in SCUBA-2 data \citep{Holland13}, a version of the same map was smoothed over the filtering scale of 8\,arcmin and subtracted from the original. 

The CHIMPS spectroscopic data were imitated using simulated CO $J=3\rightarrow2$ intensity maps in the $^{12}$CO and $^{13}$CO isotopologues generated from the model molecular cloud with {\sc radmc-3d}. The simulated data were convolved with a Gaussian kernel to represent the JCMT beam. The two convolved maps were resampled onto larger pixels to match a cloud at 5.5\,kpc. These resampled clouds were then regridded in the spatial and spectral axes to match the sampling of the respective COHRS and CHIMPS maps. The final step in the setup was to add Gaussian noise fields matching the rms of the COHRS and CHIMPS surveys. The $^{13}$CO $J=3\rightarrow2$ column densities were then calculated using the local thermodynamic equilibrium (LTE) method as described in \citet{Rigby19}. The result of this LTE analysis is displayed in the top left panel of Fig.~\ref{fakemap}.

\begin{figure*}
\begin{tabular}{ll}
\includegraphics[width=0.49\linewidth]{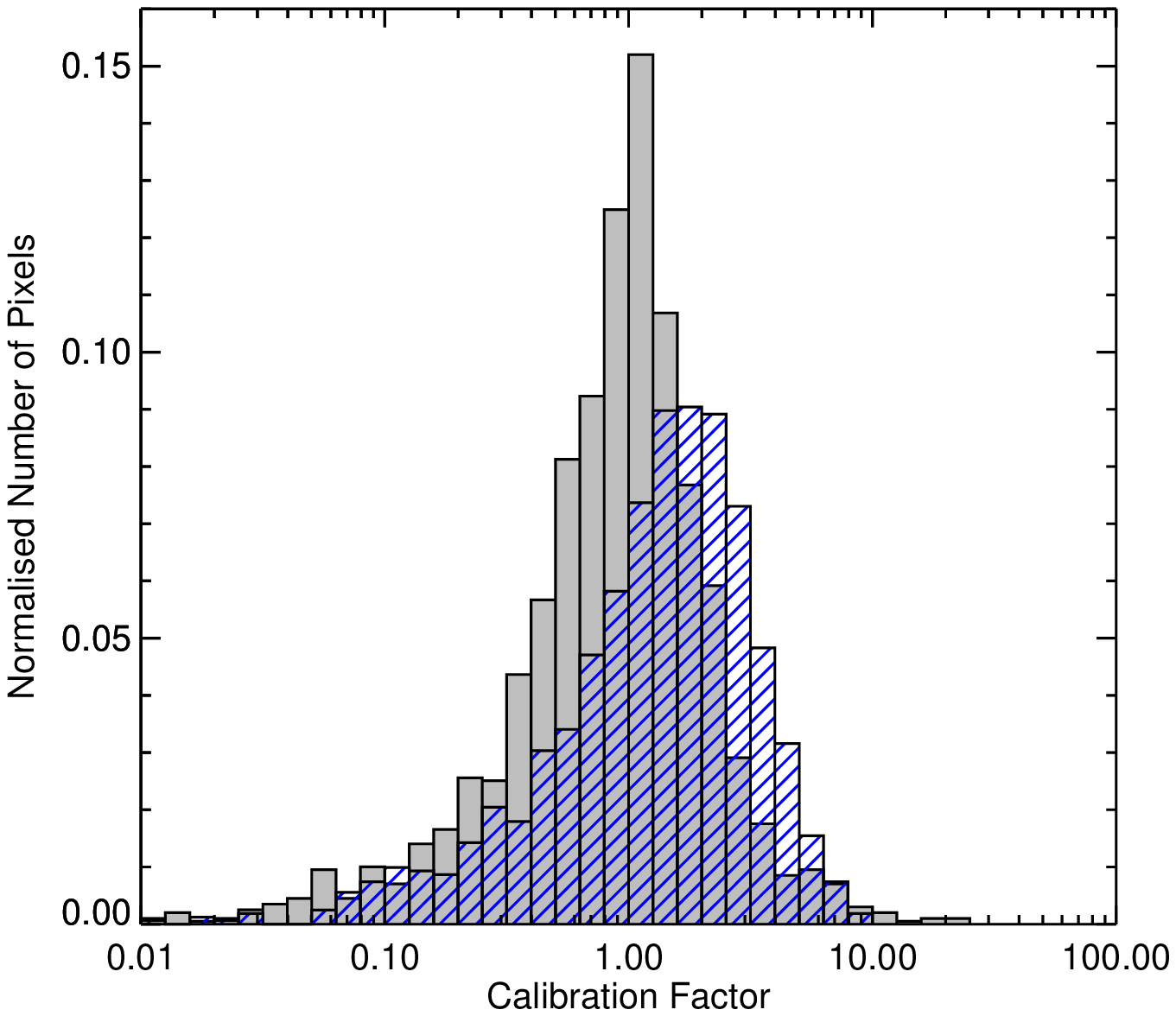} & \includegraphics[width=0.49\linewidth]{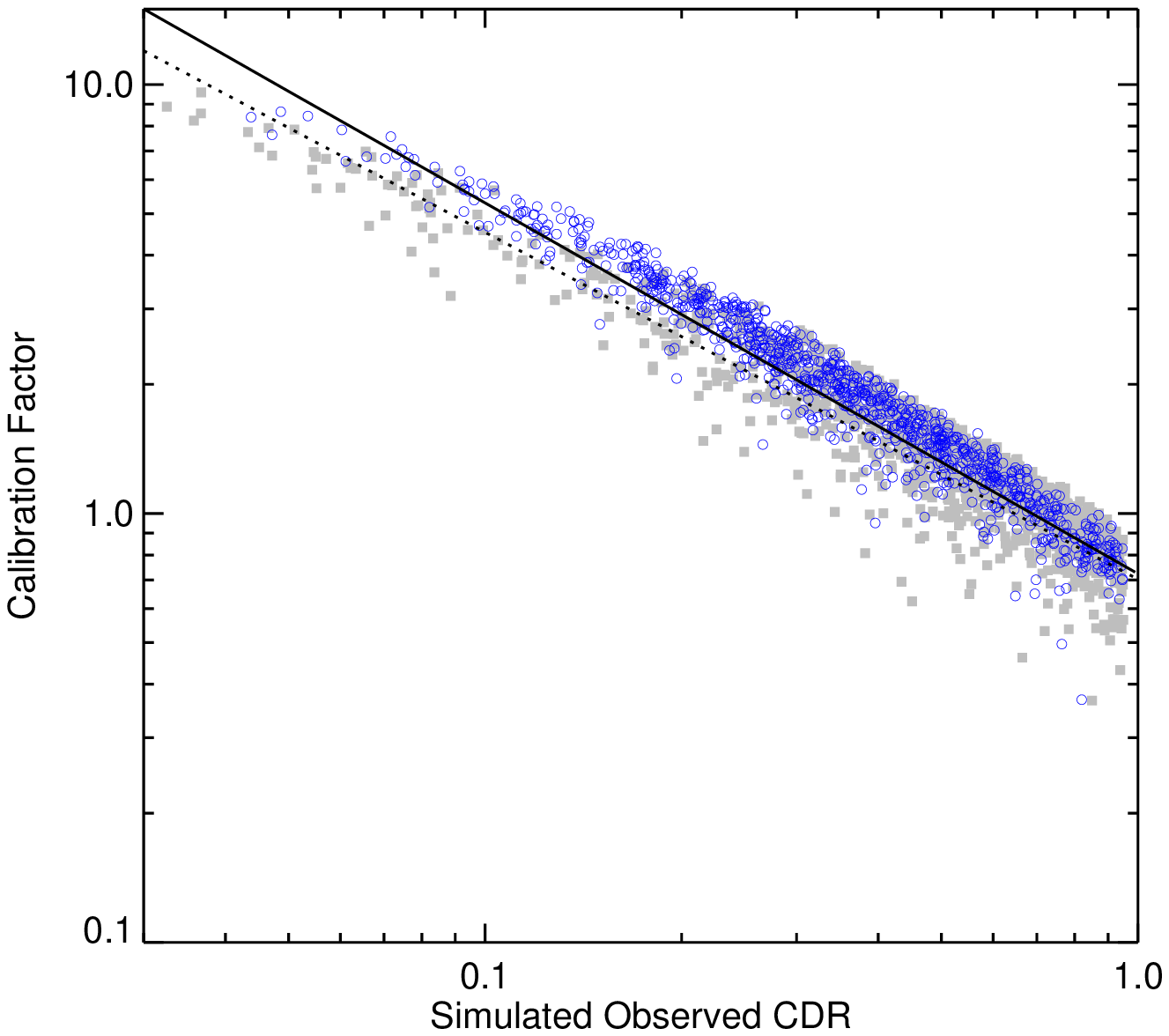} \\
\end{tabular}
\caption{Left panel: Normalised histogram of the calibration factor map shown in the bottom right panel of Fig.\ref{fakemap}, represented by the hashed blue bars. The grey histogram reflects the calibration factor derived from simulations involving $^{13}$CO $J=1\rightarrow0$. Right panel: calibration factor from the bottom right panel of Fig.\ref{fakemap} and the left panel above, plotted against the simulated observed CDR predicted by the model. The blue open circles and grey squares are from the $^{13}$CO $J=3\rightarrow2$ and $^{13}$CO $J=1\rightarrow0$ simulations, respectively. The solid line is the fit to the $^{13}$CO $J=3\rightarrow2$ distribution, with the dotted line the least-squares fit to the $^{13}$CO $J=1\rightarrow0$ distribution.}
\label{fakemaphisto}
\end{figure*}

The simulated continuum column-density map was then divided by the simulated CO column-density map, replicating the method used for the observational data. The result is displayed in the top right panel of Fig.~\ref{fakemap}.

\subsection{DGMF of the simulations}

To create the reference, a ``true'' CDR map was then produced from the simulation by integrating twice along the same axis as used above, once to find the total mass and the second for the dense gas on each line of sight. The dense-gas threshold was taken to be the critical density of CO $(J=3\rightarrow2)$ i.e., 1.6\,$\times$\,10$^{4}$\,cm$^{-3}$ \citep{Schoier05}. The value used for this threshold is arbitrary; using a lower value will increase the measured CDR, as more mass will be considered dense gas, with a higher threshold having the opposite effect.  There is a similar threshold in the observational data, the value of which is not well defined and depends on the tracers and analysis used, but what follows provides a calibration with respect to the value chosen above.

\subsection{Comparison of CDR maps}

The previous two subsections have produced two different measures of the CDR from the reference data, one that mimics the results obtained from the JPS and CHIMPS surveys, the other giving the ``true'' fraction of dense gas along a particular line of sight.
The ratio of these two measures gives us a correction or calibration relation between the observational method values and the ``true" CDR, as defined by the density threshold adopted.  More importantly for our purposes, the scatter in this relationship defines the precision of the observational method in detecting the direction and relative amplitude of variations in CDR.

The result is shown as a map in the bottom-right panel of Fig.~\ref{fakemap} and a histogram of the pixel values in this map is displayed in the left panel of Fig.~\ref{fakemaphisto}, represented by the blue bars. The modal value is 1.84, the median 1.44 and a standard deviation of 0.21 dex is estimated from a Gaussian fit. The total range of values is large but this is not due to random scatter and is mostly the result of a functional relationship with the values of the ``observational" CDR derived from the model, about which the scatter is rather small, as seen in the blue open circles in the right-hand panel of Fig.~\ref{fakemaphisto}.  A Spearman correlation analysis gives a $p$-value $< 0.001$ with a correlation coefficient of $-$0.993.

The maps in Fig.~\ref{fakemap}, namely the $^{13}$CO column-density map, the simulated observed CDR and the calibration factor, were determined from simulated maps of $^{12}$CO and $^{13}$CO $J=3\rightarrow2$. For comparison purposes, this analysis was repeated using simulated $^{12}$CO and $^{13}$CO $J=1\rightarrow0$ emission derived from the same model reference data. The results are included, as grey bars and symbols, in Fig.~\ref{fakemaphisto}. The modal value of the distribution is 1.12, a median value of 0.77 and a standard deviation of 0.42 dex. The correlation coefficient is $-$0.989, again with a $p$-value $< 0.001$.

As we have stated above, the purpose of using a simulated cloud is not to measure the true DGMF of real molecular clouds, although this can be done. The aim is to determine how reliable the CDR produced by the 850-$\upmu$m continuum and $^{13}$CO $J=3\rightarrow2$ data is in tracing and detecting spatial variations in the actual CDR. The fact that the method produces dependable values with predictable variations and low scatter indicates that we can successfully measure relative CDR values and so spatial variations in line-of-sight DGMF with $^{13}$CO $J=3\rightarrow2$ tracing the molecular content. The form of the relationship in Fig.~\ref{fakemaphisto} is the result of the choices made in the analysis and much less important than the strong correlation, which can be used as a correction function, i.e., $\log(correction) = -0.86\,\log(CDR) - 0.14$. For completeness, the correction function for CO $J=1\rightarrow0$ would be $\log(correction) = -0.81\,\log(CDR) - 0.15$.

The similarity between the distributions of the calibration factors derived from the two different molecular transitions in Fig.~\ref{fakemaphisto} shows that there is no significant advantage to using $^{13}$CO $J=1\rightarrow0$ data over $^{13}$CO $J=3\rightarrow2$. In fact, a tighter correlation is found using $J=3\rightarrow2$.

Again for reference, the total DGMF (i.e., the integrated dense-gas mass divided by the total cloud mass) in the simulated cloud was found to be 0.39 using the observational method while the ``true'' value is 0.62. The latter depends on the choice of density threshold. Using the mean number density of continuum sources from the Bolocam Galactic Plane Survey, 1.5\,$\times$\,10$^{3}$\,cm$^{-3}$, at a Galactocentric radius of 4.5\,kpc \citep{Dunham11} gives an alternative ``true'' value of 0.92.

Another consequence of these simulated maps is that the relative densities of the two tracers can also be tested. The map of number density was ``observed'' in the same fashion as the SCUBA-2 JPS map, whilst the column-densities derived in the manner of the CHIMPS observations were converted to number density using the size of the cloud, and the assumed distance of 5.5\,kpc. The mean density of the simualted JPS map was found to be 9.0\,$\times$\,10$^{3}$\,cm$^{-3}$, whilst the CHIMPS density was found to be 7.1\,$\times$\,10$^{3}$\,cm$^{-3}$. These values confirm our hypothesis that the JPS data are tracing denser material than that of the CHIMPS $^{13}$CO $J=3\rightarrow2$ data. These two numbers are fairly similar due to the critical density of CO $J=3\rightarrow2$ and the observational  analysis methods, and explain the high DGMF values found for the cloud, but also demonstrate that Fig.~\ref{fakemaphisto} can be used to calibrate the observed results.

\subsection{Summary of the simulated cloud method}

The use of the simulations of \citet{Penaloza17} have allowed the following conclusions to be drawn regarding our CDR-determination method:

\begin{enumerate}
    \item 
    The CDR values we obtain from the JPS and CHIMPS data reliably reflect the direction and magnitude of variations in the actual values in a set of reference data and, therefore, should also reflect real-world values.
    \item
    With the aid of a correction function, we can use the same data to estimate the absolute value of CDR in the molecular gas.  However, absolute values are dependent on the molecular tracer used, the details of the observing method and the analysis methods adopted.
    \item
    Since the ratio of two column densities that are effectively mass per beam with the same beam size gives a mass ratio, the CDR is equivalent to the dense-gas mass fraction, DGMF, along each line of sight.
    \item
    There is no significant disadvantage in using $^{13}$CO $J=3\rightarrow2$ data from the CHIMPS survey to estimate CDR and line-of-sight DGMF, compared to the traditional $J=1\rightarrow0$ transition, and they additionally provide a factor of $\sim$2 higher angular resolution than the currently available $^{13}$CO $J=1\rightarrow0$ data from the Galactic Ring Survey \citep{Jackson06}.
\end{enumerate}

\section{Column Density Ratio Maps}

As mentioned above, the CHIMPS column densities were smoothed to a resolution of 27.4\,arcsec during the derivation process \citep{Rigby19}.  We have smoothed the JPS column densities to the same angular scale so that this cancels in the ratio, leaving a direct measure of the mass fraction, averaged over that scale. The resulting CDR maps are displayed in Fig.~\ref{CFEmaps}. Where the JPS or CHIMPS data are zero, the resulting CDR is zero. These maps are not corrected using the relationships in Figure \ref{fakemaphisto} and Section 5.3. The map dimensions in each plot are the same, 5$\fdg$5\,$\times$\,0$\fdg$5. However, the CHIMPS survey is not complete in the $\ell$ = 30$\degr$ field, in both longitude and in latitude, due to the mapping configuration used. Details of the latter can be found in \citet{Rigby16}.

In the $\ell$ = 30$\degr$ field, one enhancement in CDR is coincident with the inner part of the W43 star-forming region, found at a longitude of $\ell \simeq 30\fdg$8. Another is associated with a filament located at a longitude of $\sim$ $\ell$ = 32$\degr$ and has a CDR of $\sim$ 0.125, with a central peak comparable to the W43 value. A YSO identified by the ATLASGAL survey \citet{Urquhart18} at $\ell \simeq 28\fdg$6 is the most extended region with a CDR greater than 0.50.

The background level of the column-density ratio in the $\ell$ = 40$\degr$ field appears to be consistent with that in the $\ell$ = 30$\degr$ field. The most significant structures found within this field are the filament highlighted in \citet{Rigby16} at $\ell \simeq 37\fdg$5 and two other local increases associated with JPS continuum sources. These have CDRs with peaks of 0.40 coincident with an ATLASGAL source ($\ell$ = 39$\fdg$2; \citealt{Urquhart14a}) and a high-mass star-forming region ($\ell$ = 38$\fdg$9; \citealt{Urquhart18}). Maps of these regions, and those mentioned in the paragraph above, are found in Fig.~\ref{closeups} in Appendix~\ref{cfes_individual}.

\begin{landscape}
\centering
\begin{figure}
\begin{tabular}{l}
\includegraphics[width=0.99\linewidth]{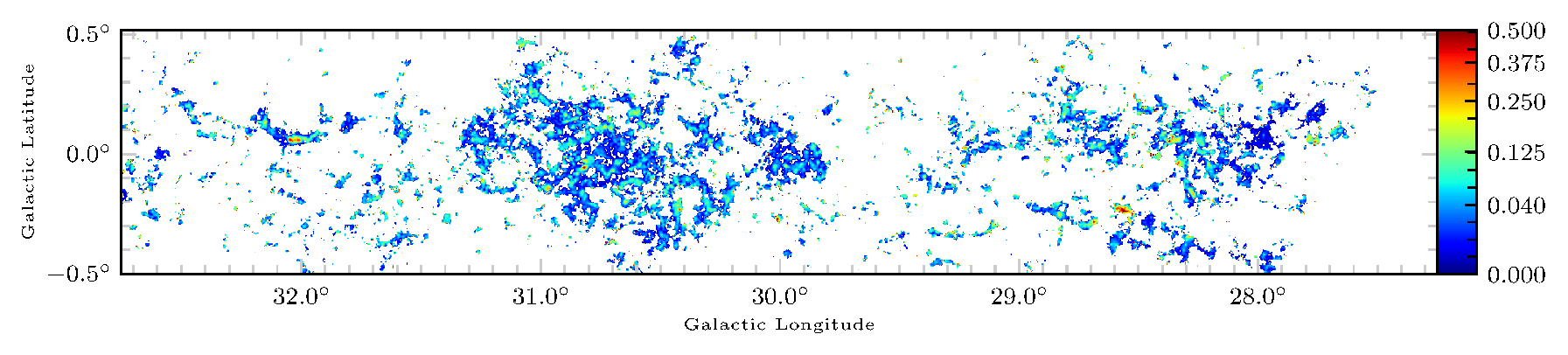}\\
\includegraphics[width=0.99\linewidth]{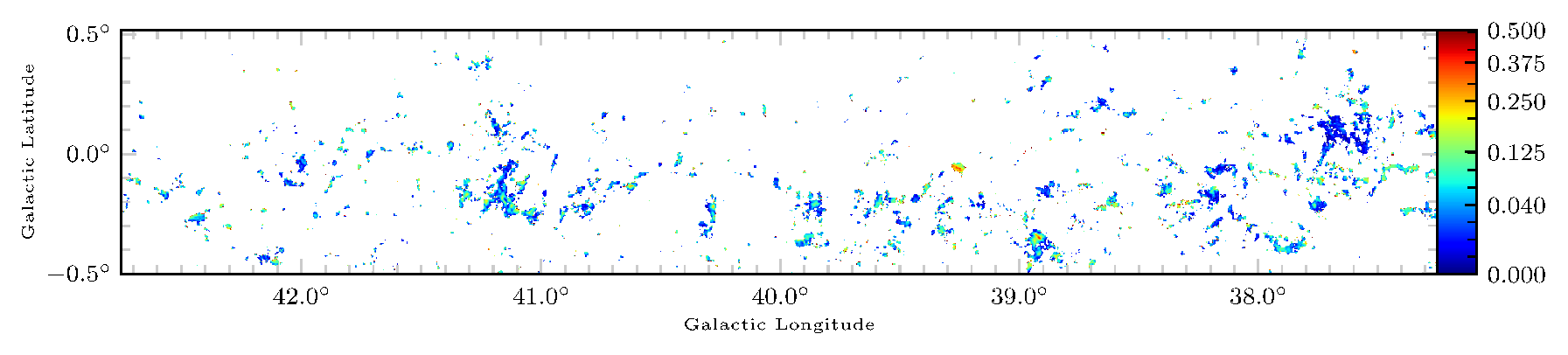}\\
\end{tabular}
\caption{The CDR distribution, as given by the ratio of the H$_2$ column densities derived from JPS continuum and CHIMPS CO data for the $\ell=30\degr$ and $\ell=40\degr$ regions, at a resolution of 27.4\,arcsec.}
\label{CFEmaps}
\end{figure}
\end{landscape}

\section{Power Spectrum Analysis}

\subsection{Power Spectrum Production}

\begin{table*}
\begin{center}
\caption{The power-law slopes and break points for the power spectra in the CDR maps, the total fields and the spiral arms in each field. The large-scale power-law slopes are those fitted above the breaks (low $n$) and vice versa. The median cloud radii are from the catalogues derived from the CHIMPS \citep{Rigby19} and GRS \citep{Roman-Duval09} surveys.}
\label{powerlaws}
\begin{tabular}{llccccccccc} \hline
Field & Spiral Arm & Assumed & Break & Break & Break & Median CHIMPS & Median GRS & Power Law & Power Law\\
 & & Distance &  & Range &  & Clump Radii & Cloud Radii & Large Scale & Small Scale\\
 & & (kpc) & (pc) & (pc) & (arcmin) & (pc) & (pc) & & \\
\hline
$\ell$\,=\,30$\degr$ & Total & 5.50 & 7.78 & 5.74 -- 11.5 & 4.86 & 1.77 & 10.1 & $-$0.80$\pm$0.22 & $-$1.91$\pm$0.23 \\
 & Scutum--Centaurus & 4.88 & 3.72 & 3.65 -- 4.27 & 2.62 & 1.70 & 11.9 & $-$1.04$\pm$0.27 & $-$2.29$\pm$0.33 \\
 & Sagittarius & 11.52 & 15.8 & 13.8 -- 19.8 & 4.72 & 3.38 & 12.8 & $-$1.32$\pm$0.24 & $-$1.85$\pm$0.24 \\
 & Perseus & 13.41 & 10.1 & 9.48 -- 12.2 & 2.58 & 3.74 & 9.10 & $-$1.12$\pm$0.29 & $-$1.93$\pm$0.35 \\
\hline
$\ell$\,=\,40$\degr$ & Total & 8.51 & 15.2 & 14.0 -- 17.8 & 6.13 & 2.32 & 7.30 & $-$0.84$\pm$0.25 & $-$1.88$\pm$0.21\\
 & Sagittarius & 9.60 & 7.20 & 6.42 -- 9.05 & 2.58 & 2.76 & 16.0 & $-$1.06$\pm$0.29 & $-$2.36$\pm$0.35\\
 & Perseus & 11.99 & 10.5 & 9.68 -- 12.0 & 3.01 & 2.76 & 7.80 & $-$0.97$\pm$0.23 & $-$2.11$\pm$0.33 \\
\hline
\end{tabular}
\end{center}
\end{table*}

Power spectra derived from maps of interstellar-medium tracers are often used to investigate the characteristics of ISM turbulence. The turbulent nature of the interstellar medium imprints itself within the tracers in the forms of density and velocity fluctuations. The shape of the power spectrum can reveal the scale at which the turbulence is injected \citep{Kowal07} and the evolution of the molecular clouds and gas \citep{Burkhart15}, with clouds containing star formation showing an altered power spectrum \citep{Federrath13}. Such studies analyse the power spectra of density and column-density maps over a number of velocity channels. Here we will be using these techniques to analyse the CDR spatial variations to identify the dominant or characteristic scale at which the variations occur, thus constraining the scale and, hence, the nature of the mechanisms that may be regulating the CDR, and thus the line-of-sight DGMF and, with the assumptions described in Section 1, the spatial variations in CFE.

The two-dimensional power spectrum of the CDR was produced by running a Fast Fourier Transform on each map in Fig.~\ref{CFEmaps}. To avoid boundary conditions, the maps need to be symmetrical in both the $x$ and $y$ directions, i.e., square, and to be continuous at the boundary. To ensure this, a border of zeroes is added such that, if the length of the map is $m$, then the total with zeroes added is $2m$ (Tahani et al., in preparation). This is to ensure that the images wrap around with no ``edge''. A similar approach is taken by \citet{Ossenkopf08} using the $\upDelta$-variance method. 

\begin{figure}
\includegraphics[width=0.99\linewidth]{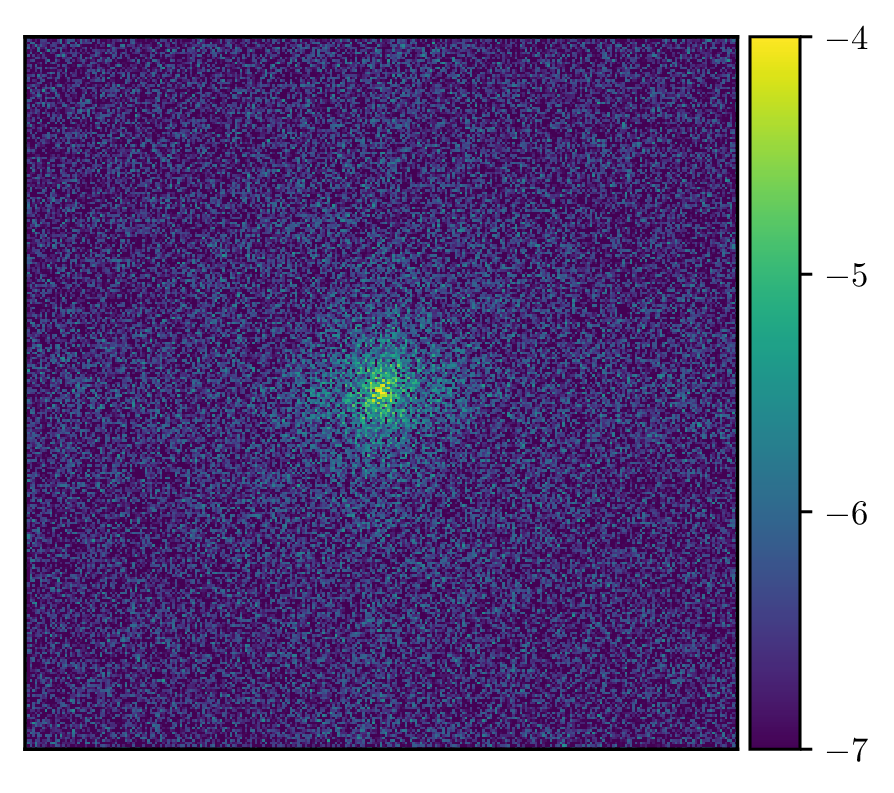}
\caption{Two-dimensional power spectrum image for the CDR map in the $\ell$ = 30$\degr$ field. The units of the map are arbitrary.}
\label{powerexample}
\end{figure}

The two-dimensional power spectrum image is centred so that the central pixel contains the power for the whole image, and can be considered to be $k\,=\,0$ (the wavenumber), or $n\,=\,1$, where $n$ is the number of wavelengths or segments across the image. An example is given in Fig.~\ref{powerexample}, being the two-dimensional power spectrum of the CDR map of the $\ell$ = 30$\degr$ field shown in Fig.~ \ref{CFEmaps}.

To convert the two-dimensional power-spectrum image into a 1-D power spectrum $P(k)$, the values of power are calculated in concentric radii outwards from the centre, with the power determined by the mean of the square of every value to fall within $n_{n-0.5}$ and $n_{n+0.5}$. As a result, small $k$ covers larger scales, whereas the higher values of $k$ indicate smaller spatial scales. This is as explained in \citet{Combes12}. The highest value of $k$ probed corresponds to the angular resolution of the column density maps, 27\,arcsec.

\subsection{Power spectra of CDR maps}

\begin{figure*}
\begin{tabular}{ll}
\includegraphics[width=0.49\linewidth]{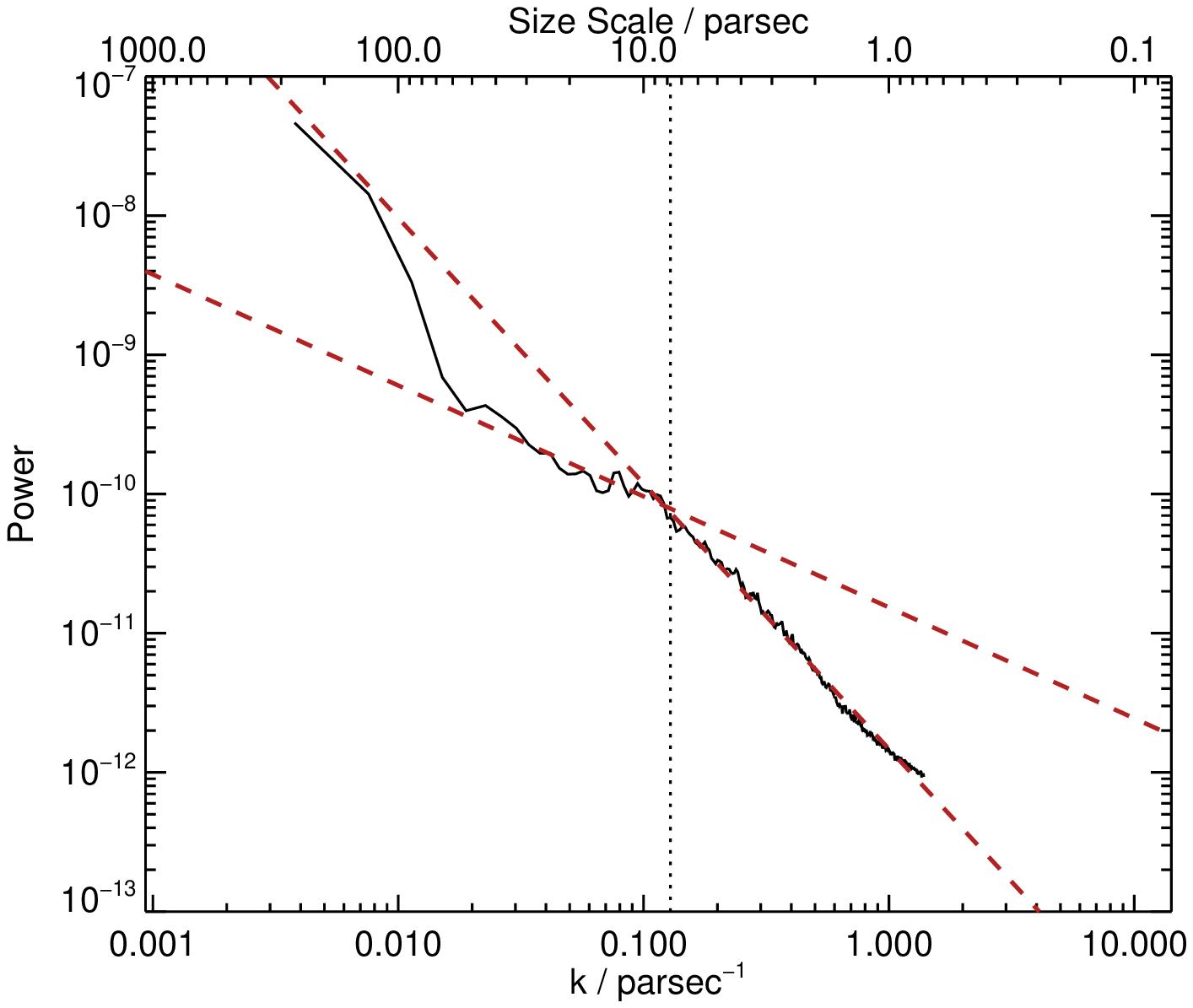} & \includegraphics[width=0.49\linewidth]{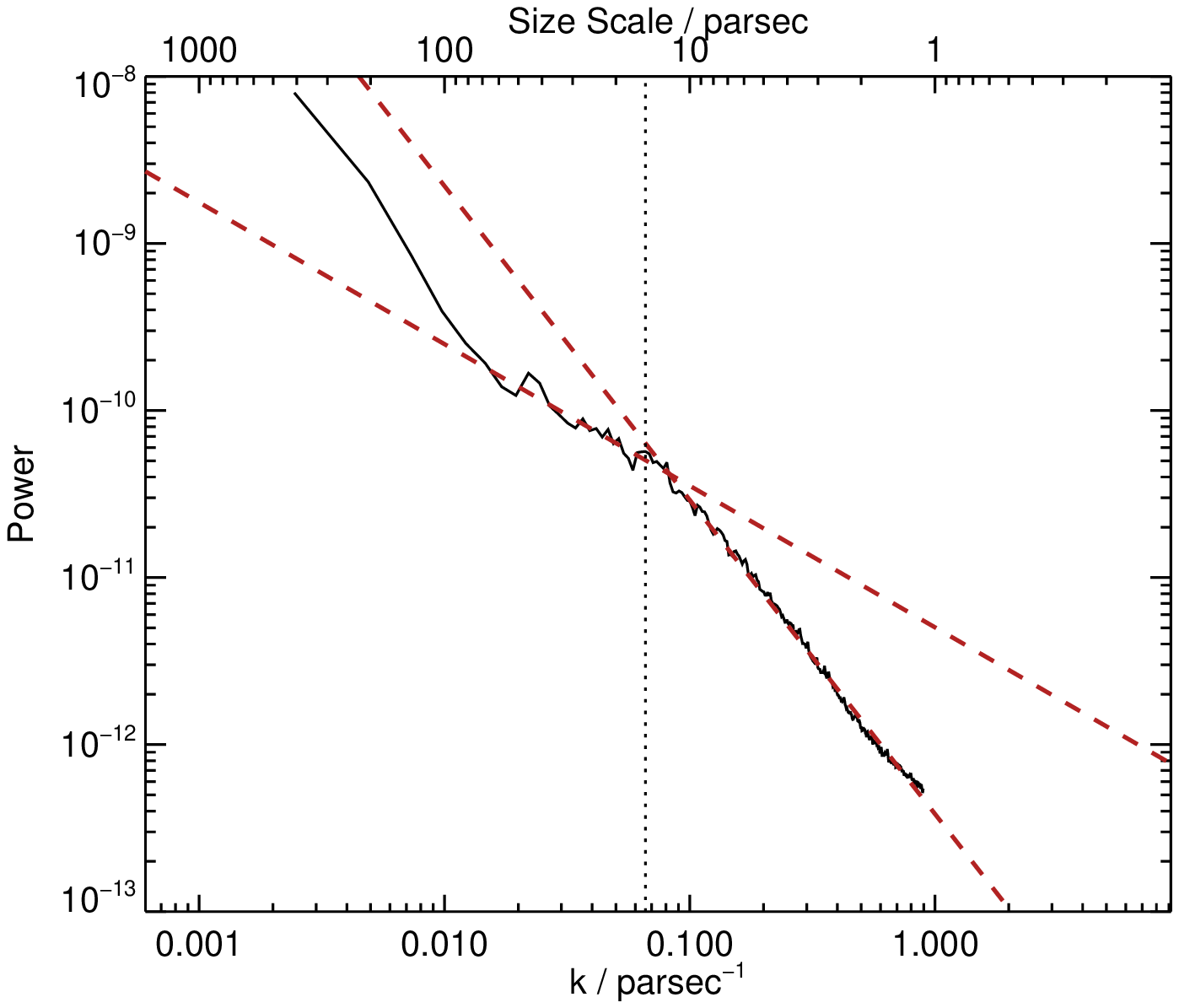} \\
\end{tabular}
\caption{The power spectra of the CDR maps in the $\ell$\,=\,30$\degr$ (left panel) and $\ell$\,=\,40$\degr$ (right panel) fields. The dashed red lines represent the power-law fits to the high $k$ and low $k$ regimes in the spectrum. The vertical dotted line indicates the break between the two power-law fits.}
\label{CFEPS}
\end{figure*}

The power spectra of the CDR maps are shown in Fig.~\ref{CFEPS}. Two power-law slopes have been fitted to each, one to the low-$k$ regime, one to the high-$k$ regime. The break between the two power laws, the characteristic scale of the CDR, and thus the DGMF or CFE mechanism, was determined by least-squares fitting for each range of $k$ above $n=28$, with the break selection occurring where the sum of the $\upchi^{2}$ values was minimised. A series of tests were performed to ensure that the location of the break was not influenced by data artefacts, as described in Appendix~\ref{pstests}. The rise in the power spectra observed at the lowest values of $k$ is due to the shape of the input map and sets the fitting range of $k$ above $n=28$, where the power spectrum becomes dominated by the shape of the underlying astronomical map, i.e. the rectangles displayed in Fig.~\ref{CFEmaps}..

We have indicated the break between the power laws with a vertical line in Fig.~\ref{CFEPS}. The corresponding $k$ value is converted to a physical scale using the modal distance within each field, i.e., 5.50\,kpc for $\ell$\,=\,30$\degr$ \citep{Russeil11,Rigby19} and 8.51\,kpc for $\ell$\,=\,40$\degr$ \citep{Rigby19}.
The break in the power spectrum of the $\ell$\,=\,30$\degr$ DGMF map is thus found at an angular scale of 4.86\,arcmin, which corresponds to a physical scale of 7.78\,pc. In the $\ell$\,=\,40$\degr$ field, the break was found at 6.13\,arcmin, or 15.16\,pc. The values of the fitted power-law exponents, break scales and the median cloud radii in each observed field are displayed in Table~\ref{powerlaws}. The range of possible break scales are also displayed. This range was calculated from the gradients of the fitted power law. The upper limit is the lowest value of $k$, or largest size scale, where the gradient of the power law fit to the small scale is within 1 sigma of the gradient at the chosen break scale. The lower limit represents the $k$ value, converted to physical size, of the highest value of $k$ where the large scale power law gradient is within 1 sigma of the break-scale larger scale gradient.

These ranges were calculated as the range of $k$, and therefore size, where both the large and small-scale power-law fits are consistent with the values listed in Table~\ref{powerlaws}.

\subsection{CDR maps of individual spiral arms}

\begin{figure*}
\begin{tabular}{l}
\includegraphics[width=0.99\linewidth]{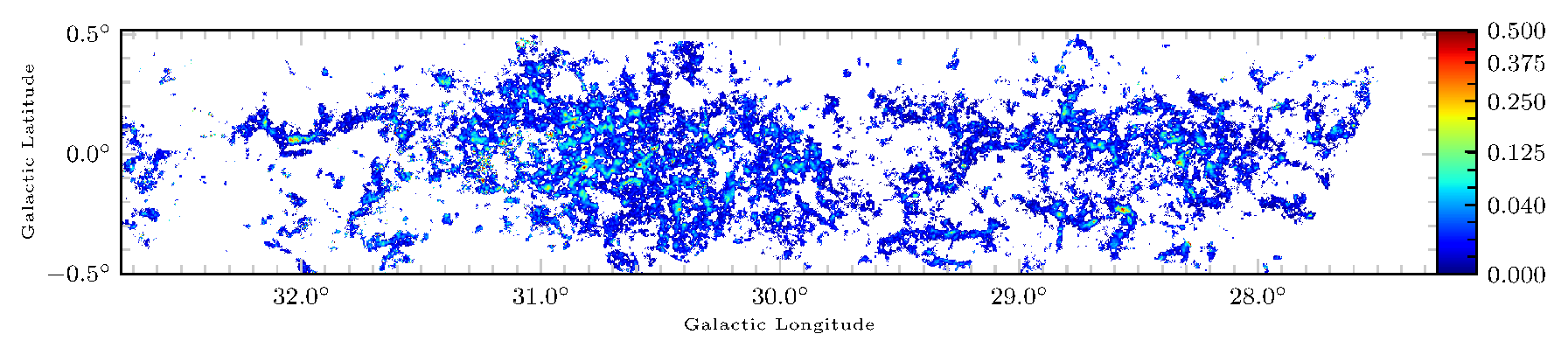}\\
\includegraphics[width=0.99\linewidth]{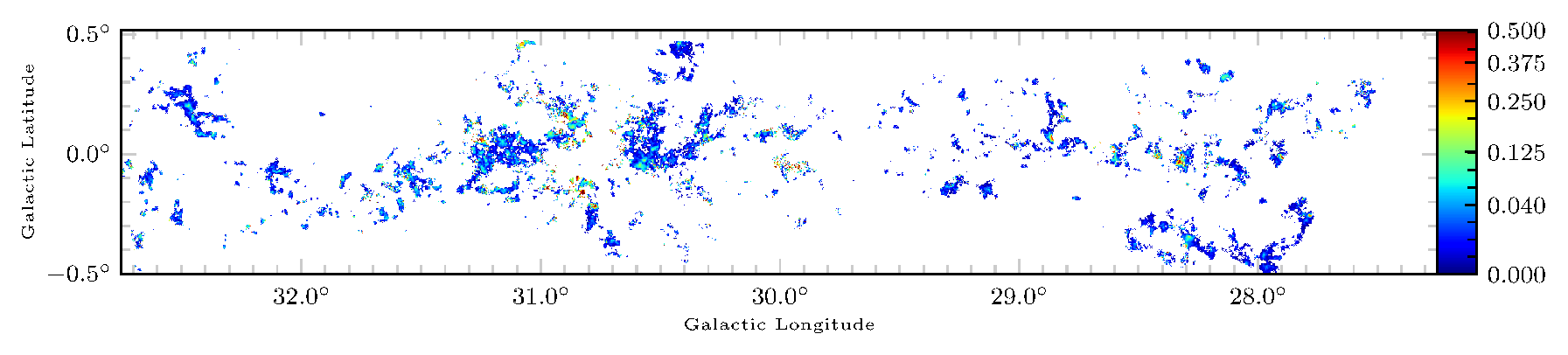}\\
\includegraphics[width=0.99\linewidth]{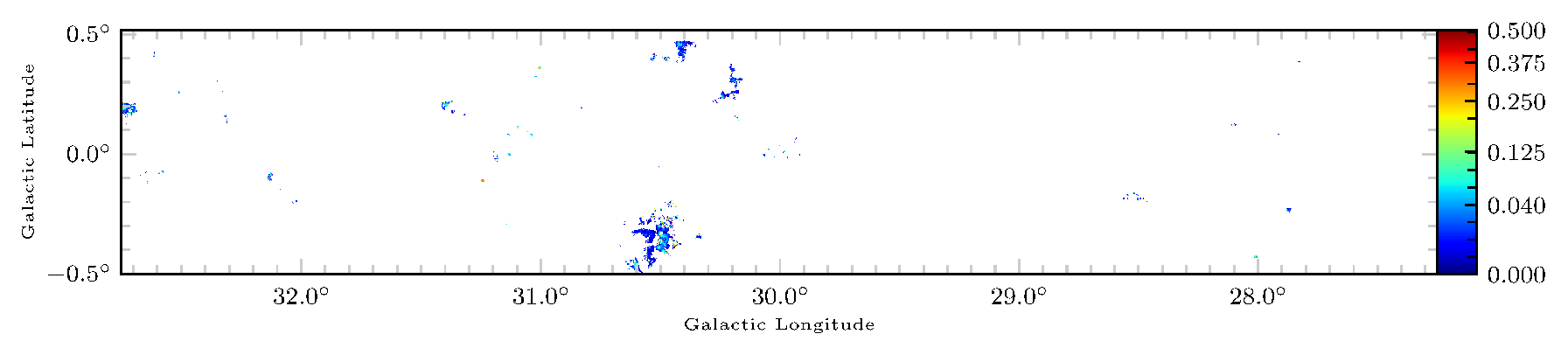}\\
\includegraphics[width=0.99\linewidth]{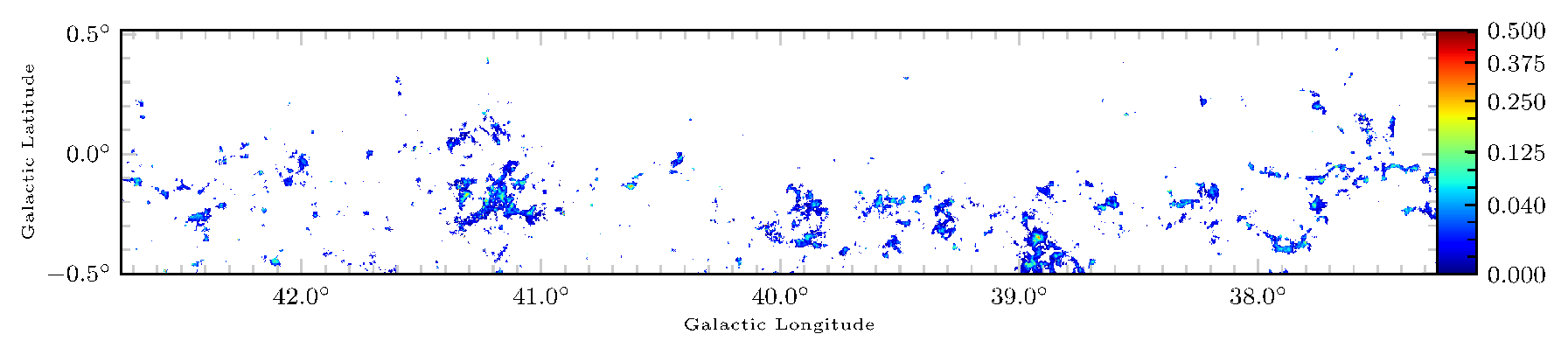}\\
\includegraphics[width=0.99\linewidth]{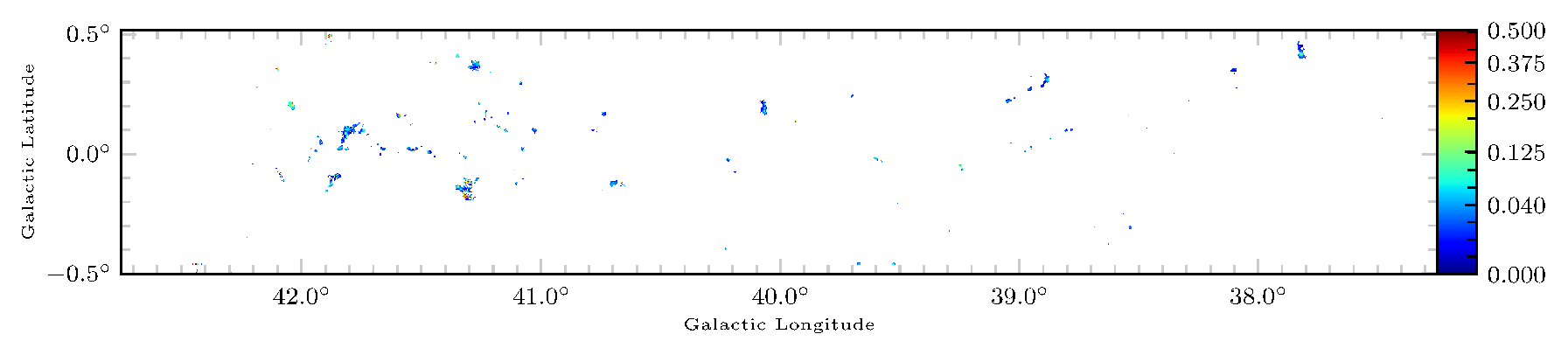}\\
\end{tabular}
\caption{The CDR     maps of the individual spiral arms within the $\ell$\,=\,30$\degr$ and $\ell$\,=\,40$\degr$ fields. In order, they are the Scutum--Centaurus, Sagittarius, and Perseus arms of the $\ell$\,=\,30$\degr$ field, and the $\ell$\,=\,40$\degr$ field Sagittarius and Perseus arms.}
\label{CFEmaps_arms}
\end{figure*}

To further refine the comparison between the power spectrum and the cloud scale, we have split the CDR maps into components corresponding to the individual spiral arms that run across the two fields. To produce these maps, the CO-derived column-density data were separated into velocity ranges and collapsed along the velocity axes. These maps were then used as a mask, and any JPS emission that lined up with that spiral arm mask was attributed to that spiral arm. The $\ell$\,=\,30$\degr$ field was split into three spiral arms: Scutum--Centaurus, Sagittarius and Perseus. The  $\ell$\,=\,40$\degr$ field contained the Sagittarius and Perseus arms. The velocities were identified from the extent of the arms in the longitude-velocity diagram of \citet{Rigby16}, which uses the spiral-arm model of \citet{Taylor93}. The ranges identified were 70--110\,km\,s$^{-1}$ for the $\ell$\,=\,30$\degr$ Scutum--Centaurus arm, 30--60\,km\,s$^{-1}$ for the $\ell$\,=\,30$\degr$ Sagittarius arm, 5--20 \,km\,s$^{-1}$ for the $\ell$\,=\,30$\degr$ Perseus arm, 30--70\,km\,s$^{-1}$ for the $\ell$\,=\,40$\degr$ Sagittarius arm, and 5--20 \,km\,s$^{-1}$ for the $\ell$\,=\,40$\degr$ Perseus arm. The results are shown in Fig.~\ref{CFEmaps_arms}.

\begin{figure*}
\begin{tabular}{ll}
\includegraphics[width=0.49\linewidth]{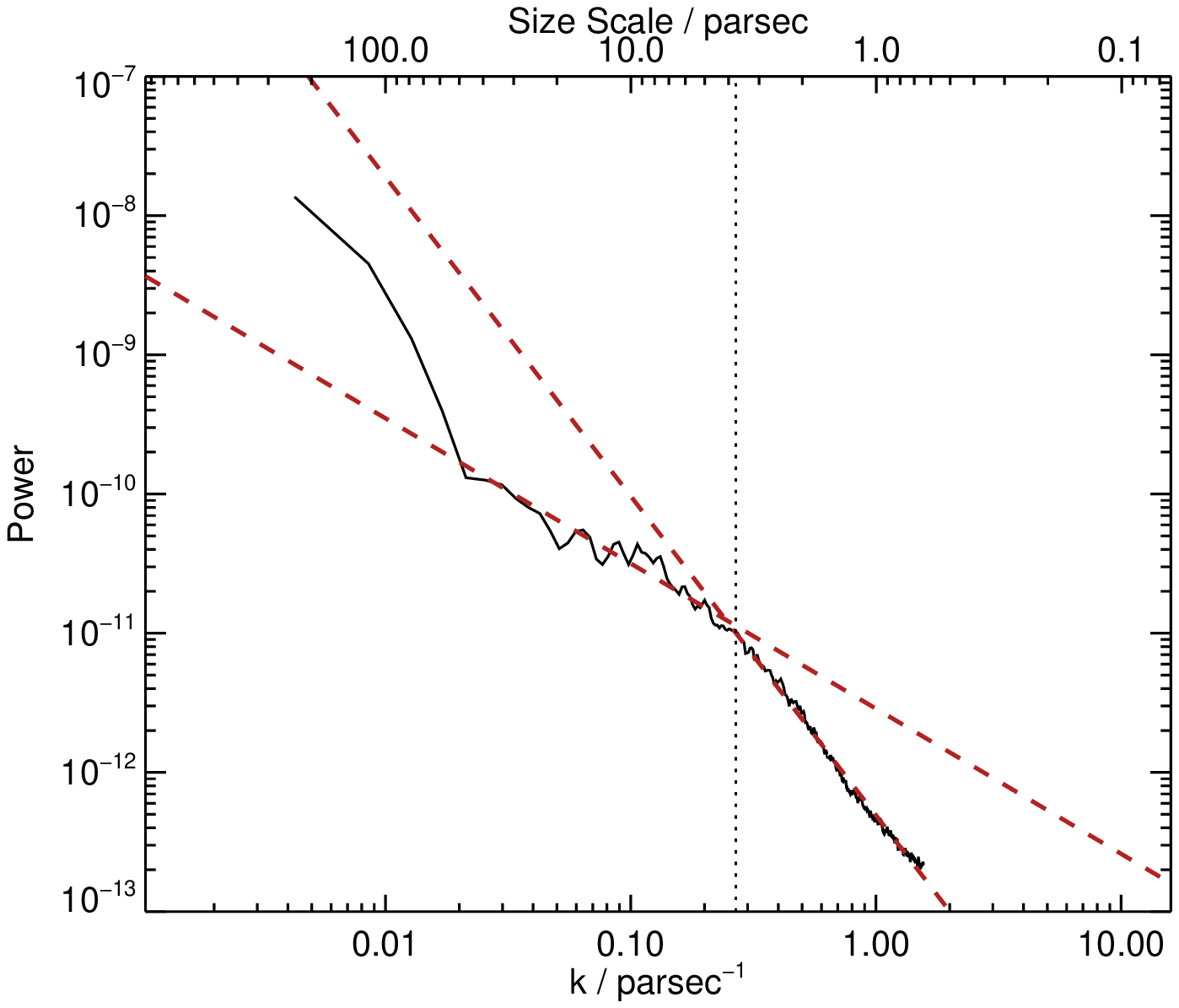} & \includegraphics[width=0.49\linewidth]{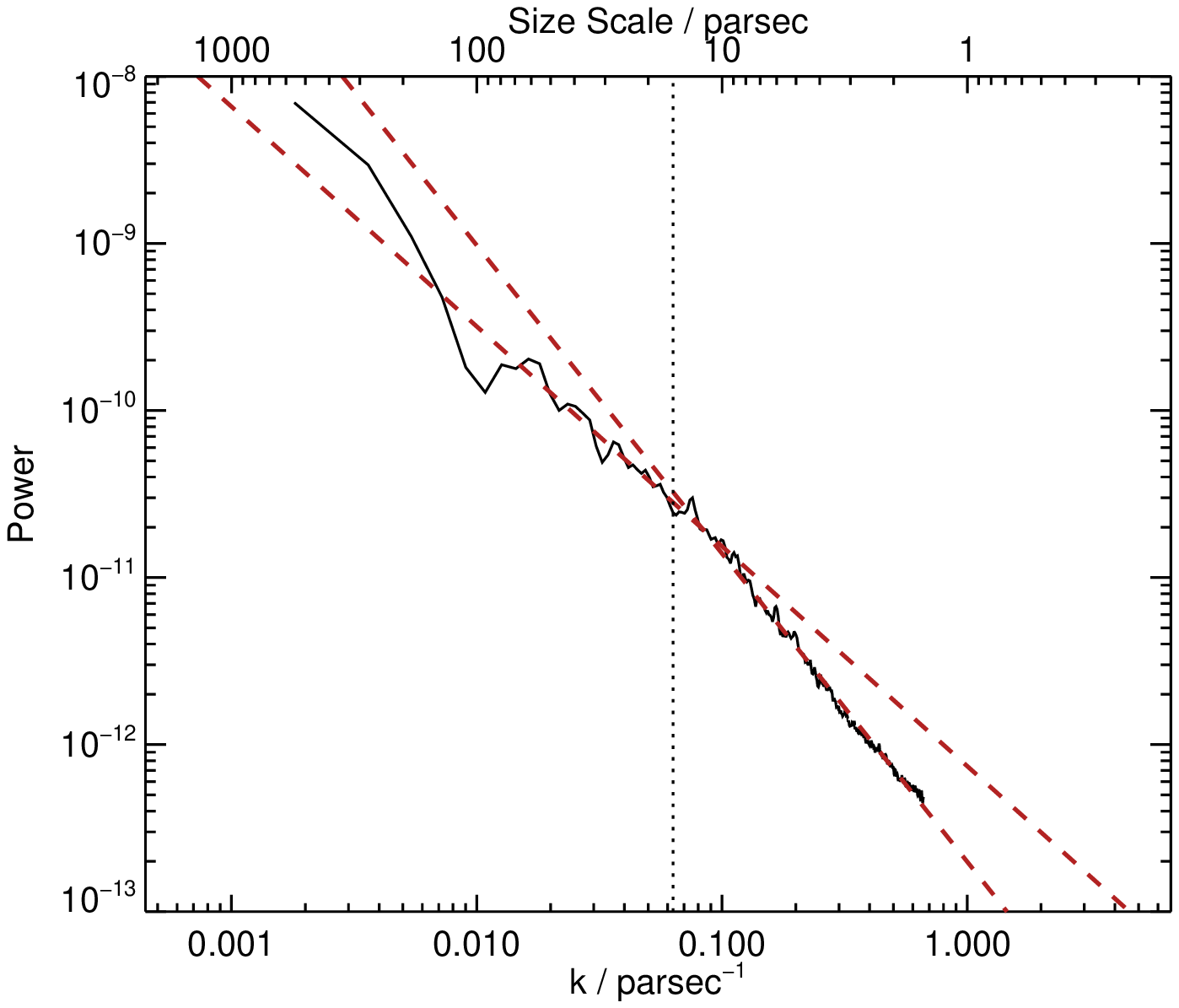} \\
\includegraphics[width=0.49\linewidth]{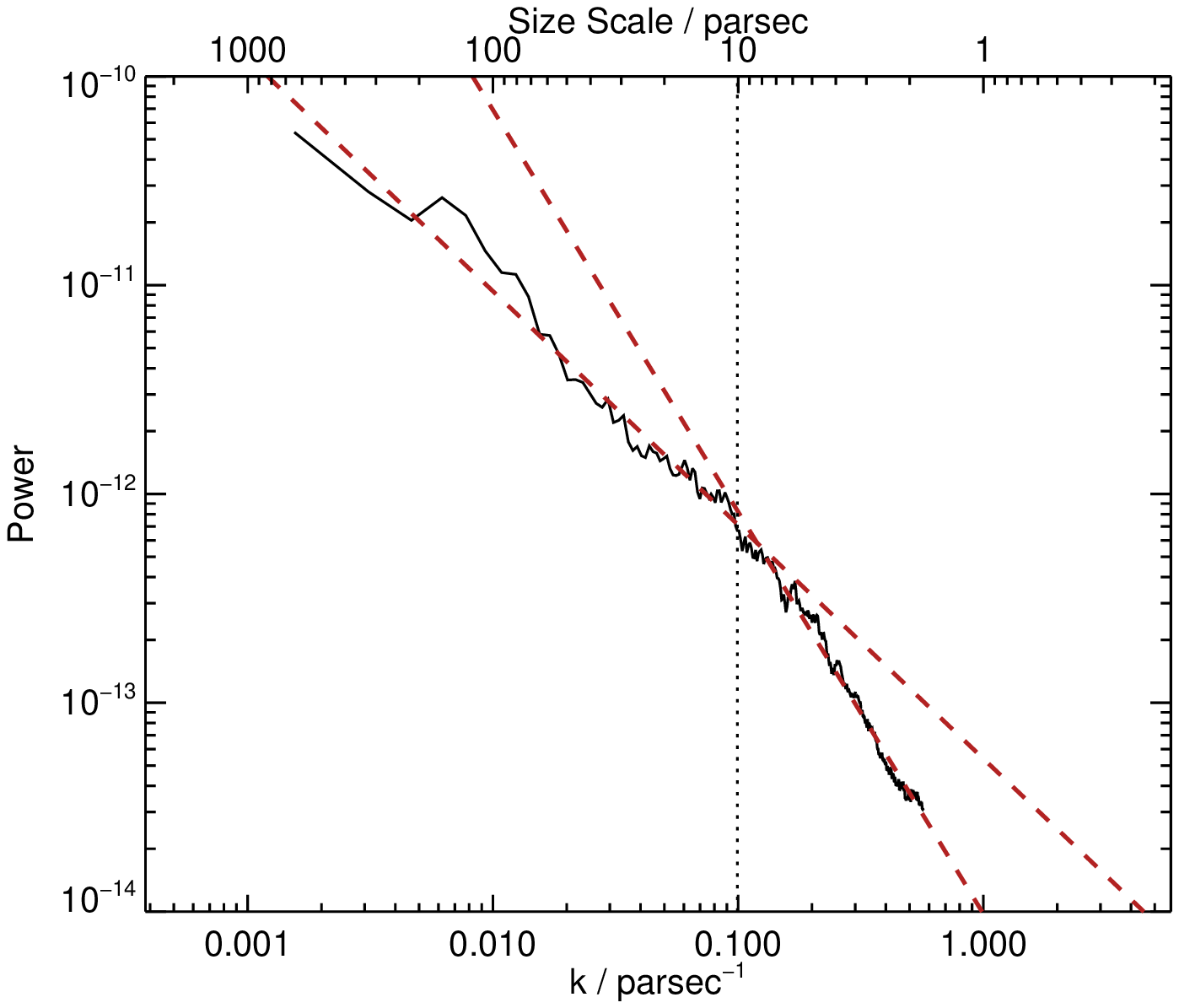} & \includegraphics[width=0.49\linewidth]{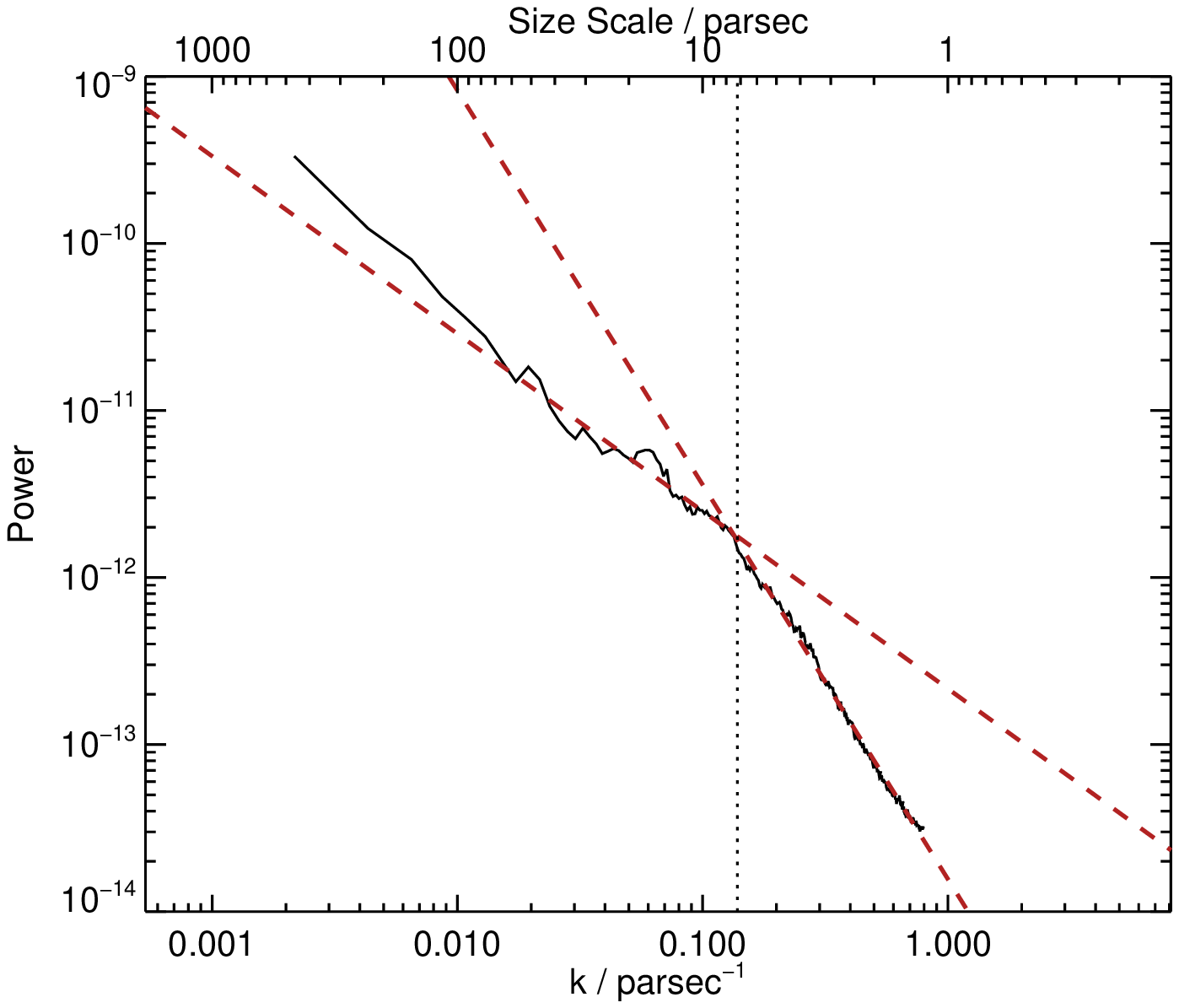} \\
\includegraphics[width=0.49\linewidth]{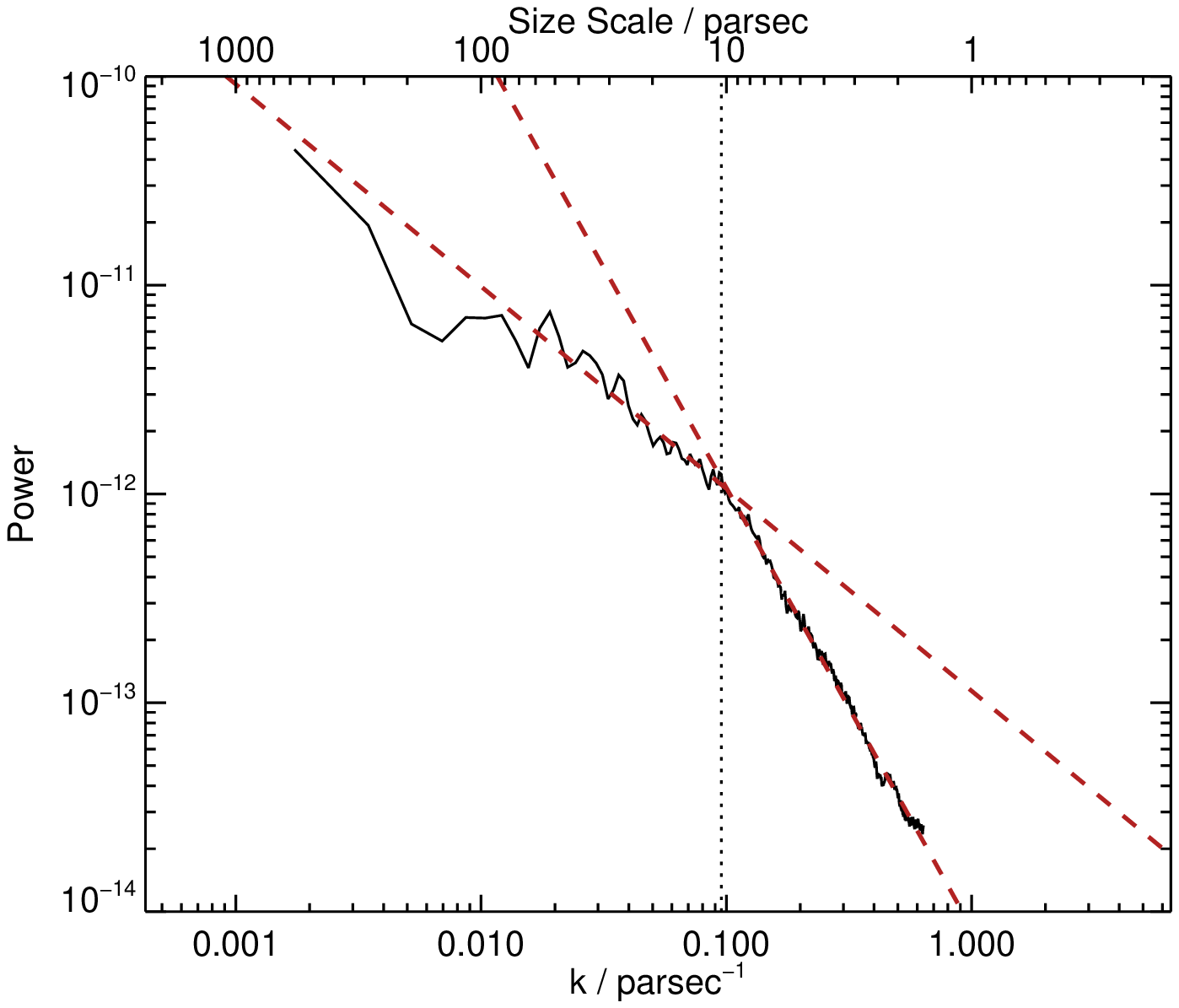} & \\
\end{tabular}
\caption{The power spectra of the CDR maps in the individual spiral arms within the $\ell$\,=\,30$\degr$ and $\ell$\,=\,40$\degr$ fields. The overplotted lines are as described in Fig.~\ref{CFEPS}. Top: $\ell$\,=\,30$\degr$ Scutum--Centaurus and $\ell$\,=\,30$\degr$ Sagittarius. Middle: $\ell$\,=\,30$\degr$ Perseus and $\ell$\,=\,40$\degr$ Sagittarius. Bottom: $\ell$\,=\,40$\degr$ Perseus.}
\label{CFEPS_arms}
\end{figure*}

The power-spectrum analysis was repeated on each of these five maps, with the distance to each arm taken as the output from the Bayesian-distance calculator of \citet{Reid16} at the central position in $(\ell,b,V)$ of each arm segment. These were 4.88\,kpc, 11.5\,kpc, and 13.4\,kpc for the $\ell$\,=\,30$\degr$ Scutum--Centaurus, Sagittarius, and Perseus arms, respectively. The $\ell$\,=\,40$\degr$ Sagittarius and Perseus arms were assigned distances of 9.60\,kpc and 12.0\,kpc, respectively. The power spectra of these five maps are shown in Fig.~\ref{CFEPS_arms}.

The individual spiral arms were found to have breaks between the two power laws at physical scales of 3.72\,pc, 15.8\,pc, 10.1\,pc, 7.20\,pc, and 10.5\,pc for the $\ell$\,=\,30$\degr$ Scutum--Centaurus, $\ell$\,=\,30$\degr$ Sagittarius, $\ell$\,=\,30$\degr$ Perseus, $\ell$\,=\,40$\degr$ Sagittarius, and $\ell$\,=\,40$\degr$ Perseus arms, respectively.

\section{Discussion}

\subsection{Scales of clump formation}

The power spectra of the total fields in Fig.~\ref{CFEPS} and of the individual arm components in Fig.~\ref{CFEPS_arms} display no features that can be related to the large-scale structure of the Milky Way spiral arms, due to the limited size of the maps. However, breaks are found at scales similar to those of the molecular clouds identified in CHIMPS \citep{Rigby19} and the GRS \citep{Roman-Duval09}. By considering the individual spiral arms, the breaks occur at a size scale of 2.19, 4.67, 2.70, 2.61, and 3.80 $\times$ the median cloud radii for the $\ell$\,=\,30$\degr$ Scutum--Centaurus, $\ell$\,=\,30$\degr$ Sagittarius, $\ell$\,=\,30$\degr$ Perseus, $\ell$\,=\,40$\degr$ Sagittarius, and $\ell$\,=\,40$\degr$ Perseus arms, respectively for the CHIMPS clouds. With regards to the GRS clouds, the breaks occur at a size scale of 0.31, 1.23, 1.11, 0.45, 1.35 $\times$ the median cloud radii for the five spiral arm samples, respectively. The sizes of the GRS clouds are somewhat more in line with the scales of the breaks, indicating that the scale of clump formation may be better traced by the $J = 1 \rightarrow 0$ emission, which detects less dense and, hence, a larger fraction of the molecular gas. The intermediate $J = 2 \rightarrow 1$ survey of SEDIGISM \citep{Schuller17}, although not observing these regions, finds a median cloud radius of 2.38\,pc (Duarte-Cabral et al., in prep). These cloud scales are more similar to those of CHIMPS than the GRS, with the clouds in CHIMPS more resembling clumps \citep{Rigby19}.

As the CDR is analogous to the instantaneous CFE, any variations in the measured CDR would reflect fluctuations in CFE. As indicated in Equation~\ref{equation}, any change in CFE is most likely to be due to an altered clump formation rate, rather than a change in clump formation timescale \citep{Eden12,Urquhart18}. Due to this relationship, we will assume that the CDR maps in Figs.~\ref{CFEmaps} \& ~\ref{CFEmaps_arms} display the CFE. 

Previous studies have found no evidence of CFE increases linked to Galactic structure \citep{Eden12,Eden13}, however CFE increases are found to be linked to local feedback sources \citep{Polychroni12}, with increased star-formation activity caused by these sources of feedback \citep{Thompson12,Kendrew12,Palmeirim17}. Other power-spectrum work has found that, in extragalactic systems, the break in the power spectrum of the continuum emission is found to be coincident with the Jeans' length \citep{Elmegreen03}, with any star-forming regions separated by greater than the break scale having no impact on each other. \citet{Elmegreen03} noted that this break was also comparable to the scale height in the Galactic disc. The scale height for high-mass stars, and therefore high-mass clumps, in the Milky Way is 30\,pc \citep[e.g.][]{Urquhart14,Urquhart18}, although no signature is found at these scales in the power spectra here. A similar result was found for the \ion{H}{i} distribution in the Galaxy \citep{Khalil06}.

Combining these results, we can expect the characteristic scale of variations in CFE to be on the cloud scale. Previous studies have found that the power spectra of turbulence is of a power-law form with variations linked to cloud environment \citep[e.g.][]{Kolmogorov41,Lazarian04,Kowal07,Collins12}. In these analyses, based on velocity and density data, a departure from a single power-law in the form of a break indicates the scale at which the turbulence is injected into a system, such as the plane thickness at $\sim$100\,pcs \citep[e.g.][]{Elmegreen03} or at which it dissipates \citep{Hennebelle12}. However, dissipation occurs at milliparsec scales \citep{Miville-Deschenes16} and the angular resolutions to detect this are too low for the JCMT. By analogy, the breaks we detect  correspond to the characteristic scale of the mechanisms responsible for the formation of dense clumps and for regulating the efficiency of that process.  This scale corresponds closely to the cloud size scale and so implicates intra-cloud turbulence and cloud formation mechanisms or initial conditions as the most likely agents determining the CFE and, hence, the star-formation efficiency in the molecular gas.

\subsection{Star formation across different spiral arms}

By splitting the CFE maps into the individual spiral arms, some of the line-of-sight ambiguities of scale are removed. The source extraction method of \citet{Rigby19} was repeated on the column-density maps of the five spiral arms in Fig.~\ref{CFEmaps_arms} using the {\sc FellWalker} algorithm (FW; \citealt{Berry15}). The masks produced by FW were then applied to the CFE maps and the pixels associated with each clump extracted. If a FW source has more than 10 pixels in a CFE map, it was counted, following the source size thresholds of \citet{Rigby19}. This extraction resulted in a total of 1619 molecular clouds with recorded CFEs. The breakdown of these molecular clouds is as follows: 960 in the $\ell$\,=\,30$\degr$ Scutum--Centaurus arm; 209 in the $\ell$\,=\,30$\degr$ Sagittarius arm; 15 in the $\ell$\,=\,30$\degr$ Perseus arm; 366 in the $\ell$\,=\,40$\degr$ Sagittarius arm; and 69 in the $\ell$\,=\,40$\degr$ Perseus arm. By combining the common spiral arms, we find 575 and 84 molecular clouds with CFE values in the Sagittarius and Perseus spiral arms, respectively.

\begin{figure}
\includegraphics[width=0.99\linewidth]{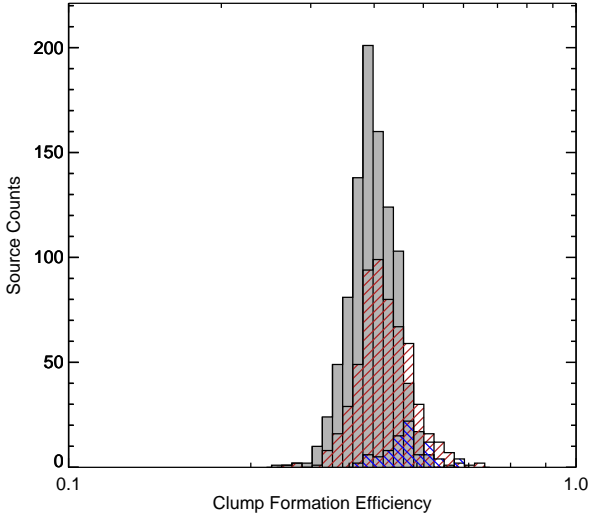}
\caption{Distribution of mean CFEs for the individual molecular clouds within the Scutum--Centaurus spiral arm (grey), Sagittarius spiral arm (red), and the Perseus spiral arm (blue).}
\label{CFEhisto}
\end{figure}

The mean CFE of each cloud was corrected using the relationship in {\color{red}Section 5.3}. The distribution of the corrected mean CFE within each of molecular clouds is shown in Fig.~\ref{CFEhisto}. Shapiro--Wilk and Anderson--Darling tests find that they are consistent with a lognormal distribution. As the samples can be considered lognormal, a Gaussian fit was performed on these samples, with the mean and standard deviation values given in Table~\ref{statistics}. The mean and median values from the data are also given in Table~\ref{statistics}. The means are significantly different from each other, but, the size of the bins are smaller than the errors on these means. Using the standard deviations, and the medians, we are not able to distinguish between these samples.

\begin{table*}
\begin{center}
\caption{CFE statistics of the molecular clouds within each spiral arm. The Gaussian mean and standard deviations are derived from the Gaussian fit to the distributions in Fig.~\ref{CFEhisto}, whereas the data mean and median are derived directly from the sample. All values are logarithmic.}
\label{statistics}
\begin{tabular}{lcccc} \hline
Spiral & Gaussian & Gaussian & Data & Data \\
Arm & Mean & Standard & Mean & Median \\
& & Deviation & & \\
\hline
Scutum--Centaurus & $-$0.401\,$\pm$\,0.001 & 0.042 & $-$0.402\,$\pm$\,0.001 & $-$0.403\,$\pm$\,0.027 \\
Sagittarius & $-$0.382\,$\pm$\,0.002 & 0.048 & $-$0.378\,$\pm$\,0.002 & $-$0.383\,$\pm$\,0.032 \\
Perseus & $-$0.335\,$\pm$\,0.004 & 0.039 & $-$0.334\,$\pm$\,0.005 & $-$0.330\,$\pm$\,0.029 \\
\hline
\end{tabular}
\end{center}
\end{table*}

The Gaussian fit mean values correspond to CFEs of $\sim$\,40, 41, and 46 per cent, for the Scutum--Centaurus, Sagittarius, and Perseus spiral arms, respectively. The absolute values of CFE found here are different to those in \citet{Eden12}, \citet{Eden13}, and Urquhart et al. (submitted), however, the trend between the different spiral arms is the same. These higher values are likely to be due to the choice of density threshold as the $^{13}$CO $J=3\rightarrow2$ critical density. By using this as the threshold, both elements of the CFE ratio have the same threshold in the model, pushing the predictions closer to unity. The variations from cloud to cloud are much greater than the averages over larger scales, a result seen in \citet{Eden15}. This is a result of the central-limit theorem, with a well defined mean over a large sample.

The SFE of each spiral-arm segment can also be estimated using the power spectra of the column-density maps, as displayed in Fig.~\ref{CHIMPSCDPSarms}, and the models of \citet{Federrath13}. These models predict the SFE from the index of the density spectrum, which can be derived from the column-density power spectrum as $\alpha\,=\,\beta + 1$ where $\alpha$ is the index of the density power spectrum and $\beta$ is the index of the column-density power spectrum as the star formation alters the density field. The indices for each spiral-arm segment are found in Table~\ref{psteststable}. We choose the models with a Mach number of 3 as these models were the best match for CHIMPS clouds that have a mean Mach number of 5.6 \citep{Rigby19}. We find upper limits for the predicted SFEs in a solenoidal turbulent cloud of 0.28 per cent, 0.82 per cent, 1.62 per cent, 1.73 per cent, and 2.59 per cent for the $\ell$\,=\,30$\degr$ Scutum--Centaurus, $\ell$\,=\,30$\degr$ Sagittarius, $\ell$\,=\,30$\degr$ Perseus, $\ell$\,=\,40$\degr$ Sagittarius, and $\ell$\,=\,40$\degr$ Perseus arms, respectively. The compressive turbulence model gives SFEs of 2.48 per cent, 3.16 per cent, 4.18 per cent, 4.33 per cent, and 5.51 per cent, respectively.

This trend in the predicted SFEs follows that of the measured CFE, in that the Perseus arm is found to be the most efficient at forming stars. This is consistent with the results of \citet{Eden15}, who found an increased ratio of $L/M$, a proxy of SFE \citep[e.g.][]{Moore12,Eden15,Eden18}, in the Perseus arm compared to the Scutum--Centaurus and Sagittarius spiral arms. However, this was found to be due to distinctly different time gradients for star formation across the spiral arms, with the Perseus star formation at a more advanced stage. If the star formation in the Perseus arm is at a more advanced stage, it is sensible that the SFE would be higher, especially as the SFE is defined as the ratio of stellar mass to molecular cloud mass. This is further evidence that the physics within molecular clouds is the most important regulator of the star-formation process since on scales larger than individual molecular arms, there is no evidence of any difference between the spiral arms.

\section{Summary and Conclusions}

We have produced maps of the ratio of column density from the JCMT Plane Survey (JPS; \citealt{Eden17}) and $^{13}$CO/C$^{18}$O $(J=3\rightarrow2)$ Heterodyne Inner Milky Way Plane Survey (CHIMPS; \citealt{Rigby16,Rigby19}), analogous to the clump-formation efficiency (CFE), and dense-gas mass fraction (DGMF), in two fields of the plane of the Milky Way centred at $\ell$\,=\,30$\degr$ and $\ell$\,=\,40$\degr$. We confirmed that the ratio of 850-$\upmu$m emission-derived column density to $^{13}$CO $J=3\rightarrow2$ column density was tracing the DGMF by using the simulated molecular clouds of \citet{Penaloza17,Penaloza18} to imitate the JPS and CHIMPS observations, and to determine the ``true'' DGMF. The ratio of these two methods has a very defined mean, and the two methods are well correlated.

We performed a power-spectrum analysis of these maps, and found breaks at size scales of 7.8\,pc and 15.2\,pc in $\ell$\,=\,30$\degr$ and $\ell$\,=\,40$\degr$ fields respectively. We split the two fields into the individual spiral arms that run across each field, and found breaks at size scales of 3.7\,pc, 15.8\,pc, 10.1\,pc, 7.2\,pc, and 10.5\,pc for the $\ell$\,=\,30$\degr$ Scutum--Centaurus, $\ell$\,=\,30$\degr$ Sagittarius, $\ell$\,=\,30$\degr$ Perseus, $\ell$\,=\,40$\degr$ Sagittarius, $\ell$\,=\,40$\degr$ Perseus spiral arms, respectively. The breaks in the spectra are determined to be the characteristic scale of CFE variations. This corresponds to the molecular-cloud scale. The power spectra of turbulent environments are in the form of a power law, and breaks indicate the scale at which turbulence is injected into a system. By corresponding to this scale, we can confirm that the largest variations in CFE, and star-formation efficiency, occur from cloud-to-cloud.

We extracted the DGMF/CFE of each molecular cloud within each individual spiral arm and find that the distributions for the individual spiral arms are consistent with a lognormal distribution. The three arms (Scutum--Centaurus, Sagittarius, and Perseus) had mean CFE values of 40, 41, and 46 per cent, respectively. The power spectra of the $^{13}$CO CHIMPS column-density maps were compared with the simulations of \citet{Federrath13}, and star-formation efficiency values were found for the spiral-arm segments to range from 0.28 -- 2.59 per cent for solenoidal-turbulence dominated systems, and 2.48-- 5.81 per cent for compressive turbulent systems. These trends are consistent with previous work \citep[e.g.][]{Eden15} and validate the use of the $L/M$ ratio as a proxy for star formation.

\section*{Acknowledgements}

The authors would like to thank the referee, Michael Burton, for a careful reading of the manuscript, and also Chris Brunt and Ivan Baldry for constructive discussions. DJE is supported by a STFC postdoctoral grant (ST/R000484/1). The JCMT has historically been operated by the Joint Astronomy Centre on behalf of the Science and Technology Facilities Council of the United Kingdom, the National Research Council of Canada and the Netherlands Organization for Scientific Research. Additional funds for the construction of SCUBA-2 were provided by the Canada Foundation for Innovation. This research has made use of NASA's Astrophysics Data System. The Starlink software \citep{Currie14} is currently supported by the East Asian Observatory. The data used in this paper are available in \citet{Eden17} and \citet{Rigby19}.

\section*{Data Availability}

The data used for this paper are available from the archives of the JCMT Plane Survey \citep{Eden17}\footnote{http://dx.doi.org/10.11570/17.0004} and the CO Heterodyne Inner Milky Way Plane Survey (CHIMPS) \citep{Rigby19}\footnote{https://doi.org/10.11570/19.0028}.

\bibliographystyle{mnras} 
\bibliography{cfe_map}

\appendix

\section{CDR of individual regions}
\label{cfes_individual}

\begin{figure*}
\begin{center}
\begin{tabular}{ll}
\includegraphics[width=0.49\textwidth]{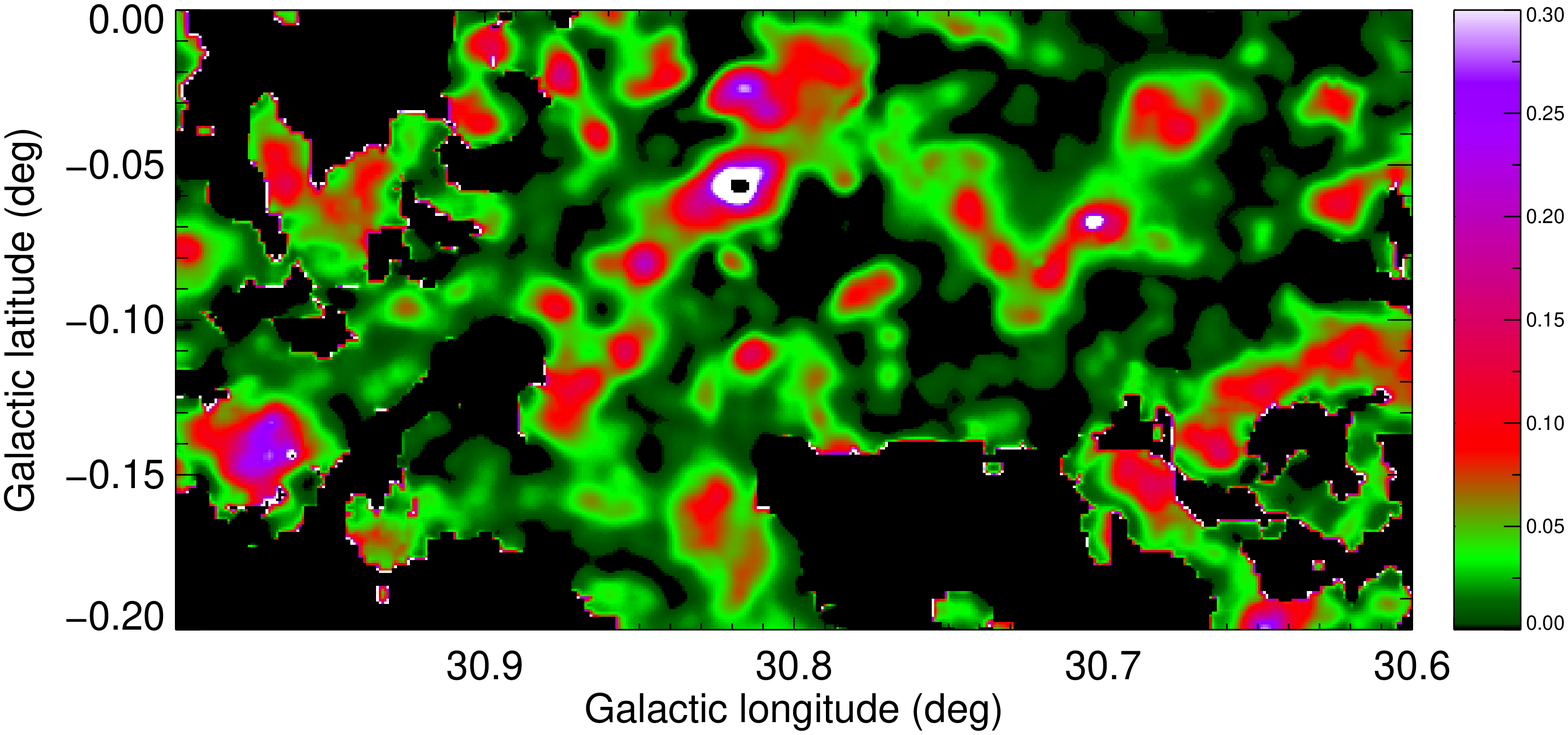} & \includegraphics[width=0.49\textwidth]{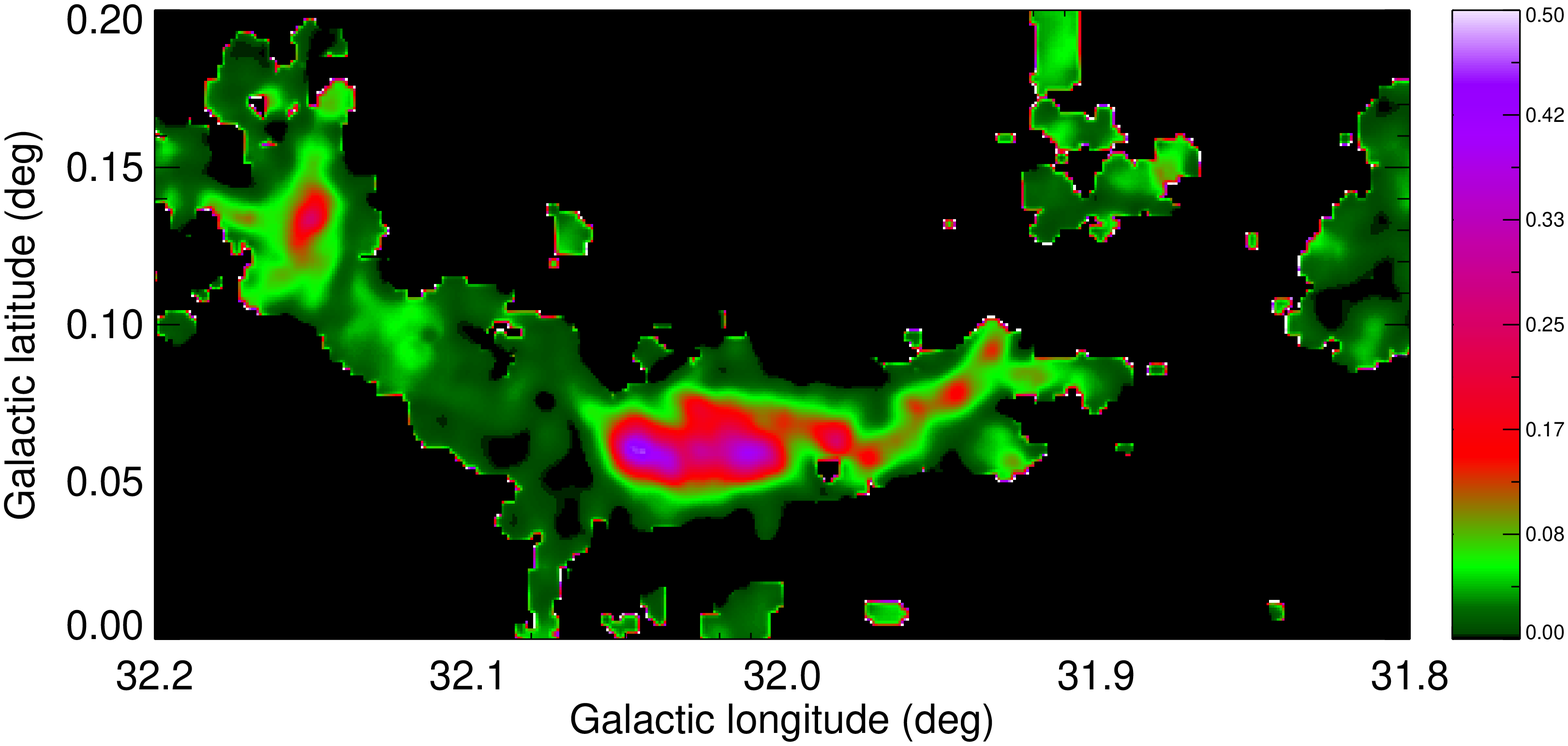}\\ \includegraphics[width=0.49\textwidth]{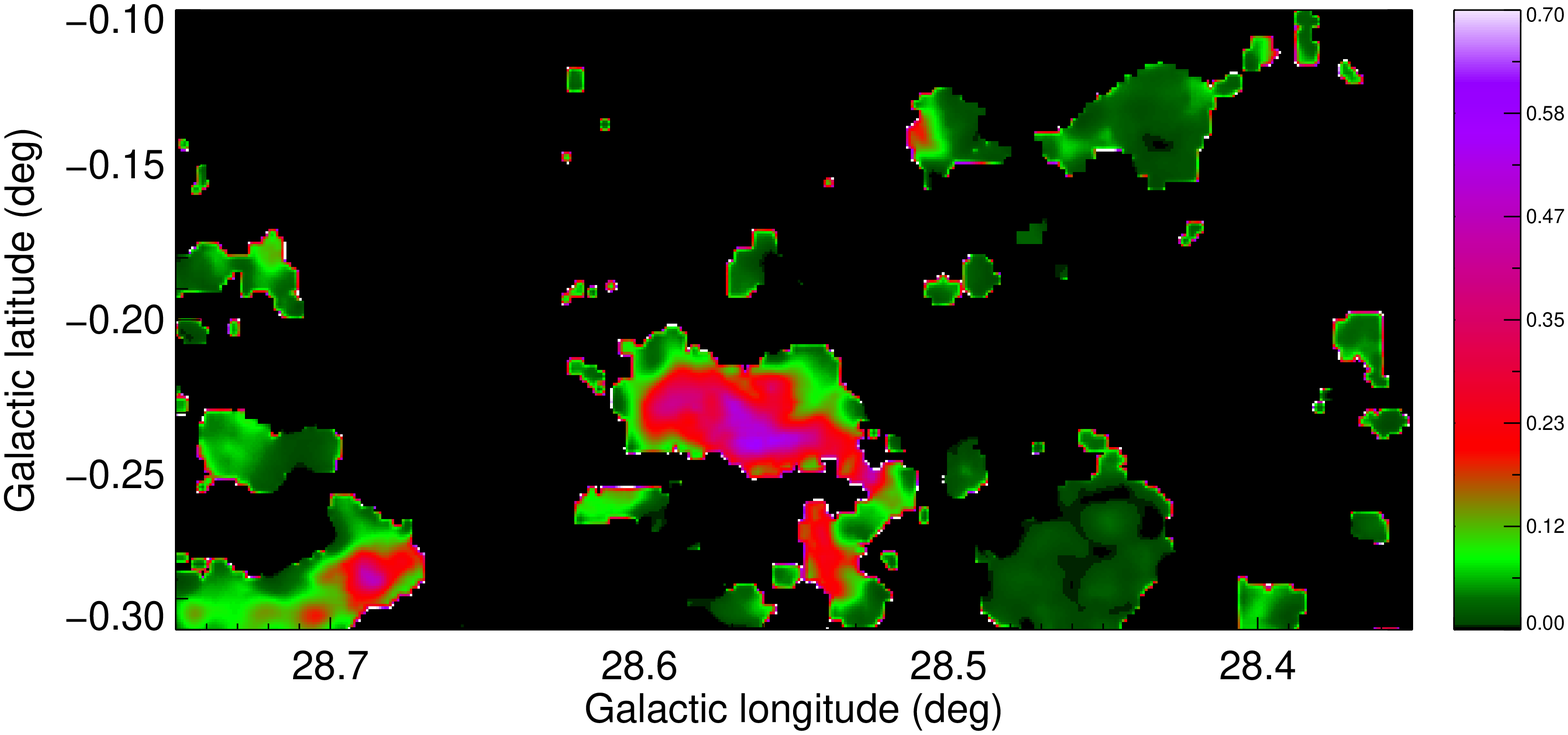} &
\includegraphics[width=0.49\textwidth]{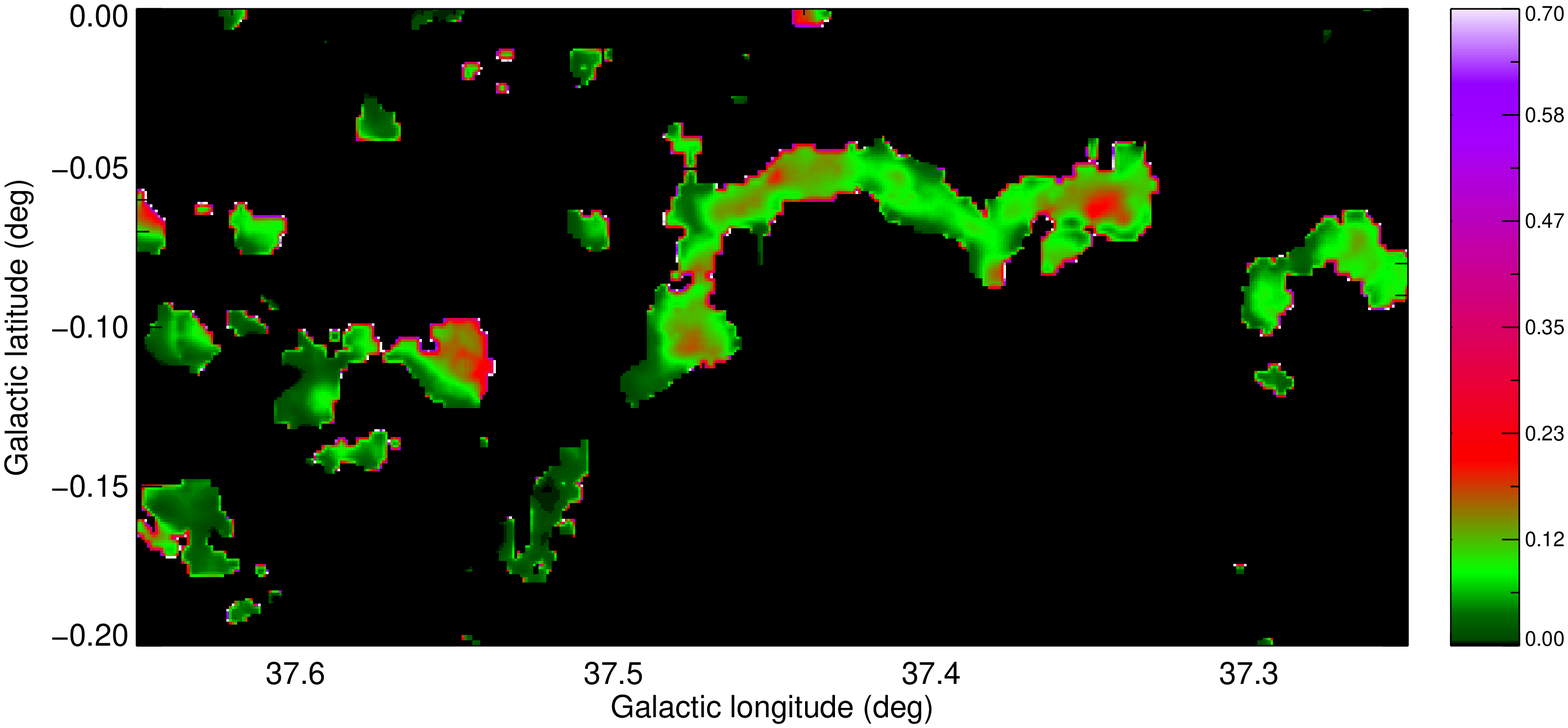} \\ \includegraphics[width=0.49\textwidth]{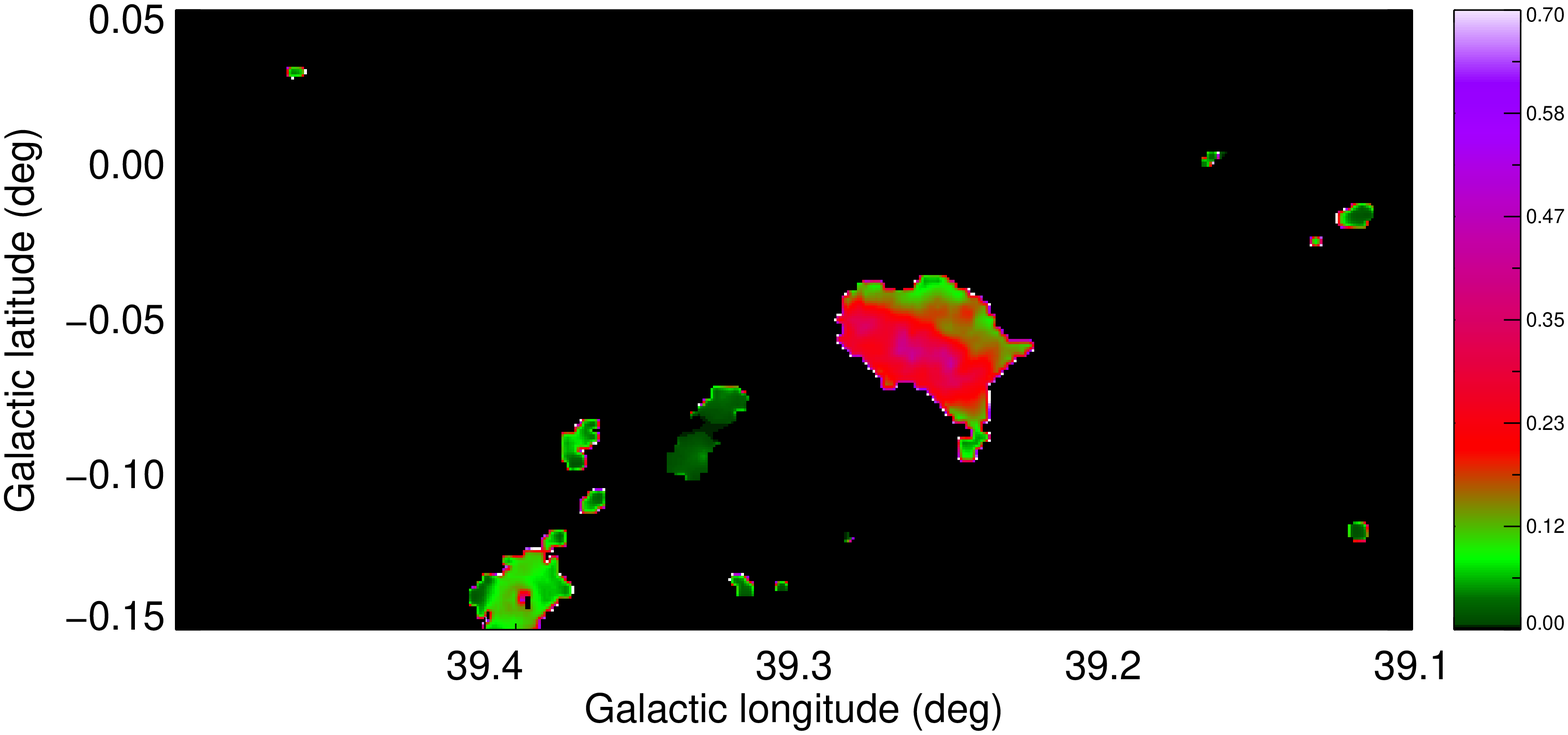} & \includegraphics[width=0.49\textwidth]{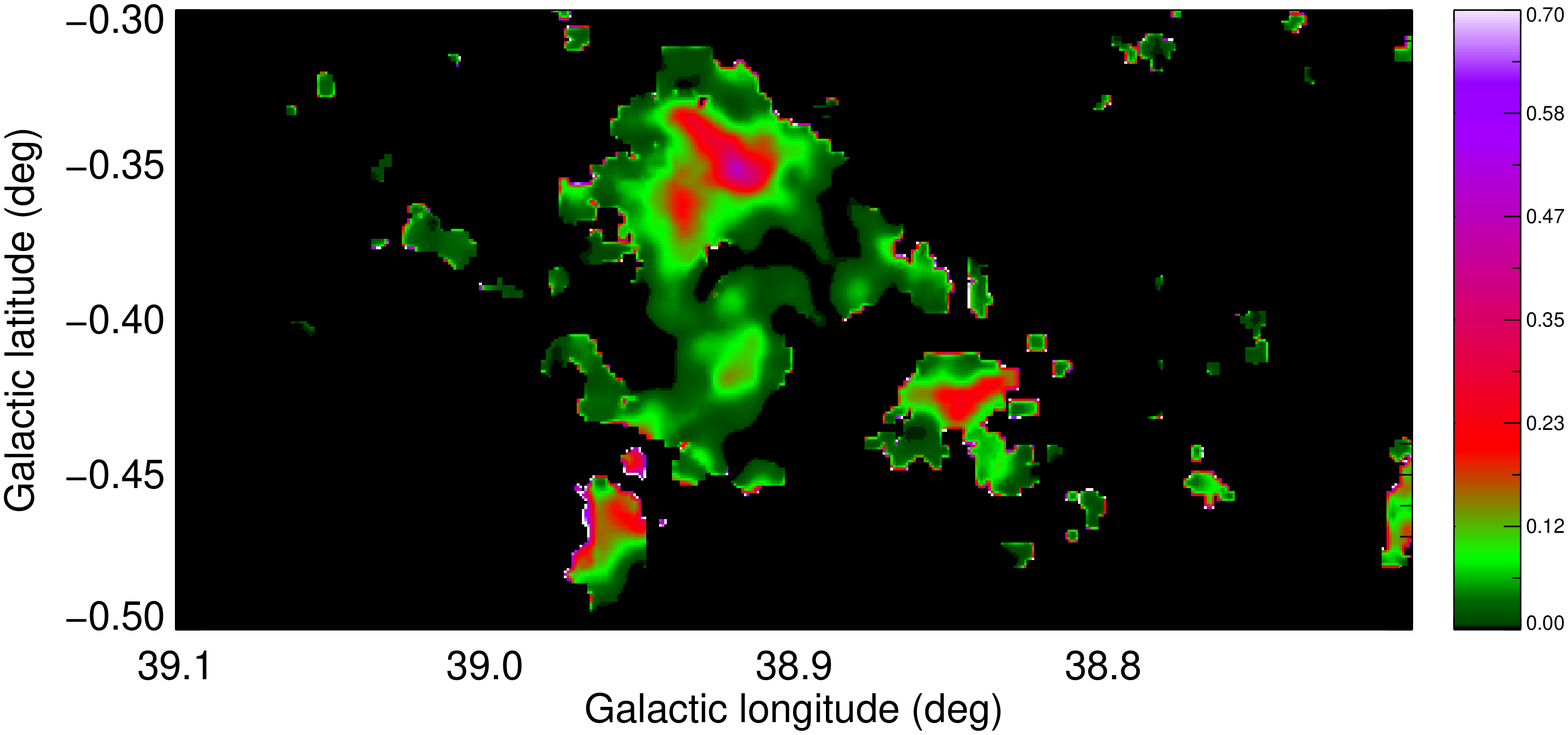} \\
\end{tabular}
\caption{CDR maps of elevated regions within the $\ell$\,=\,30$\degr$ and $\ell$\,=\,40$\degr$ fields. Top row: the W43 star-forming region and a giant filament containing an IRDC, in the left and right panel, respectively. Middle row: a star-forming clump from ATLASGAL and the molecular filament identified by \citet{Rigby16},  in the left and right panel, respectively. Bottom row: an ATLASGAL clump and an ATLASGAL cluster containing signatures of massive star formation, in the left and right panel, respectively.}
\label{closeups}
\end{center}
\end{figure*}

The CDR maps of individual regions within the $\ell$\,=\,30$\degr$ and $\ell$\,=\,40$\degr$ fields are displayed in Fig.~\ref{closeups}. These represent regions in which the CDR is elevated over the background level within each field.

\subsection{$\ell$\,=\,30$\degr$ regions}

\subsubsection{W43 star-forming region}

W43 is one of the most prominent star-forming regions in the Milky Way. It is located at the near end of the Galactic Long Bar \citep{NguyenLuong11} and contains a massive amount of gas and dust $M\,\sim\,6.5\,\times10^{6}$ \citep{NguyenLuong11}. The region is commonly referred to as a mini-starburst system due to the amount of star-forming material available and the predicted high future star-formation rate \citep{Motte03}. However, the global CFE and SFE of the region is found to be consistent with the rest of the Galaxy \citep{Eden12,Eden15}.

The individual sources of the W43 star-forming region do, though, show some variation in CDR, with the highest CDR region corresponding to the location of the UC\ion{H}{ii} region G30.667$-$0.209 \citep[e.g.][]{Bally10}, and a local CFE peak found by \citet{Eden12}.

\subsubsection{$\ell$ = 32$\degr$ filament}

A filament located at a longitude of $\sim$ $\ell$ = 32$\degr$ shows a CDR of $\sim$ 0.125, with a central peak comparable to the W43 value. This filament is as identified by \citet{Battersby14}, is likened to the giant molecular filaments described by \citet{Ragan14} with which it has a consistent CDR. It is also the site of an IRDC, which is determined to be above the threshold mass for high-mass star formation \citep{Zhou19}. 

\subsubsection{$\ell$ = 29$\degr$ massive YSO}

The most extended region with a CDR greater than 0.50 is found at a longitude of $\ell$ = 28$\fdg$6 and latitude of $\emph{b}$ = $-0\fdg$22. This region contains three ATLASGAL sources, one classed as protostellar and two as quiescent via SED fitting \citep{Konig17}, with those sources that were mid-IR weak but far-IR bright considered to be in the early stages of star formation \citep{Urquhart18}. The two quiescent sources had no evidence of 70-$\upmu$m emission, the presence of which is often used as a signpost for ongoing star formation \citep[e.g.][]{Ragan12,Traficante15}.

\subsection{$\ell$\,=\,40$\degr$ regions}

\subsubsection{$\ell$ = 37$\degr$ filament}

A filament identified in the CHIMPS survey located at a longitude of $\sim$ $\ell$ = 37$\fdg$5 \citep{Rigby16} is found to have an elevated CDR. Across its 20-pc length \citep{Li16}, the mean CDR is $\sim$\,0.12, with a peak of 0.25 at the western end. This peak corresponds to the position of a \ion{H}{ii} region \citep{Johnston09}.

\subsubsection{$\ell$ = 39$\degr$ ATLASGAL source}

A peak CDR of $\sim$\,0.35-0.40 is found at a longitude of $\ell$ = 39$\fdg$2 and corresponds to three ATLASGAL clumps, each classed as mid-IR bright and so assumed to be housing a YSO \citep{Konig17}. These ATLASGAL clumps have $L/M$ values below the median for the entire ATLASGAL survey and those clumps housing at least one YSO \citep{Urquhart18}, therefore these clumps are relatively cool and be at an early-stage of evolution.

\subsubsection{$\ell$ = 39$\degr$ massive star-forming region}

The elevated CDR at a longitude of $\sim$\,$\ell$ = 38$\fdg$9 is coincident with a cluster of eight ATLASGAL clumps \citep{Urquhart18} consisting of one protostellar source, five YSOs and two sites of massive-star formation containing luminous YSOs from the RMS survey \citep{Urquhart14}.

\section{Power Spectrum Tests}
\label{pstests}

The CDR maps make use of multiple data sets and two column density maps. Therefore, any power spectrum produced from these data and column density maps could include artefacts and features that may introduce breaks in the DGMF maps. To test this, we have run the same power-spectrum analysis on these maps to search for characteristic scales in the input data and so rule out spurious breaks.

\subsection{Column Density Maps}

\begin{figure*}
\begin{tabular}{ll}
\includegraphics[width=0.49\linewidth]{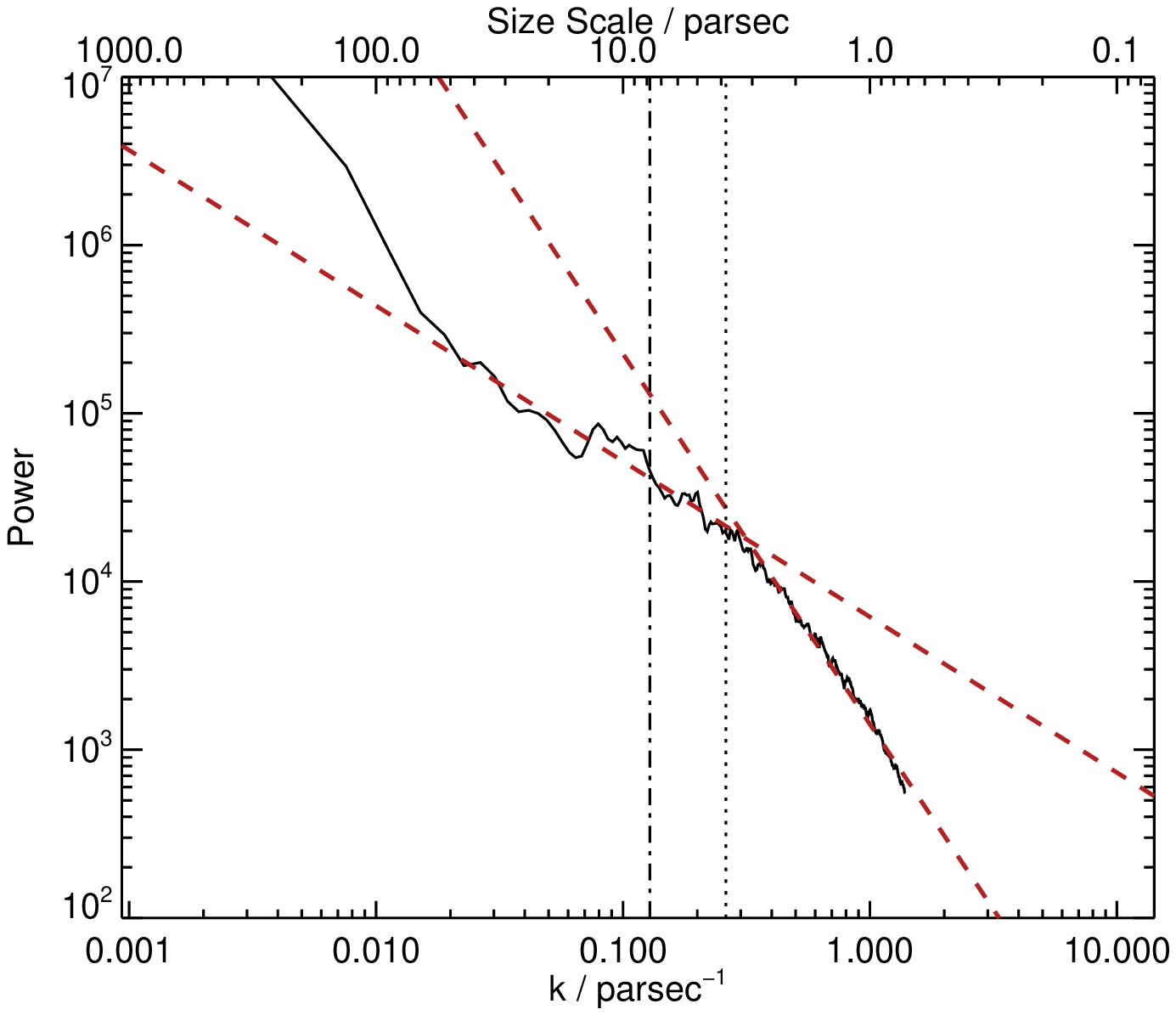} & \includegraphics[width=0.49\linewidth]{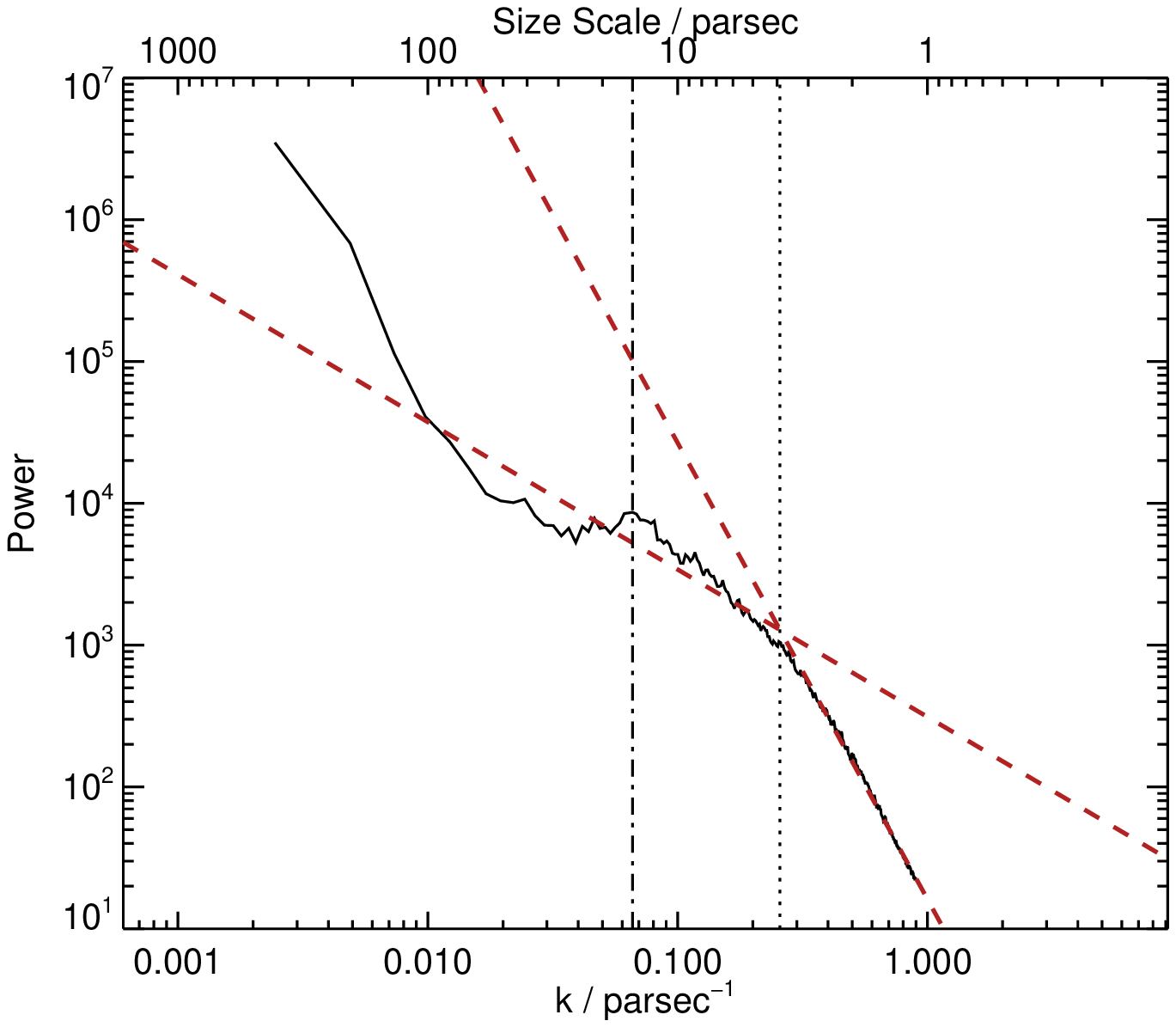} \\
\end{tabular}
\caption{The power spectra of the JPS column-density maps in the $\ell$\,=\,30$\degr$ (left panel) and $\ell$\,=\,40$\degr$ (right panel) fields. The dashed red lines represent the power-law fits to the high $k$ and low $k$ regimes in the spectrum. The vertical dotted line indicates the break between the two power-law fits, whereas the dash-dot line represents the break in the CDR power spectrum.}
\label{JPSCDPS}
\end{figure*}

\begin{figure*}
\begin{tabular}{ll}
\includegraphics[width=0.49\linewidth]{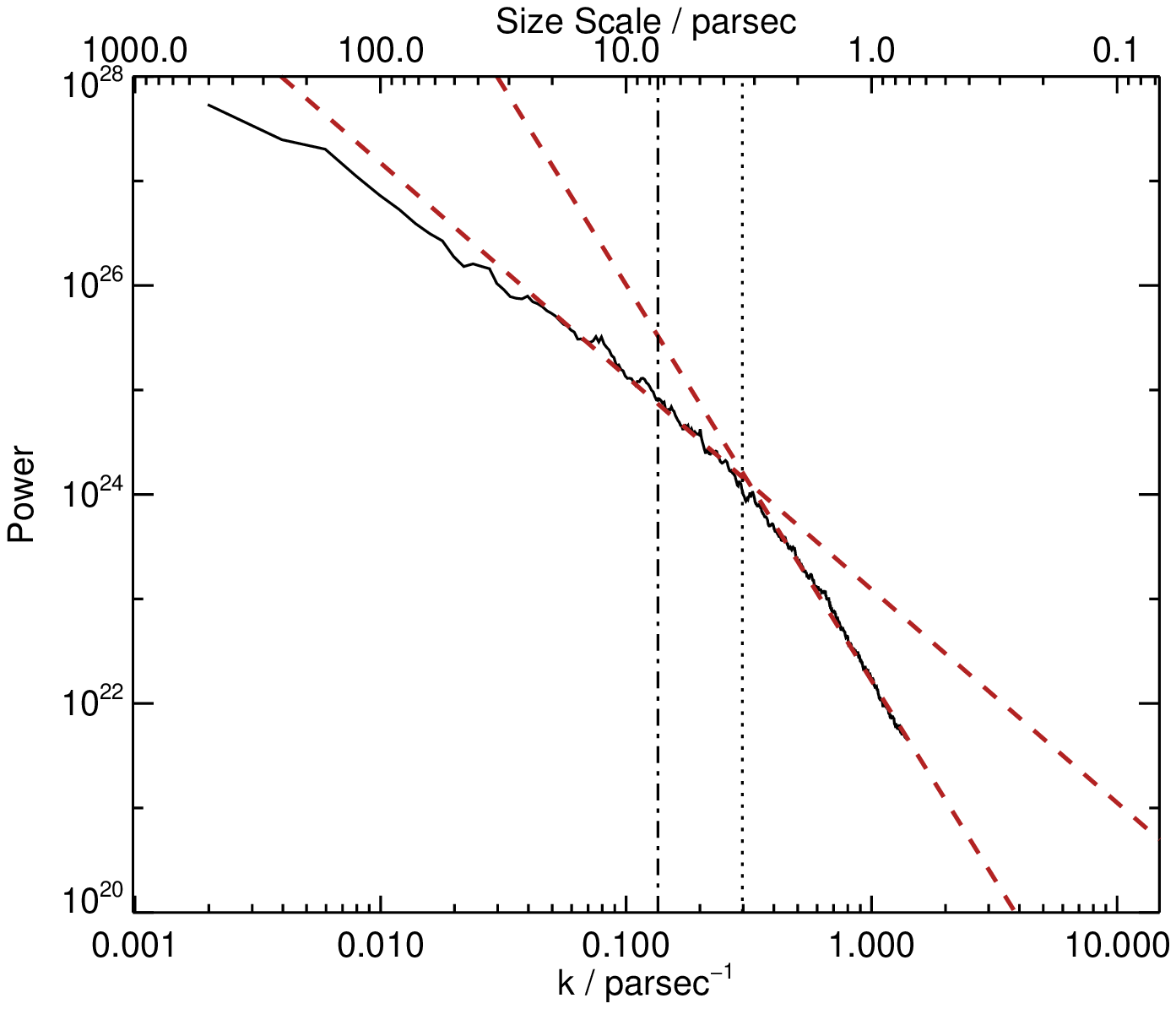} & \includegraphics[width=0.49\linewidth]{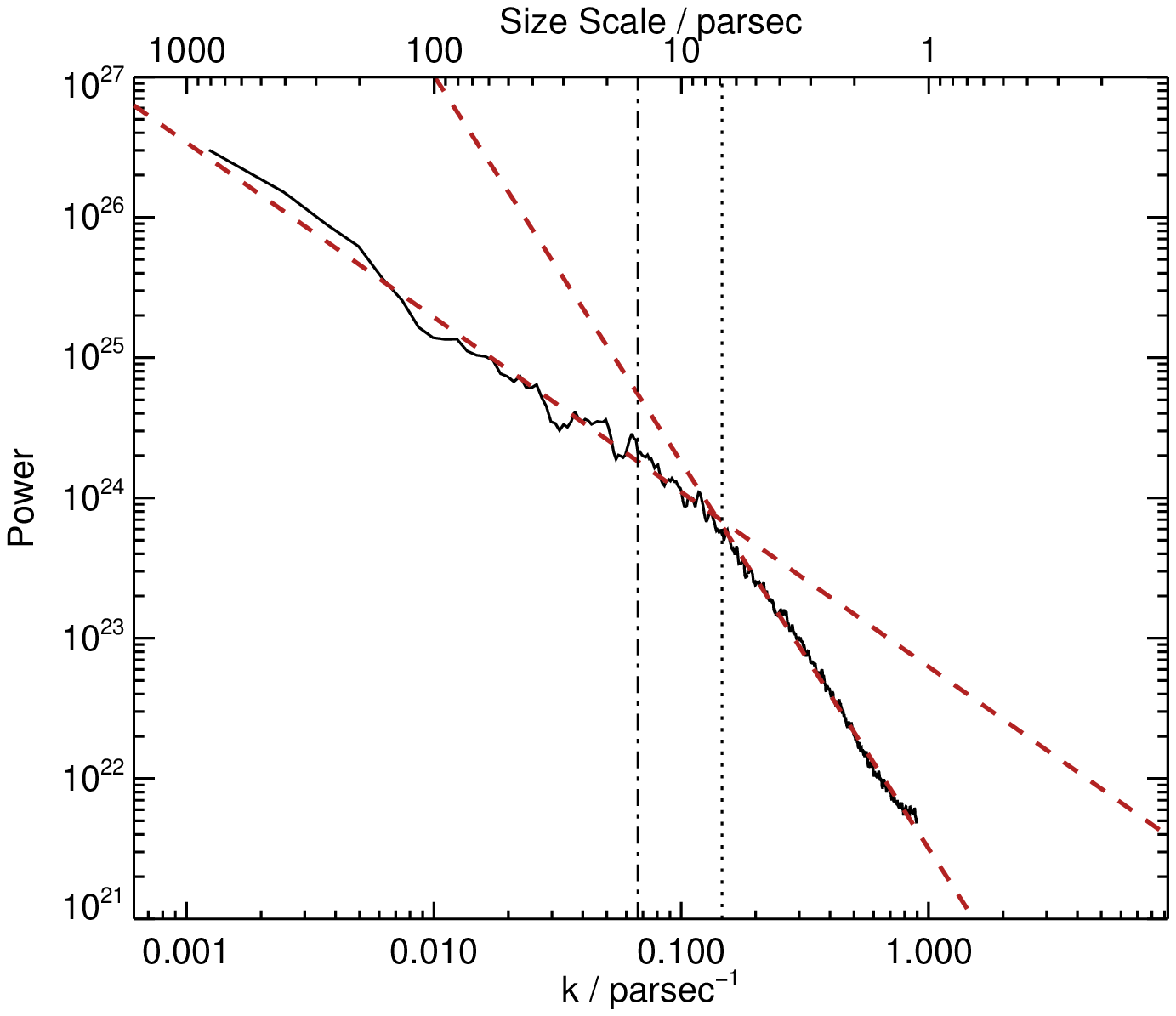} \\
\end{tabular}
\caption{Same as Fig.~\ref{JPSCDPS} but the power spectra of the CHIMPS column density maps.}
\label{CHIMPSCDPS}
\end{figure*}

The column-density maps of both the JPS and CHIMPS data are the direct input into the CDR calculations. The JPS column densities are produced using the \ppmap-derived temperatures and the JPS data. The JPS data are thresholded to include only emission above 3 $\upsigma$, whereas the CHIMPS column-density maps only include values above 3\,$\times$\,10$^{21}$\,cm$^{-2}$. These two limits cause the white space within the CDR maps seen in Figs~\ref{CFEmaps} and \ref{CFEmaps_arms}. One potential source of the break in the maps is their sparse nature. If this is responsible for the breaks, similar features should also be present in the column-density map power spectra. The power spectra of the JPS maps are shown in Fig.~\ref{JPSCDPS} and those of the CHIMPS data in Fig.~\ref{CHIMPSCDPS}. The break scale for the spectrum, along with the corresponding CDR break scale is displayed on each figure. The break scales found for the four maps are 3.84\,pc and 3.36\,pc for the JPS and CHIMPS column-density maps, respectively, in the $\ell$\,=\,30$\degr$ field, and 3.91\,pc and 6.85\,pc in the $\ell$\,=\,40$\degr$ field. These are compared to the CDR break scales of 7.78\,pc and 15.2\,pc. These break scales and the fits to the power laws are displayed in Table~\ref{psteststable}.

\begin{figure*}
\begin{tabular}{ll}
\includegraphics[width=0.49\linewidth]{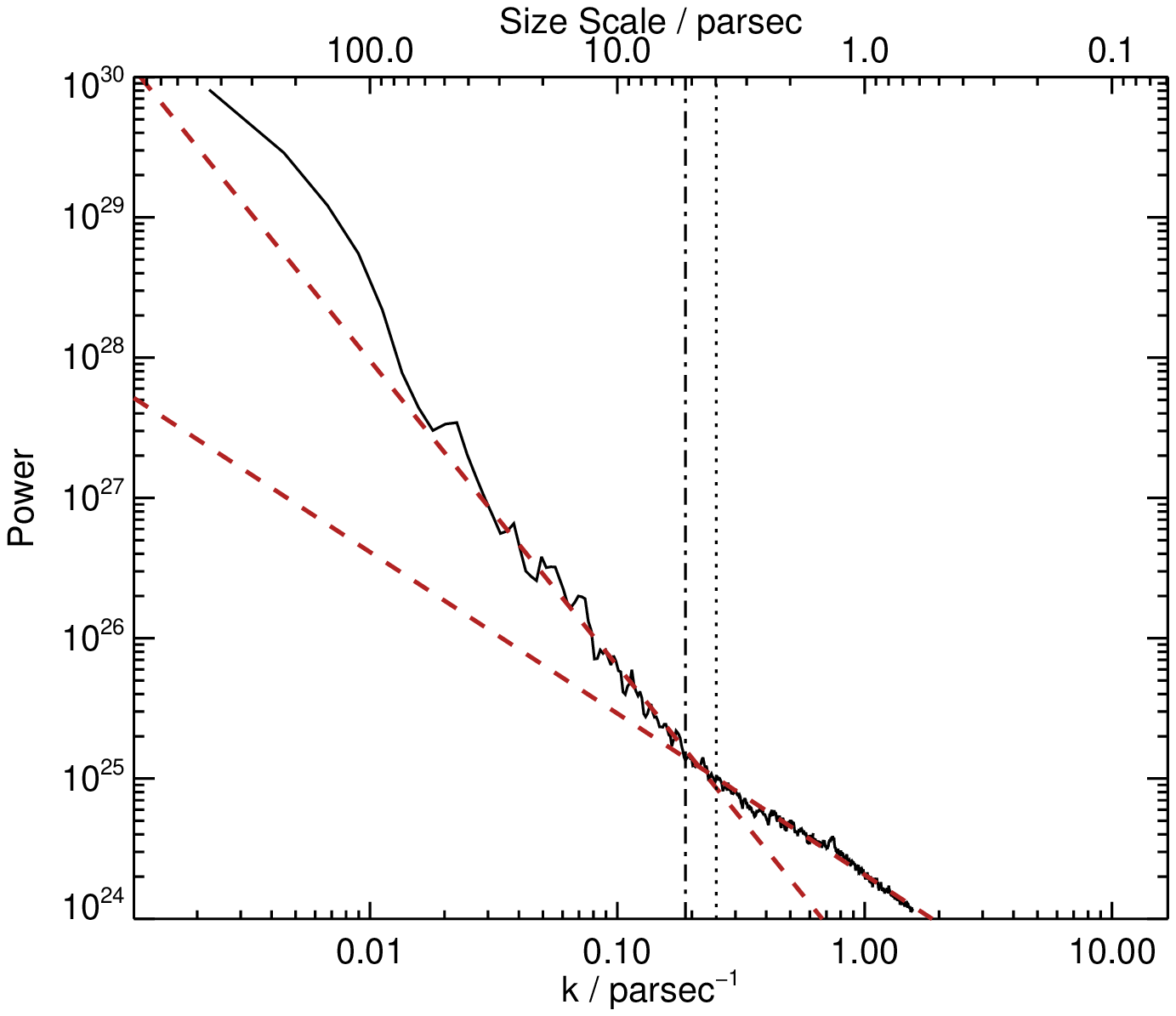} & \includegraphics[width=0.49\linewidth]{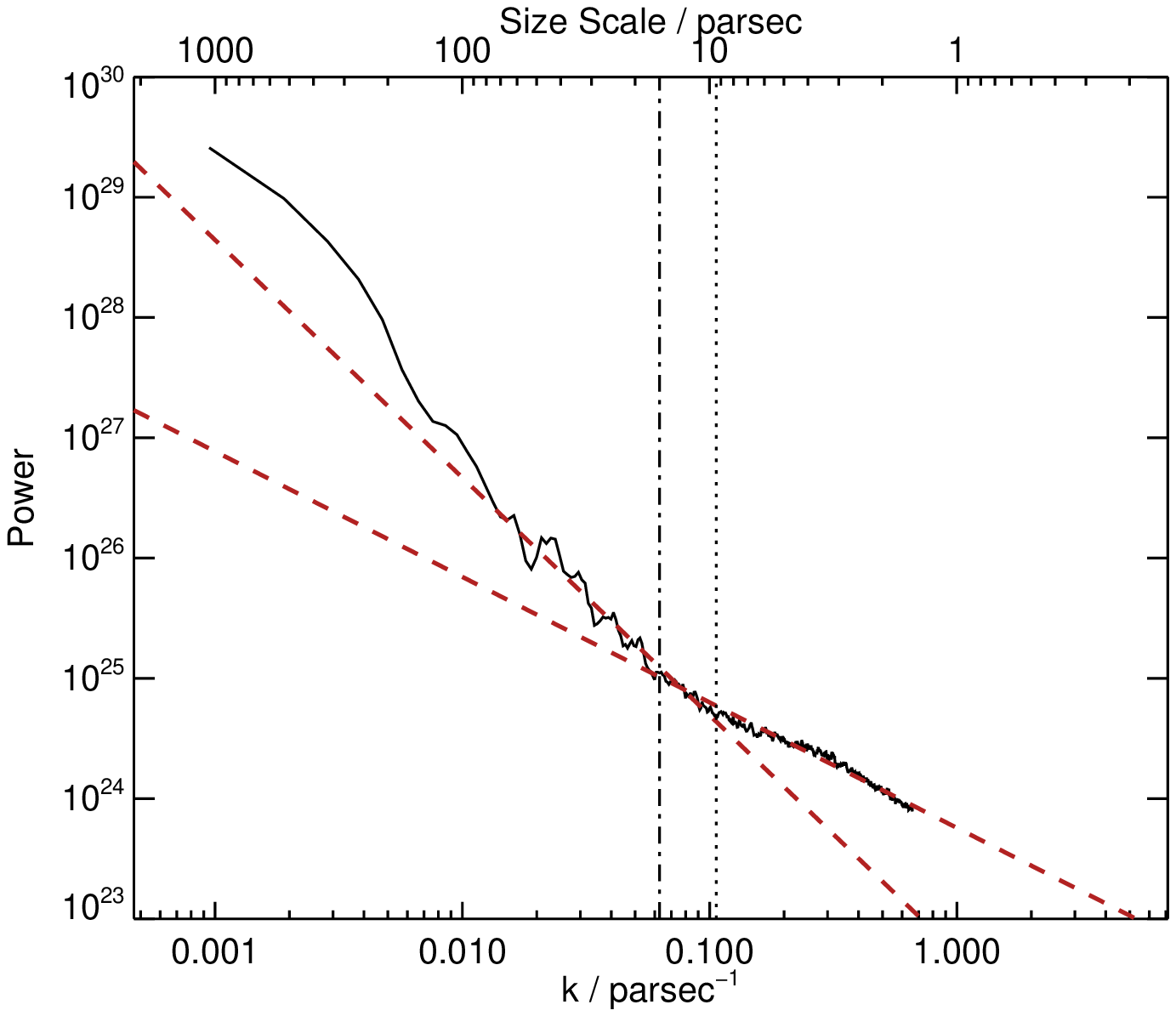} \\
\includegraphics[width=0.49\linewidth]{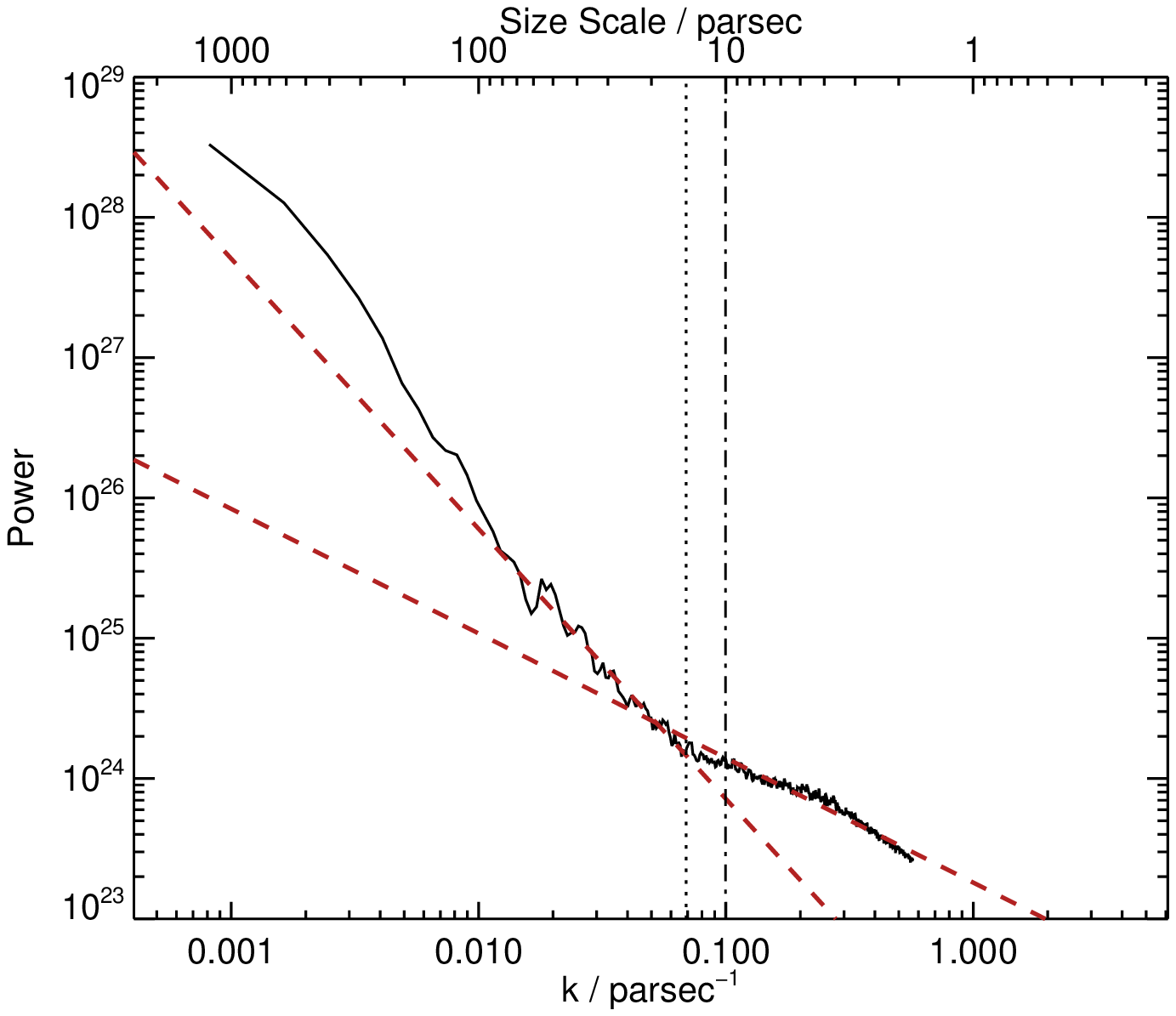} & \includegraphics[width=0.49\linewidth]{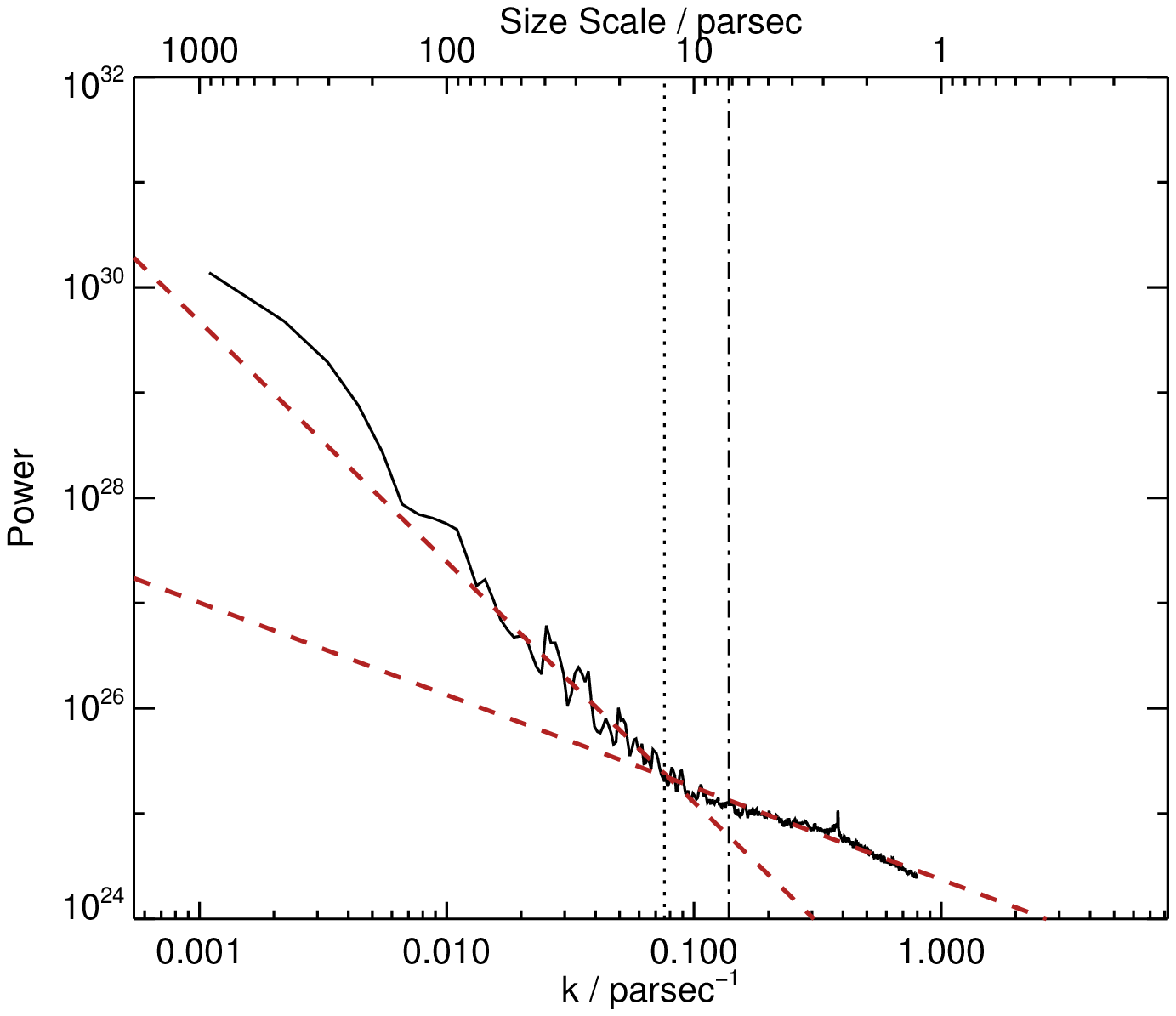} \\
\includegraphics[width=0.49\linewidth]{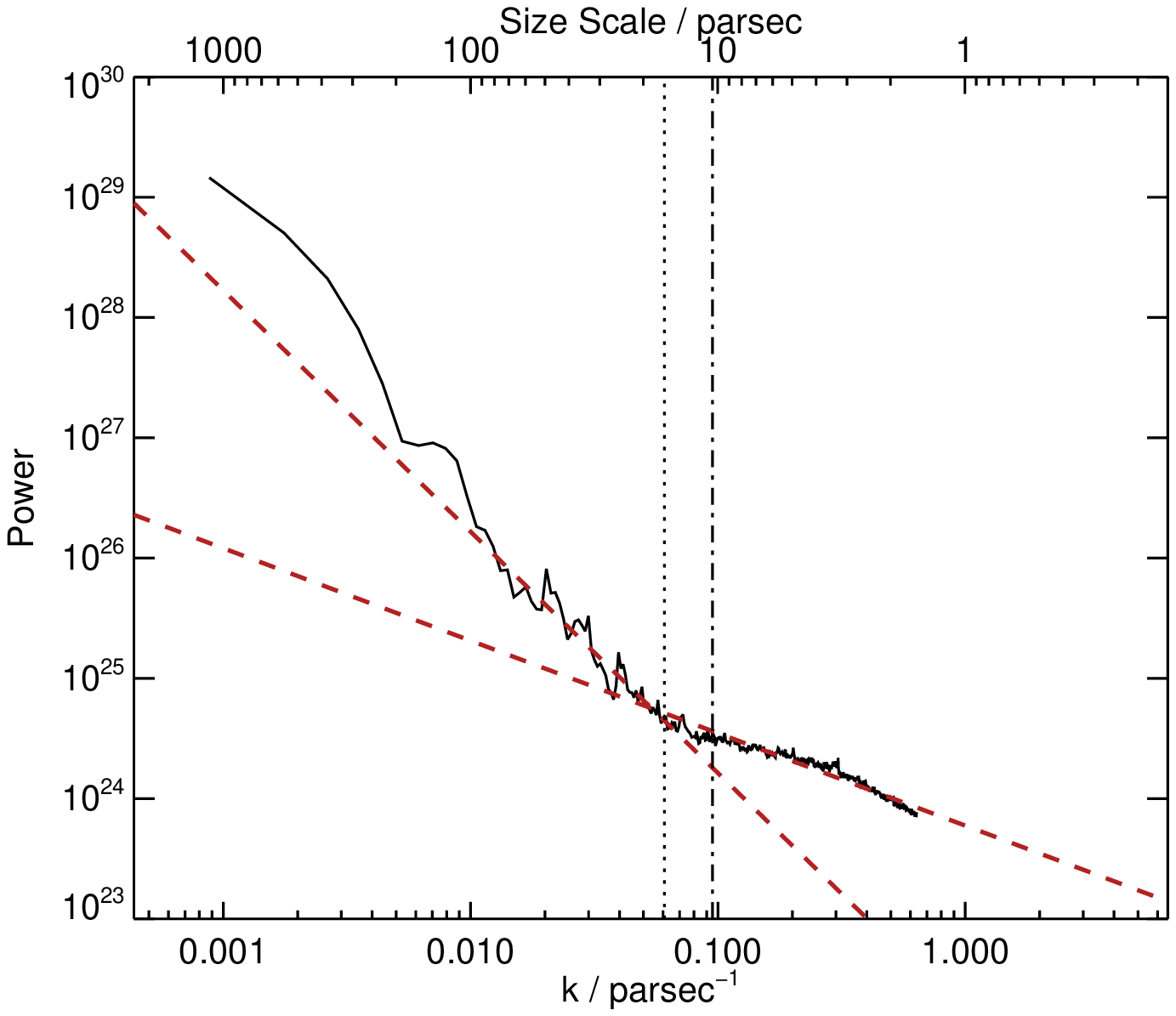} & \\
\end{tabular}
\caption{The power spectra of the CHIMPS column-density maps in the individual spiral arms within the $\ell$\,=\,30$\degr$ and $\ell$\,=\,40$\degr$ fields. The overplotted lines are as described in Fig.~\ref{JPSCDPS}. Top: $\ell$\,=\,30$\degr$ Scutum--Centaurus and $\ell$\,=\,30$\degr$ Sagittarius. Middle: $\ell$\,=\,30$\degr$ Perseus and $\ell$\,=\,40$\degr$ Sagittarius. Bottom: $\ell$\,=\,40$\degr$ Perseus.}
\label{CHIMPSCDPSarms}
\end{figure*}

We also tested the velocity slices used to produce the individual spiral-arm CDR power spectra. These CHIMPS column-density power spectra are displayed in Fig.~\ref{CHIMPSCDPSarms}. These maps, other than the $\ell$\,=\,30$\degr$ Scutum--Centaurus spiral arm, are very sparsely populated. However, as with the total field power spectra, the breaks are not coincident with those in the CDR power spectra. The breaks are found at scales of 3.97\,pc, 9.32\,pc, 14.6\,pc, 13.2\,pc, and 16.5\,pc for the $\ell$\,=\,30$\degr$ Scutum--Centaurus, $\ell$\,=\,30$\degr$ Sagittarius, $\ell$\,=\,30$\degr$ Perseus, $\ell$\,=\,40$\degr$ Sagittarius, and $\ell$\,=\,40$\degr$ Perseus arms, respectively, compared to the breaks in the CDR power spectra of 3.72\,pc, 15.8\,pc, 10.1\,pc, 7.20\,pc, and 10.5\,pc.

The shapes of the power spectra in Fig.~\ref{CHIMPSCDPSarms} are markedly different to those of the total field column density maps. The shape of these are consistent with isolated star-forming turbulent material, such as the simulations of \citep{Federrath13}.

\begin{figure*}
\begin{tabular}{ll}
\includegraphics[width=0.49\linewidth]{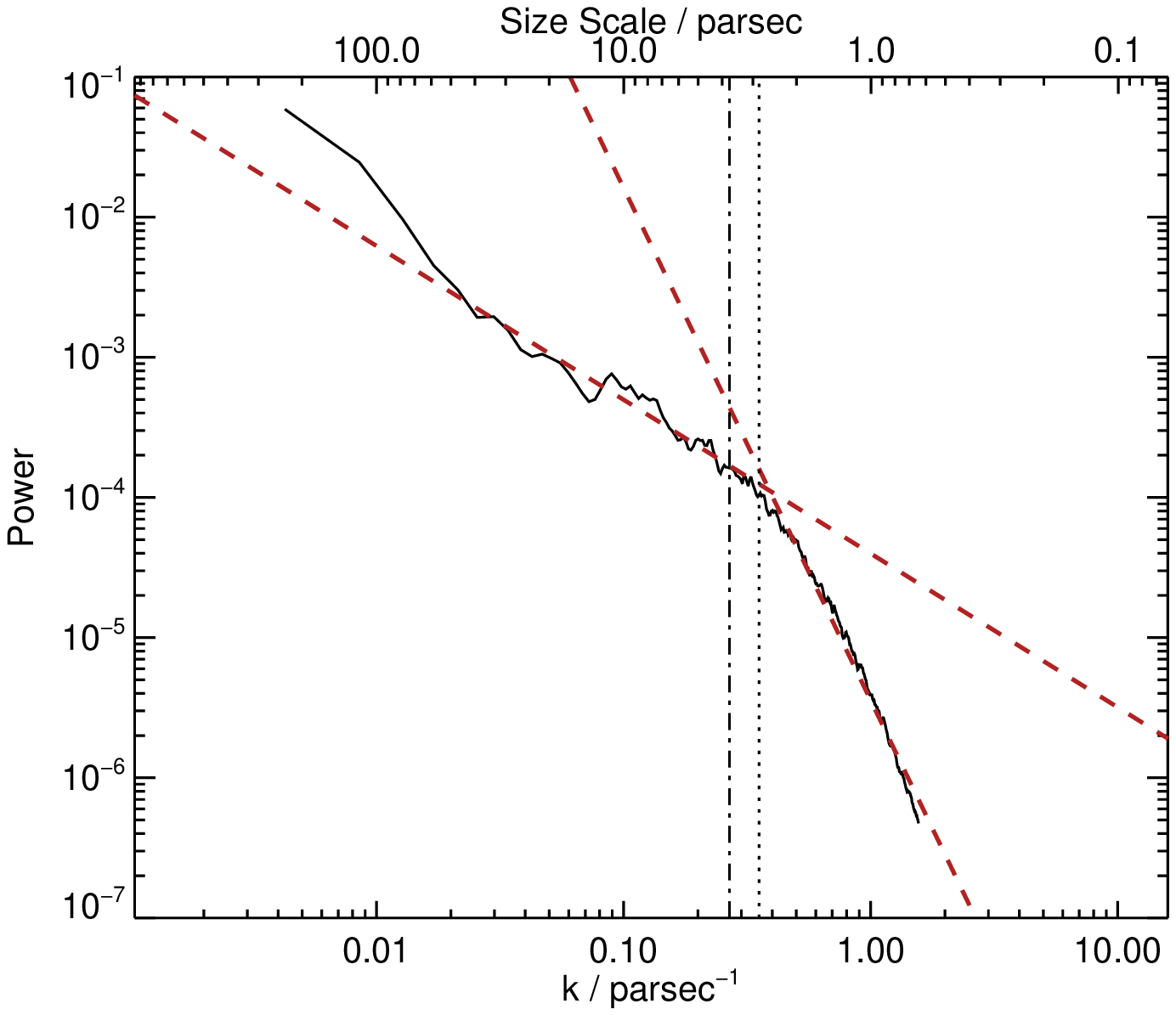} & \includegraphics[width=0.49\linewidth]{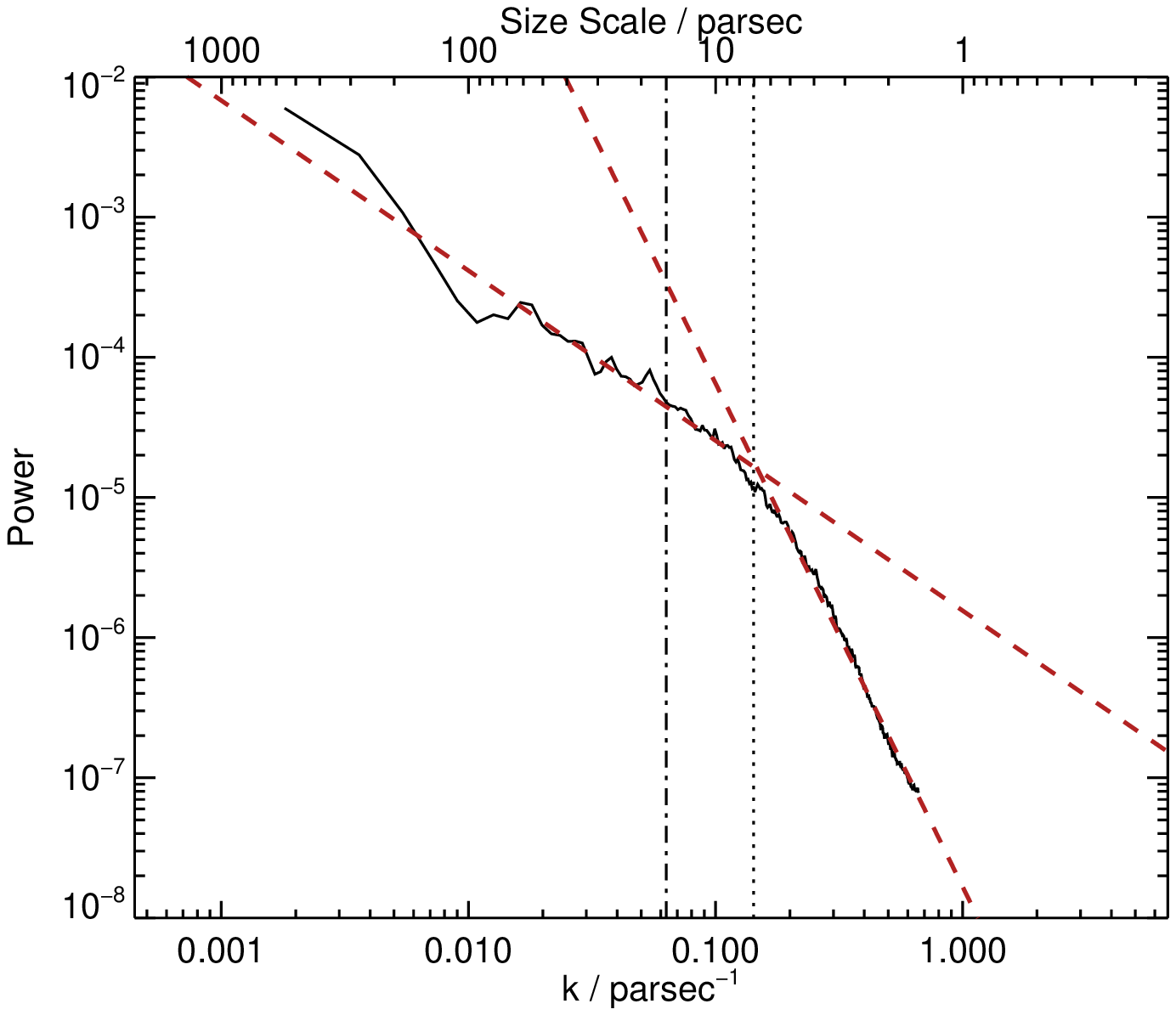} \\
\includegraphics[width=0.49\linewidth]{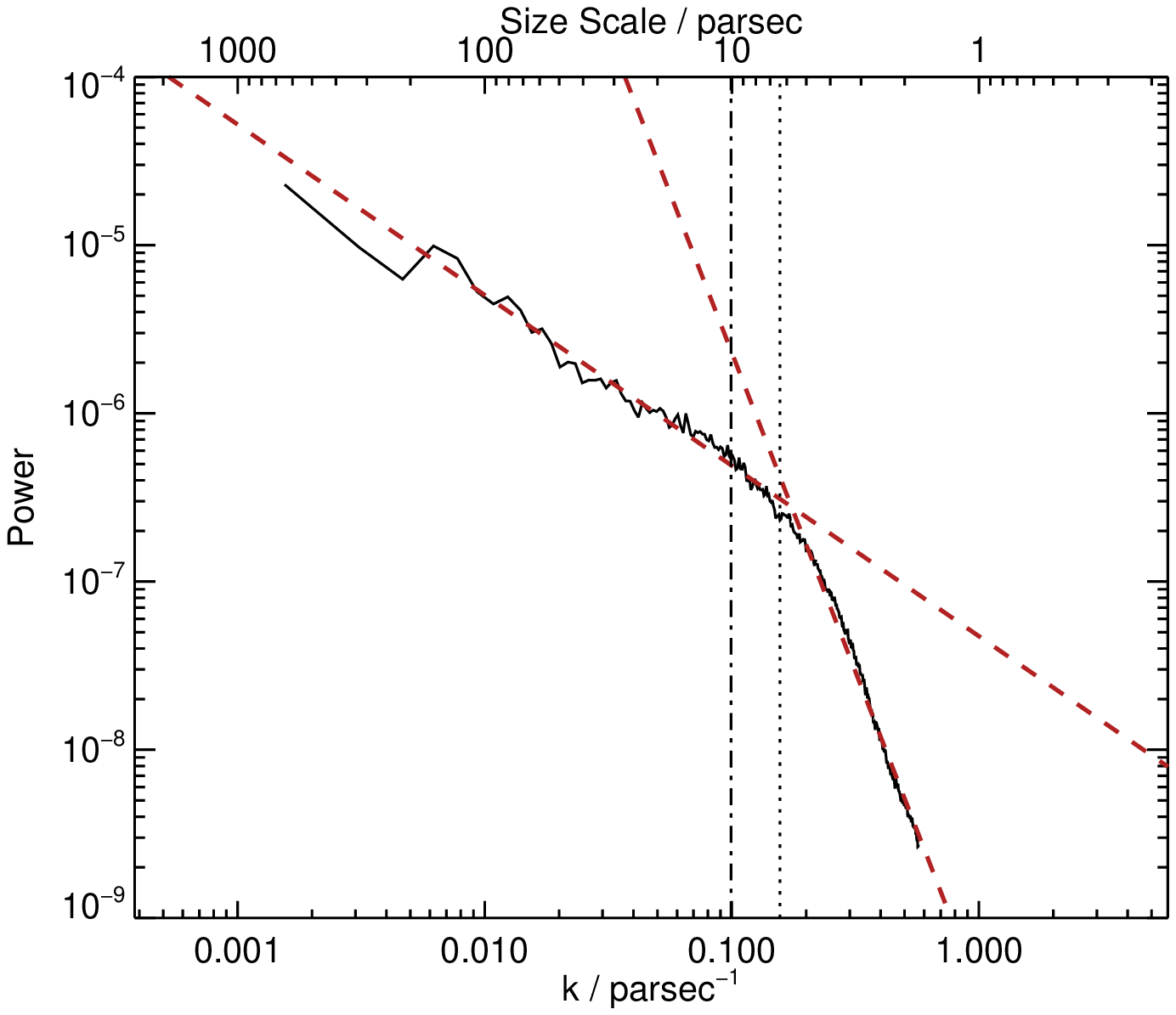} & \includegraphics[width=0.49\linewidth]{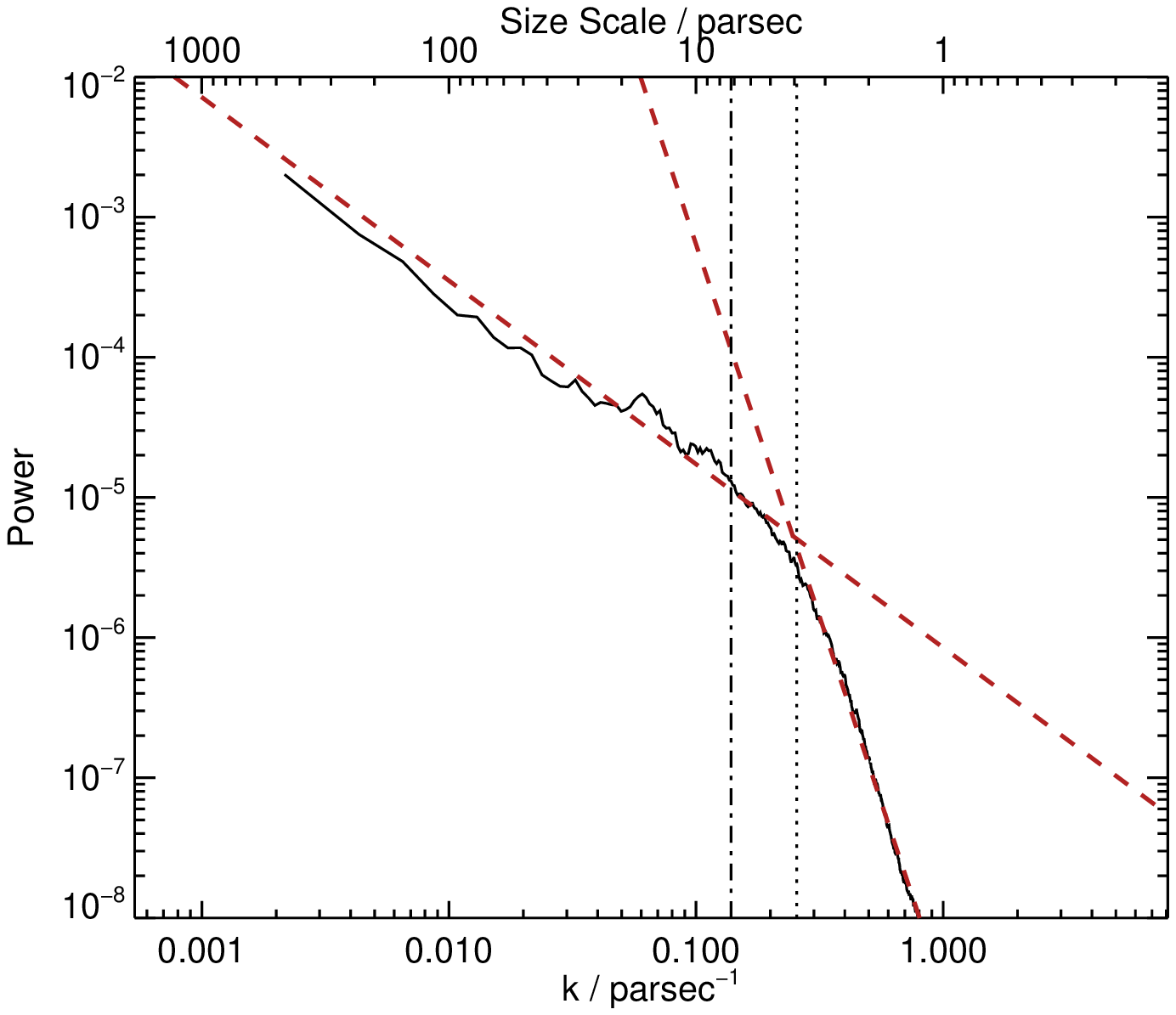} \\
\includegraphics[width=0.49\linewidth]{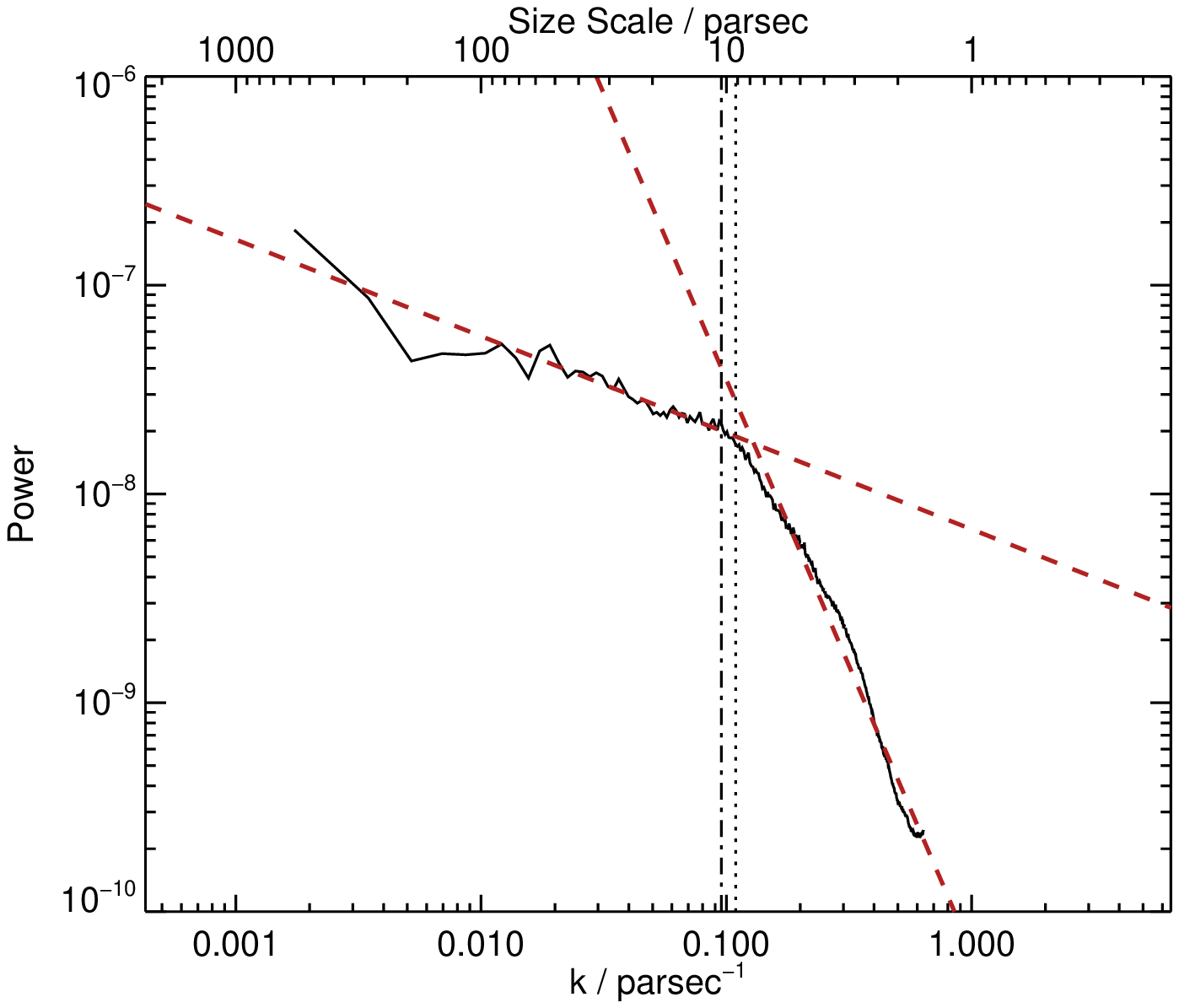} & \\
\end{tabular}
\caption{The power spectra of the JPS column-density maps in the individual spiral arms within the $\ell$\,=\,30$\degr$ and $\ell$\,=\,40$\degr$ fields. The overplotted lines are as described in Fig.~\ref{JPSCDPS}. Top: $\ell$\,=\,30$\degr$ Scutum--Centaurus and $\ell$\,=\,30$\degr$ Sagittarius. Middle: $\ell$\,=\,30$\degr$ Perseus and $\ell$\,=\,40$\degr$ Sagittarius. Bottom: $\ell$\,=\,40$\degr$ Perseus.}
\label{JPSCDPSarms}
\end{figure*}

The corresponding JPS power spectra are displayed in Fig.~\ref{JPSCDPSarms}. The breaks are not coincident to the breaks in the CDR power spectra with breaks at 2.78\,pc, 7.00\,pc, 6.40\,pc, 3.91\,pc, and 9.17\,pc for the $\ell$\,=\,30$\degr$ Scutum--Centaurus, $\ell$\,=\,30$\degr$ Sagittarius, $\ell$\,=\,30$\degr$ Perseus, $\ell$\,=\,40$\degr$ Sagittarius, and $\ell$\,=\,40$\degr$ Perseus arms, respectively.

\begin{table*}
\begin{center}
\caption{The power-law slopes and break points for the power spectra in the tests. the total fields and the spiral arms in each field. The large-scale power-law slopes represent the larger physical scales and the lowest values of $n$, whilst the small-scale power scales represent the inverse.}
\label{psteststable}
\begin{tabular}{llcccccc} \hline
Field & Data Set & Break & Break & CDR & CDR Break & Power Law & Power Law\\
 & & (pc) & Range (pc) & Break (pc) & Range (pc) & Large Scale & Small Scale\\
\hline
$\ell$\,=\,30$\degr$ & JPS Column Density & 3.84 & 3.21 -- 3.49 & 7.78 & 5.74 -- 11.5 & $-$0.92$\pm$0.28 & $-$2.16$\pm$0.25\\
$\ell$\,=\,40$\degr$ & JPS Column Density & 3.91 & 3.78 -- 3.98 & 15.2 & 14.0 -- 17.8 & $-$1.04$\pm$0.12 & $-$3.22$\pm$0.09\\
\hline
$\ell$\,=\,30$\degr$ & CHIMPS Column Density & 3.36 & 3.30 -- 3.47 & 7.78 & 5.74 -- 11.5 & $-$2.25$\pm$0.25 & $-$3.79$\pm$0.27\\
$\ell$\,=\,40$\degr$ & CHIMPS Column Density & 6.85 & 6.16 -- 7.76 & 15.2 & 14.0 -- 17.8 & $-$1.25$\pm$0.25 & $-$2.75$\pm$0.22\\
\hline
$\ell$\,=\,30$\degr$ & CHIMPS CD Scutum--Centaurus & 3.97 & 3.07 -- 3.43 & 3.72 & 3.65 -- 4.27 & $-$2.18$\pm$0.27 & $-$1.15$\pm$0.24\\
$\ell$\,=\,30$\degr$ & CHIMPS CD Sagittarius & 9.32 & 8.28 -- 10.7 & 15.8 & 13.8 -- 19.8 & $-$1.98$\pm$0.27 & $-$1.04$\pm$0.24\\
$\ell$\,=\,30$\degr$ & CHIMPS CD Perseus & 14.6 & 12.8 -- 16.3 & 10.1 & 9.48 -- 12.2 & $-$1.93$\pm$0.23 & $-$0.89$\pm$0.20\\
$\ell$\,=\,40$\degr$ & CHIMPS CD Sagittarius & 13.2 & 9.38 -- 13.6 & 7.20 & 6.42 -- 9.05 & $-$2.28$\pm$0.30 & $-$0.87$\pm$0.28\\
$\ell$\,=\,40$\degr$ & CHIMPS CD Perseus & 16.5 & 12.1 -- 17.2 & 10.5 & 9.68 -- 12.0 & $-$2.01$\pm$0.29 & $-$0.77$\pm$0.36\\
\hline
$\ell$\,=\,30$\degr$ & JPS CD Scutum--Centaurus & 2.78 & 2.71 -- 2.82 & 3.72 & 3.65 -- 4.27 & $-$1.10$\pm$0.14 & $-$3.65$\pm$0.16\\
$\ell$\,=\,30$\degr$ & JPS CD Sagittarius & 7.00 & 6.84 -- 7.10 & 15.8 & 13.8 -- 19.8 & $-$1.21$\pm$0.18 & $-$3.59$\pm$0.19\\
$\ell$\,=\,30$\degr$ & JPS CD Perseus & 6.40 & 6.24 -- 6.44 & 10.1 & 9.48 -- 12.2 & $-$1.01$\pm$0.21 & $-$3.79$\pm$0.29\\
$\ell$\,=\,40$\degr$ & JPS CD Sagittarius & 3.91 & 3.85 -- 3.99 & 7.20 & 6.42 -- 9.05 & $-$1.31$\pm$0.14 & $-$5.31$\pm$0.30\\
$\ell$\,=\,40$\degr$ & JPS CD Perseus & 9.17 & 9.00 -- 9.45 & 10.5 & 9.68 -- 12.0 & $-$0.46$\pm$0.13 & $-$2.74$\pm$0.24\\
\hline
$\ell$\,=\,30$\degr$ & JPS Data & 1.81 & 1.79 -- 1.84 & 7.78 & 5.74 -- 11.5 & $-$1.28$\pm$0.09 & $-$3.04$\pm$0.28\\
$\ell$\,=\,40$\degr$ & JPS Data & 2.60 & 2.52 -- 2.63 & 15.2 & 14.0 -- 17.8 & $-$1.17$\pm$0.11 & $-$3.16$\pm$0.38\\
\hline
$\ell$\,=\,30$\degr$ & CHIMPS Data & 5.06 & 4.82 -- 5.10 & 7.78 & 5.74 -- 11.5 & $-$2.25$\pm$0.25 & $-$3.05$\pm$0.14\\
$\ell$\,=\,40$\degr$ & CHIMPS Data & 13.4 & 12.4 -- 13.8 & 15.2 & 14.0 -- 17.8 & $-$1.02$\pm$0.25 & $-$1.79$\pm$0.11\\
\hline
\end{tabular}
\end{center}
\end{table*}

\subsection{Data}

\begin{figure*}
\begin{tabular}{ll}
\includegraphics[width=0.49\linewidth]{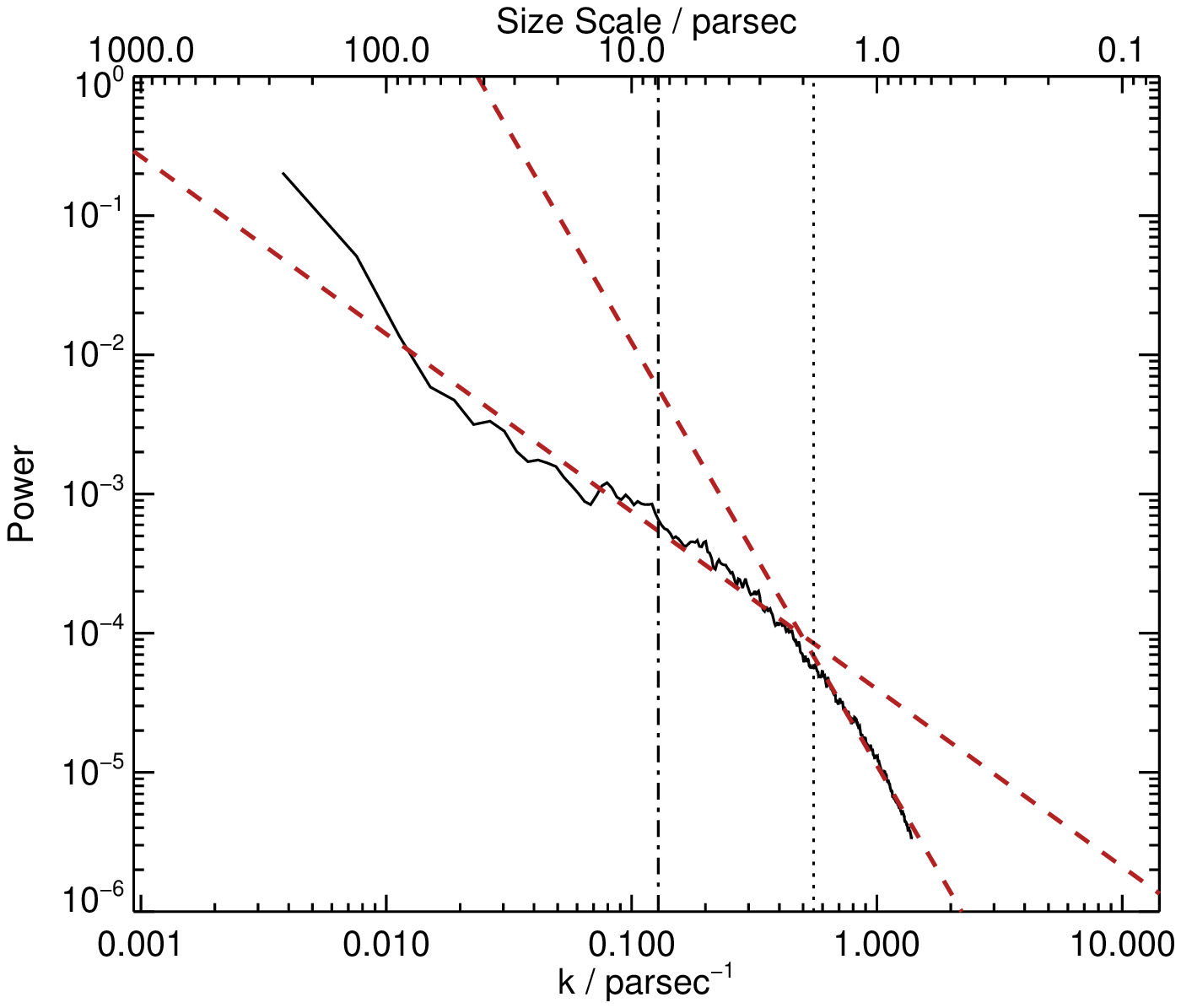} & \includegraphics[width=0.49\linewidth]{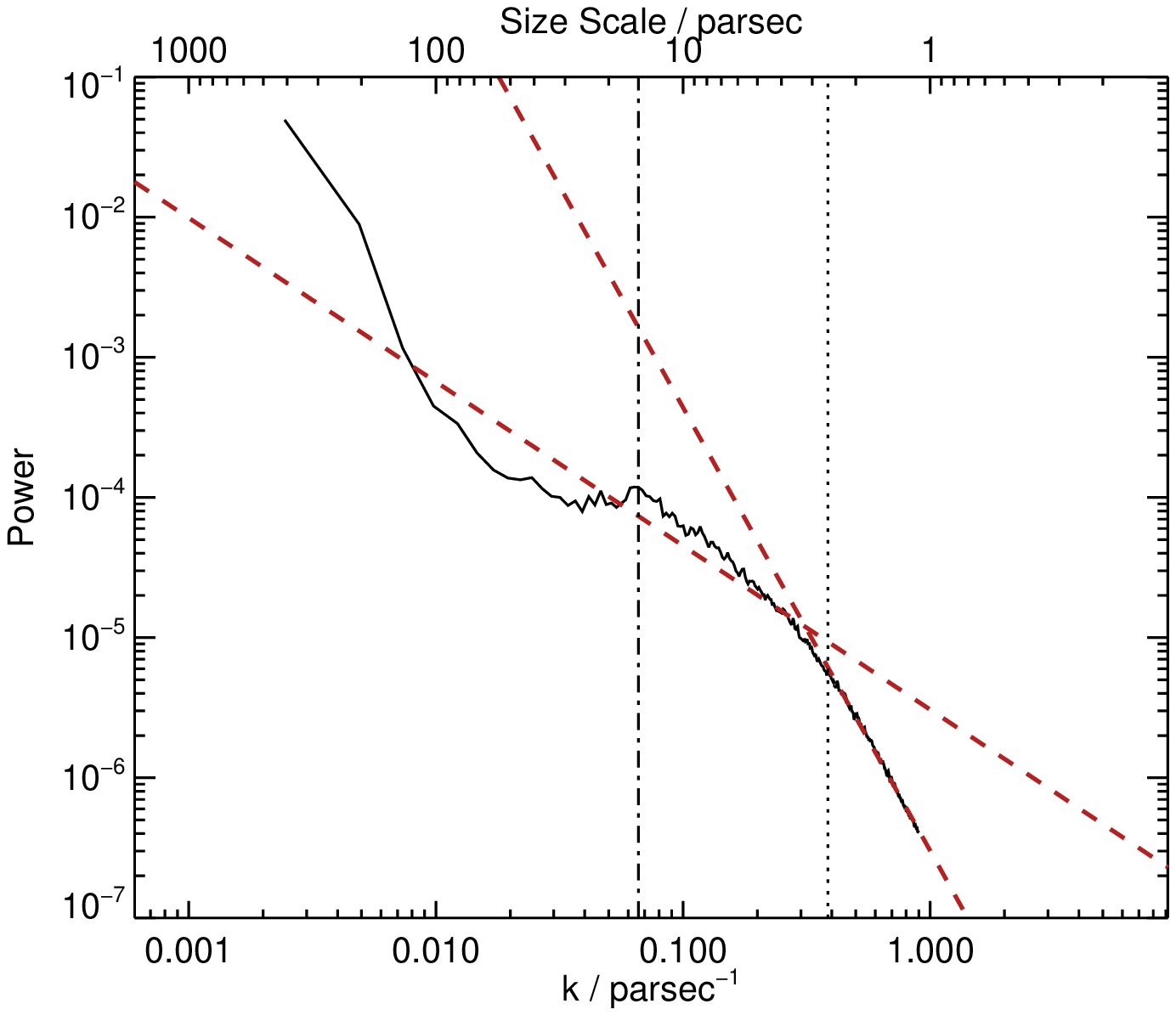} \\
\end{tabular}
\caption{The power spectra of the JPS data in the $\ell$\,=\,30$\degr$ (left panel) and $\ell$\,=\,40$\degr$ (right panel) fields. The dashed red lines represent the power-law fits to the high $k$ and low $k$ regimes in the spectrum. The vertical dotted lines indicate the breaks between the two power-law fits, whereas the dash-dot lines represent the positions of the breaks in the CDR power spectrum.}
\label{JPSDataPS}
\end{figure*}

As the JPS column-density maps were made by scaling the data by the \ppmap\ temperatures, the original data also needs to be investigated. The power spectra are shown in Fig.~\ref{JPSDataPS}. As with the column-density maps, the break scale in the spectrum, along with the corresponding CDR break scale, is displayed on each figure. The breaks in these spectra occur at size scales that do not correspond to the CDR breaks, with sizes of 1.81\,pc and 2.60\,pc in the $\ell$\,=\,30$\degr$ and $\ell$\,=\,40$\degr$ fields, respectively, compared to the 7.78\,pc and 15.2\,pc breaks in the CDR maps. These breaks are included in Table~\ref{psteststable}.

There are other features in the JPS data that may also be present in the power spectra, that would cause the breaks. The JPS fields were constructed as a series of $\emph{pong3600}$ maps, which have a diameter of 1 degree \citep{Bintley14}. At the distances assumed for the $\ell$\,=\,30$\degr$ and $\ell$\,=\,40$\degr$ fields, this would refer to a size scale of 96\,pc and 148\,pc. The second scale that could be reflected in the power spectra is the large-scale filtering by SCUBA-2, which occurs at 8 arcmin \citep{Chapin13}. This is also not coincident with either the break of the CDR power spectra, or that of the JPS data.

The CHIMPS maps are not subject to the spatial filtering of the JPS data and so would not be expected to show any signatures that would be reflected in the CDR distributions, other than the size of the individual component map elements employed in the survey of 22 arcmins \citep{Rigby16}. However, since the breaks in the CFE maps do not occur near 22 arcmins (4.86 arcmins and 6.13 arcmins, respectively), this scale is not reflected in the CDR spectra. Other features within the data, such as the pixel scale and beam size (14.5 arcsec) are below the lowest scales probed, and lie at higher values of $k$.

The power spectra of the data can indicate the nature of the turbulence that is injected into a system. The CHIMPS intensity data have slopes of $-3.05\,\pm\,0.14$ and $-1.79\,\pm\,0.11$ in the $\ell$\,=\,30$\degr$ and $\ell$\,=\,40$\degr$ fields, respectively.  Following the advice of \citet{Lazarian00}, the three-dimensional density field can be inferred if the velocity dispersions are smaller than the width of the integration. Since the smallest integration occurring in this study is 15\,km\,s$^{-1}$, and the largest velocity dispersion found in the CHIMPS survey is less than 5\,km\,s$^{-1}$ \citep{Rigby19}, we can assume that this applies here. It is found that the integrated-intensity power spectra have a universal slope of $-3$ in optically thick media \citep{Lazarian04,Burkhart13}. The value for the $\ell$\,=\,30$\degr$ slope is consistent with $-3$, suggesting that the data are subject to optical-depth issues, which are independent of Mach number. However, this is surprising in $^{13}$CO $J=3\rightarrow2$ emission, which is thought to be optically thin \citep{Rigby19}. If it is not subject to these effects, the power spectrum has a slope that is somewhat shallower than Kolmogorov turbulence \citep{Kolmogorov41}, which would have an index of $-11/3$, and this is consistent with the Mach numbers found in the CHIMPS data, which are greater than 1 in all sources \citep{Rigby19} and would indicate that the clouds are supersonic but without self-gravity. \citet{Burkhart13} find that a slope of $-3$ is consistent with sub-Alfv\'{e}nic turbulence. The $\ell$\,=\,40$\degr$ field slope is significantly shallower than Kolmogorov turbulence. The shallower the slope becomes, the more gravity takes over in the supersonic structures \citep[e.g.][]{Collins12,Burkhart15}, which would resemble the sparse nature of the field, which resembles individual, isolated, self-gravitating clouds.

Within the JPS data, slopes are found of $-3.04\,\pm\,0.28$ and $-3.16\,\pm\,0.28$. These values are consistent with the predictions of \citet{Padoan97} and \citet{Burkhart15} for the power laws that would be found in dust-continuum images of self-gravitating, supersonic turbulence.

\begin{figure*}
\begin{tabular}{ll}
\includegraphics[width=0.49\linewidth]{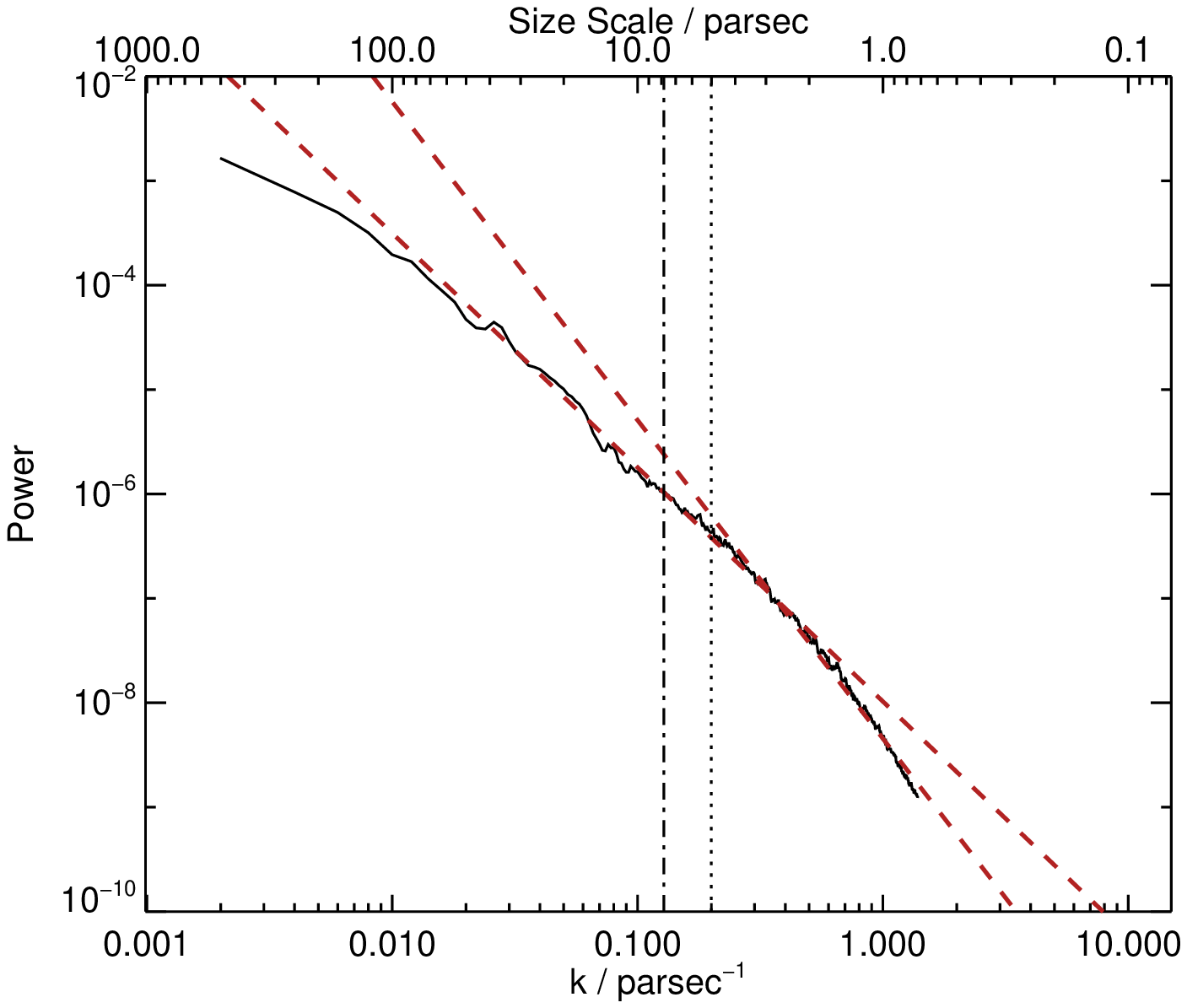} & \includegraphics[width=0.49\linewidth]{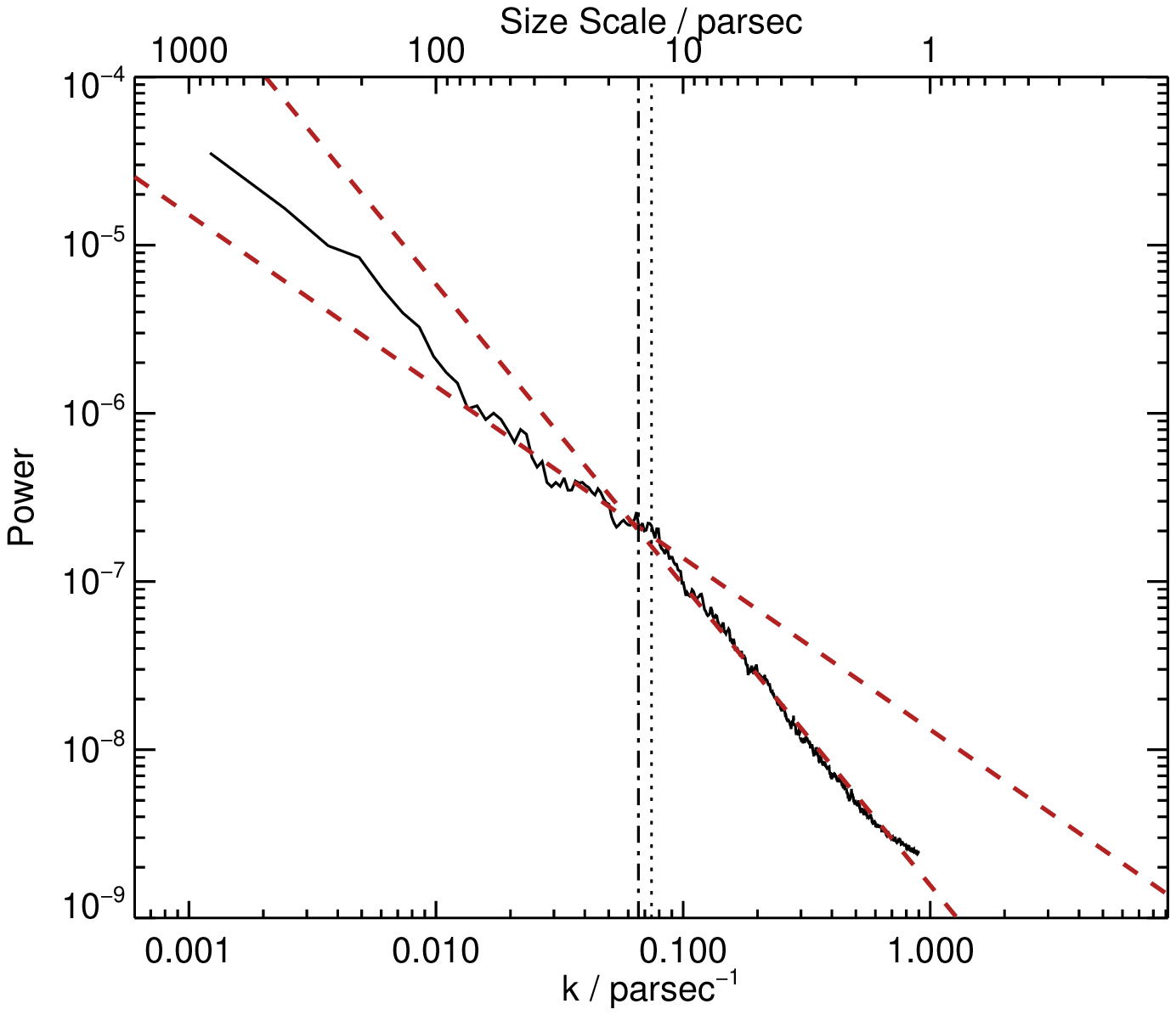} \\
\end{tabular}
\caption{Same as Fig.~\ref{JPSDataPS} but the power spectra of the CHIMPS data.}
\label{CHIMPSDataPS}
\end{figure*}

\subsection{Noise Maps}

\begin{figure*}
\begin{tabular}{ll}
\includegraphics[width=0.49\linewidth]{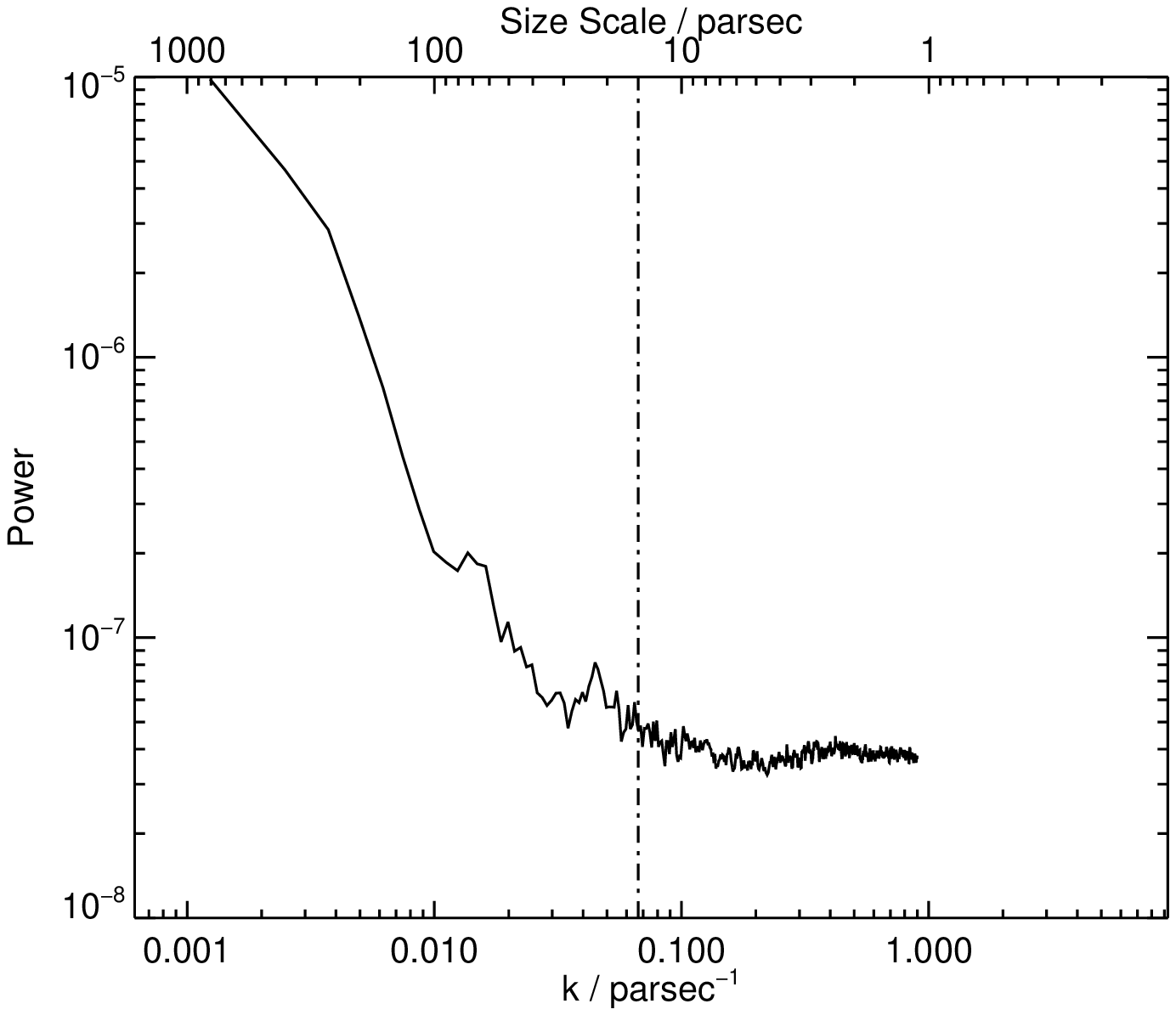} & \includegraphics[width=0.49\linewidth]{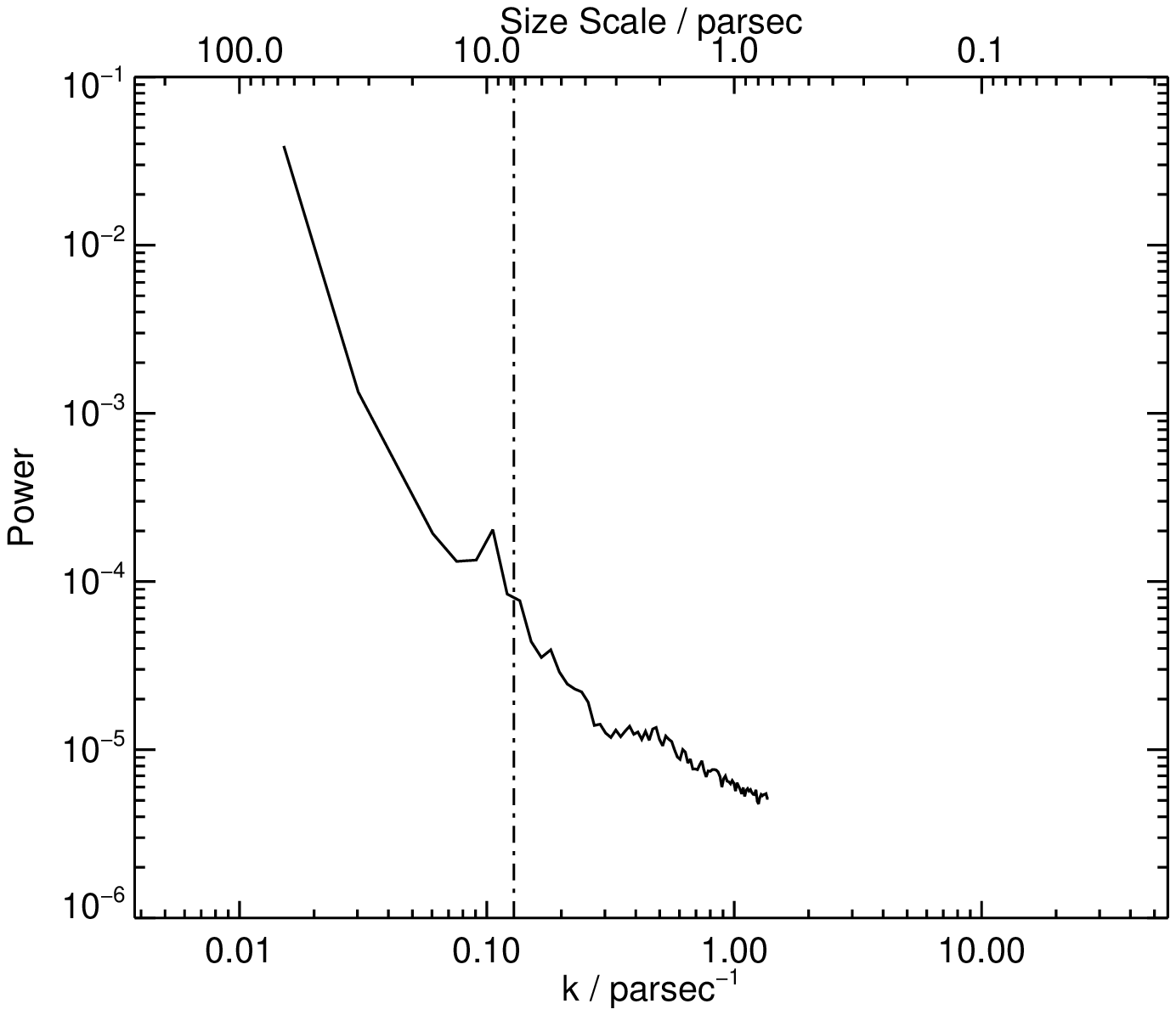} \\
\end{tabular}
\caption{Power spectra of the noise in the CHIMPS and JPS maps. Left panel: the noise in the CHIMPS data, scaled to the distances of the $\ell$\,=\,40$\degr$ field. The dash-dot line represents the break in the DGMF power spectrum. Right panel: Power spectrum of the noise in the JPS data, scaled to the distances of the $\ell$\,=\,30$\degr$ field. The dash-dot line represents the break in the CDR power spectrum.}
\label{noise}
\end{figure*}

\citet{Feddersen19} warned against interpreting the power spectra of molecular gas without first looking at the noise spectrum. In the $\ell$\,=\,40$\degr$ field, emission-free channels are found from 90 -- 110 km\,s$^{-1}$ and this range was collapsed to produce a noise image. The noise spectrum of the CHIMPS data (scaled to the distance of the $\ell$\,=\,40$\degr$ field) is shown in the left panel of Fig.~\ref{noise}. There is no peak or break at the scale of the DGMF break, which is indicated on the figure.

The JPS noise field was extracted from a smaller, emission-free region in the $\ell$\,=\,60$\degr$ JPS field. Since the JPS fields were observed and reduced in the same manner \citep{Eden17}, they can be assumed to be consistent with each other. The power spectrum of this noise field is shown in the right panel of Fig.~\ref{noise}, scaled to the $\ell$\,=\,30$\degr$ field. No peak or break is found consistent with the break in the CDR power spectrum.

\bsp
\label{lastpage}

\end{document}